\documentclass[12pt,english,floatfix,preprint,amsmath,amssymb,aps,prd,superscriptaddress, titlepage]{revtex4-2}

\usepackage{graphicx}
\usepackage[utf8]{inputenc}
\usepackage[T1]{fontenc} 
\usepackage{microtype} 
\usepackage{physics}    
\usepackage{tensor}
\usepackage{tensor} 
\usepackage[svgnames]{xcolor} 
\usepackage[obeyFinal, 
    color       = LightGray,
    bordercolor = LightGray,
    textsize    = footnotesize,
    figwidth    = 0.99\linewidth,
    prependcaption]{todonotes} 
\usepackage[allcolors=blue,colorlinks,pdftitle={Shadow and Quasinormal Modes of a Radiating Compact Object},pdfauthor={Caio C. Rodrigues and A. C. L. Santos}]{hyperref} 
\usepackage{orcidlink} 

\begin{document}
\title{Shadows and gravitational perturbations of black bounces}
\author{A. C. L. Santos\,{\orcidlink{0000-0002-0080-5699}}}
\email{alanasantos@fisica.ufc.br}
\affiliation{Universidade Federal do Ceará (UFC), Departamento de Física,
 Campus do Pici, Fortaleza- CE, C.P. 6030, 60455-760- Brazil}
\affiliation{Instituto de Física Corpuscular (IFIC), CSIC‐Universitat de Val\`{e}ncia, Spain} 
\author{L. A. Lessa\, {\orcidlink{0009-0009-1961-9819} }}
\email{leandrophys@gmail.com}
\affiliation{Faculdade de Física, Campus Salinópolis, Universidade Federal do Pará, 68721-000, Salinópolis, Pará, Brazil}
\author{R. V. Maluf\,{\orcidlink{0000-0002-9952-4589}}}
\email{r.v.maluf@fisica.ufc.br}
\affiliation{Universidade Federal do Ceará (UFC), Departamento de Física,
 Campus do Pici, Fortaleza- CE, C.P. 6030, 60455-760- Brazil}
\author{G. J. Olmo\,{\orcidlink{0000-0001-9857-0412}}}
\email{gonzalo.olmo@uv.es}
\affiliation{Instituto de Física Corpuscular (IFIC), CSIC‐Universitat de Val\`{e}ncia, Spain}  
\affiliation{Universidade Federal do Ceará (UFC), Departamento de Física,
 Campus do Pici, Fortaleza- CE, C.P. 6030, 60455-760- Brazil}

\date{\today}

\begin{abstract}
We investigate axial gravitational perturbations and shadow formation in black-bounce geometries supported by anisotropic fluids within general relativity. Two distinct classes of solutions are analyzed: a symmetric Simpson--Visser-like black bounce and an asymmetric deformation of the Reissner--Nordström metric, which can feature either a bounded interior region or an unbounded wormhole-like extension. Our findings indicate that configurations with horizons, whether symmetric or asymmetric, exhibit single-barrier effective potentials and quasinormal frequencies nearly indistinguishable from those of standard black holes. In contrast, horizonless symmetric configurations may develop multiple potential barriers, giving rise to gravitational-wave echoes whose amplitude and separation depend sensitively on the bounce parameter $a$ and the density parameter $\rho_0$. Horizonless asymmetric solutions, on the other hand, exhibit regularized effective potentials that do not produce echoes and remain phenomenologically close to the Reissner--Nordström case. From the optical perspective, both classes of solutions with horizons exhibit a single relevant photon sphere and shadow profiles nearly indistinguishable from those of their black-hole counterparts, while horizonless symmetric configurations can sustain multiple photon rings, whose number and relative brightness vary with the bounce parameters. Horizonless asymmetric configurations may instead support either a single photon ring or none. Finally, we recover the standard eikonal correspondence between quasinormal modes and the instability of circular null geodesics for configurations with horizons. For horizonless geometries, however, the emergence of multiple potential barriers prevents a straightforward extension of this relation, although the strong similarity between the wave and optical sectors suggests that a meaningful phenomenological correspondence may still persist.
\end{abstract}

\maketitle

\section{Introduction}\label{sec: intro}
General relativity (GR) provides a consistent theoretical framework that explains with remarkable precision classical deviations from Newtonian predictions, such as the perihelion precession of Mercury and the gravitational deflection of light \cite{Will:2014kxa}. Nonetheless, the theory remains incomplete, as it faces unresolved fundamental problems, including incompatibilities between theoretical predictions and cosmological observations \cite{Peebles:2002gy, Planck:2018vyg, Bull:2015stt, Verde:2019ivm, Clifton:2011jh}, as well as the inevitable formation of singularities \cite{Penrose:1964wq,Geroch:1968ut}. The formulation of a fundamental theory of gravitation therefore remains one of the central open problems in both theoretical and experimental physics.

Due to the limited availability of observational data in the strong-gravity regime, considerable attention has recently focused on two main probes: gravitational waves from binary black hole \cite{LIGOScientific:2016aoc,LIGOScientific:2016sjg,LIGOScientific:2017bnn,LIGOScientific:2017ycc} and neutron star mergers \cite{LIGOScientific:2017vwq}, and the black hole shadows imaged in M87 \cite{EventHorizonTelescope:2019dse,EventHorizonTelescope:2019uob,EventHorizonTelescope:2019jan,EventHorizonTelescope:2019ths,EventHorizonTelescope:2019pgp,EventHorizonTelescope:2019ggy} and SgrA* \cite{EventHorizonTelescope:2022wkp,EventHorizonTelescope:2022apq,EventHorizonTelescope:2022wok,EventHorizonTelescope:2022exc,EventHorizonTelescope:2022urf,EventHorizonTelescope:2022xqj}. The gravitational waves observed in these merging processes exhibit three distinct phases: inspiral, merger, and ringdown. The latter is characterized by quasinormal modes (QNMs), in which the real part encodes the oscillation frequency and the imaginary part governs the exponential damping of the signal. This frequency is essentially determined by the fundamental parameters of the compact object — mass, charge, and angular momentum \cite{Berti:2009kk}. Consequently, it has stimulated increasing attention to quasinormal modes in recent studies \cite{Lo:2025njp,Pezzella:2024tkf,Konoplya:2025afm,Konoplya:2025mvj,Antoniou:2024jku,Macedo:2016wgh}. 

On the other hand, since the advent of horizon-scale imaging, and indeed even earlier, considerable theoretical effort has been devoted to modeling such observations within different spacetime geometries \cite{Olmo:2025ctf,Koch:2025gaw,Olmo:2023lil}. Such analyses are primarily based on solving the null geodesic equations that govern photon trajectories, taking into account different accretion disk models \cite{Guerrero:2022qkh}. The main challenges include the pronounced sensitivity of the shadow to the adopted accretion disk model, as well as the extremely subtle distinctions produced by different spacetime geometries \cite{Guerrero:2021ues}. A well-established connection between these two phenomenological approaches exists for solutions with horizons, since in the eikonal limit a direct relation emerges between QNM frequencies and the photon sphere \cite{Cardoso:2008bp}. However, the horizonless case requires further investigation \cite{Duran-Cabaces:2025sly}.

Mapping, and consequently distinguishing, geometries through their phenomenological imprints is one of the approaches employed to address the previously mentioned problems. From a theoretical perspective, several proposals have been advanced. An excellent compilation can be found in \cite{CANTATA:2021asi}. In particular, black bounces are a class of regular spacetimes introduced by Simpson and Visser \cite{Simpson:2018tsi}, characterized by an areal function that develops a non-vanishing minimum, the so-called bounce. This feature (usually) ensures geodesic completeness and finite curvature invariants, thus removing the central singularity present in classical black holes \cite{Carballo-Rubio:2019fnb,LimaJunior:2025uyj}.

The possibility of the existence of such structures has been investigated in several scenarios in recent years, including linear electrodynamics \cite{Junior:2026ism}, dark matter halos \cite{B:2026jtm}, self-interacting 3-form fields \cite{Lobo:2026flk}, and quantum corrections \cite{Alencar:2026qeb}. Since these geometries generally interpolate smoothly between black holes and wormholes, their phenomenology has been widely explored, including gravitational lensing \cite{Nascimento:2020ime, Cheng:2021hoc, Islam:2021ful, Ghosh:2022mka}, black hole shadows \cite{Guerrero:2021ues, Olmo:2023lil, Guerrero:2022qkh}, and gravitational-wave echoes \cite{Silva:2024fpn, Ou:2021efv, Yang:2021cvh}, among other phenomena \cite{Franzin:2022iai, Zhang:2022zox, Yang:2022ryf, Javed:2023iih, daSilva:2023jxa, Ovejero-Bermudez:2026jja, Siqueira:2026uzu}.

Considering general relativity, it was shown in \cite{Lessa:2024erf} that such compact objects can be supported by anisotropic fluids through the imposition of a minimum for the areal function in a general spherically symmetric geometry, therefore relaxing the usual requirement $g_{tt}=g_{rr}^{-1}$. One of the geometries constructed by the authors corresponds to the symmetric Simpson–Visser-like, in which the radial coordinate is modified as $r\to \sqrt{a^2+r^2}$. Another geometry represents an asymmetric black bounce, constructed as a deformation of the Reissner–Nordström metric. In this case, the exterior region remains identical to the charged black hole, but the interior develops a nontrivial structure governed by a parameter $l_0$. For $l_0>0$, the bounce leads to a bounded universe hidden inside the horizon, whereas for $l_0 <0$, the geometry extends into an unbounded wormhole-like region. 

Our analysis of scalar perturbations in these geometries shows that horizon configurations, either symmetric or asymmetric, are characterized by an effective potential with a single barrier, leading to a QNM spectrum that is basically indistinguishable from that of standard black holes. For horizonless solutions, the symmetric case may develop multiple potential barriers, producing gravitational wave echoes whose properties depend on the parameters $a$ and $\rho_0$. In contrast, horizonless asymmetric configurations do not generate echoes and only exhibit small deviations from the Reissner–Nordström geometry, which remain observationally indistinguishable \cite{Santos:2025xbk}.

To contribute to the study of the relation between quasinormal modes and shadows in horizonless configurations, and to clarify previous results obtained from scalar perturbations by properly examining the propagation of gravitational waves and photons, in this work we investigate axial gravitational perturbations and analyze the corresponding shadow formation. It is worth mentioning that this question was initially investigated by the authors in Ref. \cite{Eiroa:2025mws}. Here, however, we focus primarily on the astrophysical contributions, including those associated with the accretion disk.

The paper is organized as follows. In Sec. \ref{secII}, we introduce the symmetric and asymmetric black-bounce geometries obtained from GR coupled to an anisotropic fluid. Sec. \ref{secIII} describes the formalism for axial perturbations in these backgrounds, including the derivation of the effective potential and the numerical scheme employed. Sec. \ref{secV} is dedicated to the modeling of black hole shadows. Sec. \ref{secVI} provides a discussion of our results, with emphasis on the differences between horizon and horizonless configurations. Sec. \ref{seccorres} we analyze the correspondence between quasinormal modes and black hole shadows. Finally, Sec. \ref{secVII} summarizes our conclusions and future perspectives.

\section{BOUNCE MODELS}\label{secII}
In this section, we analyze two distinct classes of regular black-bounce solutions supported by anisotropic fluids, i.e., $T^{\mu}{}_{\nu} =\frac{1}{8\pi G}\text{diag}(-\rho,p_r,p_t,p_t)$, where $\rho$, $p_r$, and $p_t$ represent the energy density, the radial pressure and the tangential pressure of the fluid, respectively. The geometries are constructed assuming the equation of state $\rho + p_r = 0$ and $p_t = \omega \rho$, which can be naturally satisfied by nonlinear electrodynamics sources for example, and a general spherically symmetric spacetime \cite{Lessa:2024erf}
\begin{equation}\label{eq1}
ds^2 = - A(r) dt^2 + \frac{1}{B(r)} dr^2 + \Sigma(r)^2 d\Omega^2,     
\end{equation}
where the area function $\Sigma(r)$ is assumed to be non-monotonic and to have a minimum. Under these assumptions, it is straightforward to show that the area function is given by
\begin{equation}\label{sigmageral}
    \Sigma(r)={\Sigma_0}/{\rho(r)^{\frac{1}{2(\omega+1)}}} \ ,
\end{equation}
where $\Sigma_0$ is an integration constant. The determination of the metric functions $A(r)$ and $B(r)$ follows from the Einstein field equations 
\begin{equation}
R_{\mu\nu} - \frac{1}{2}Rg_{\mu\nu} =  8\pi G T_{\mu\nu}.   
\end{equation}
 
\subsection{Model I: Symmetric Black Bounce}

The first configuration corresponds to a symmetric black bounce, in which the areal function is defined as \cite{Lessa:2024erf}
\begin{equation}\label{symetric}
\Sigma_I(r) = \sqrt{a^2 + r^2},
\end{equation}
featuring a regular minimum at $r = 0$, where $\Sigma(0) = a$ denotes the radius of the throat. For this choice, the energy density is given by
\begin{equation}
\rho(r) = \frac{\rho_0}{(a^2 + r^2)^{1 + \omega}},
\end{equation}
and the metric functions take the form
\begin{align}
A_I(r) &= 1 - \frac{2M}{\sqrt{a^2 + r^2}} + \frac{\rho_0}{(2\omega - 1)(a^2 + r^2)^{\omega}}, \\
B_I(r) &= \left(1 + \frac{a^2}{r^2}\right) A_I(r),
\end{align}
where $M$ is an integration constant that represents the asymptotic ADM mass. This geometry generalizes the Kiselev solution and reproduces the Reissner--Nordström form for $\omega = 1$. The structure of horizons depends on the model parameters and includes regular black holes, extremal configurations and traversable wormholes.

\subsection{Model II: Asymmetric (Un)Bounded Black Bounce}

The second solution features an asymmetric geometry defined only for $r \geq 0$, and supported by a modified areal function \cite{Lessa:2024erf}:
\begin{equation}\label{asymetric}
\Sigma_{II} (r) = \left(|l_0|e^{\frac{l_0 \gamma+\text{r0}}{l_0}}e^{-\chi(r)}\right)^{\alpha^2}-(l_0+r_0),    
\end{equation}
where
\begin{equation}
\chi(r) = \mathrm{Ei}\left(-\frac{l_0}{r}\right) + \frac{r_0}{l_0} e^{-l_0/r},
\end{equation}
with $\mathrm{Ei}(r)$ denoting the exponential integral, $\alpha$, $r_0$, $l_0$ and $\gamma$ are constants. For $l_0 > 0$, this configuration admits a \textit{bounded internal region}, where the minimal surface is located at $r = r_0$, and the areal radius attains a constant value as $r \to 0^+$. For $l_0<0$,  the resulting structure would not be bounded and the internal region represents an asymptotically Minkowskian space-time. The energy density reads
\begin{equation}
\rho(r) = \tilde{\rho}_0 \left( \frac{\tilde{\Sigma}_0}{\Sigma_{II}(r)} \right)^{2(1 + \omega)},
\end{equation}
and the metric functions become
\begin{equation}
A_{II}(r) = 
\begin{cases}
\displaystyle 
1 - \frac{2\tilde{M}}{\Sigma_{II}(r)}
 - \frac{\tilde{\rho}_0\,\tilde{\Sigma}_0^{\,2}}{(1-2\omega)} 
 \left( \frac{\tilde{\Sigma}_0}{\Sigma_{II}(r)} \right)^{2\omega},
& \text{if } \omega \neq \tfrac{1}{2}, \\[1.2em]
\displaystyle 
1 - \frac{2\tilde{M}}{\Sigma_{II}(r)}
 - \frac{\tilde{\rho}_0\,\tilde{\Sigma}_0^{\,3}}{\Sigma_{II}(r)}
 \ln\!\left( \frac{\Sigma_{II}(r)}{\tilde{\Sigma}_0} \right),
& \text{if } \omega = \tfrac{1}{2}.
\end{cases}
\end{equation}
\begin{equation}
B_{II}(r) = \frac{A_{II}(r)}{\Sigma_{II}'(r)^2},    
\end{equation}
where $\tilde{M}$ is the ADM mass.

\section{AXIAL GRAVITATIONAL PERTURBATIONS}\label{secIII}
Based on the line element~(\ref{eq1}), we will adopt an approach similar to that developed in \cite{Duran-Cabaces:2025sly}, introducing linear perturbations around the background metric, $\bar{g}_{\mu\nu}$
\begin{equation}
g_{\mu\nu} = \bar{g}_{\mu\nu} + h_{\mu\nu},   
\end{equation}
with  $\lvert h_{\mu\nu}\rvert \ll 1$. Considering that the energy–momentum tensor undergoes a perturbation of the same order, $T_{\mu\nu}= \bar{T}_{\mu\nu}+\delta T_{\mu\nu}$, the perturbed Einstein field equations can be written as
\begin{align}\label{eqlinear}
\nabla^{\lambda} \nabla_{\mu} h_{\lambda\nu} + \nabla^{\lambda} \nabla_{\nu} h_{\lambda\mu} - \Box h_{\mu\nu} - \nabla_{\nu}\nabla_{\mu}h + \Box h \bar{g}_{\mu\nu} \nonumber\\ - \nabla_{\alpha}\nabla_{\beta} h^{\alpha\beta}\bar{g}_{\mu\nu} - \bar{R}h_{\mu\nu} + \bar{g}_{\mu\nu}h^{\alpha\beta}\bar{R}_{\alpha\beta} = 16 \pi G  \delta T_{\mu\nu}, 
\end{align}
where $\bar{R}$ and $\bar{R}_{\alpha\beta}$ are the background Ricci curvature scalar and Ricci tensor, respectively. The spherical symmetry of the background metric ensures that the axial (odd-parity) and polar (even-parity) sectors remain decoupled under parity transformations, allowing each type of perturbation to be treated independently. Restricting our analysis to the axial sector and adopting the Regge–Wheeler gauge, we decompose the metric and energy–momentum tensor perturbations into tensor spherical harmonics \cite{Duran-Cabaces:2025sly}
\begin{equation}
h_{\mu\nu} = \sum_{l,m}
\left(
\begin{array}{cccc}
0 & 0 & \frac{h_{lm}^{Bt}(t,r)}{\sin\theta} \partial_{\varphi} & -\sin\theta\, h_{lm}^{Bt}(t,r)\, \partial_{\theta} \\
0 & 0 & \frac{h_{lm}^{B1}(t,r)}{\sin\theta} \partial_{\varphi} & -\sin\theta\, h_{lm}^{B1}(t,r)\, \partial_{\theta} \\
\frac{h_{lm}^{Bt}(t,r)}{\sin\theta} \partial_{\varphi} & \frac{h_{lm}^{B1}(t,r)}{\sin\theta} \partial_{\varphi} & 0 & 0 \\
-\sin\theta\, h_{lm}^{Bt}(t, r)\, \partial_{\theta} & -\sin\theta\, h_{lm}^{B1}(t,r)\, \partial_{\theta} & 0 & 0
\end{array}
\right)
Y_{lm}(\theta,\varphi),
\end{equation}
\begin{equation}\label{mperturbation}
\delta T_{\mu\nu} = \sum_{l,m}
\left(
\begin{array}{ccccc}
0 & 0 & \frac{s_{lm}^{Bt}(t, r)}{\sin\theta}\,\partial_{\varphi} & -\sin\theta\, s_{lm}^{Bt}(t, r)\,\partial_{\theta}  \\
0 & 0 & \frac{s_{lm}^{B1}(t,r)}{\sin\theta}\,\partial_{\varphi} & -\sin\theta\, s_{lm}^{B1}(t,r)\,\partial_{\theta}  \\
\frac{s_{lm}^{Bt}(t, r)}{\sin\theta}\,\partial_{\varphi} & \frac{s_{lm}^{B1}(t, r)}{\sin\theta}\,\partial_{\varphi} & -\dfrac{1}{\sin\theta}\, s_{lm}^{B2}(t,r)\, X & \sin\theta\, s_{lm}^{B2}(t,r)\, W \\
-\sin\theta\, s_{lm}^{Bt}(t,r)\,\partial_{\theta} & -\sin\theta\, s_{lm}^{B1}(t,r)\,\partial_{\theta} & \sin\theta\, s_{lm}^{B2}(t,r)\, W & \sin\theta\, s_{lm}^{B2}(t,r)\, X 
\end{array}
\right)
Y_{lm}(\theta,\varphi),
\end{equation}
where,  $h^{Bt}_{lm}(t,r), h^{B1}_{lm}(t,r), s^{Bt}_{lm}(t,r), s^{B1}_{lm}(t,r),  s^{B2}_{lm}(t,r)$ are the coefficients of the expansion in tensor harmonics and 
\begin{equation}
X = 2\,\partial_{\theta}\partial_{\varphi} - 2\cot\theta\,\partial_{\varphi},    
\end{equation}
\begin{equation}
W = \partial_{\theta}^{2} - \cot\theta\,\partial_{\theta} - \frac{1}{\sin^{2}\theta}\,\partial_{\varphi}^{2}.   
\end{equation}
The three equations involving the axial perturbations are obtained from the components ($t \phi$), ($r \phi$) and ($\theta \phi$)
\begin{equation}\label{eq:1}
\frac{B(r)\,\partial_{r}\bigl(h_{lm}^{B1}(t,r)\,A(r)\bigr) +A(r)\,\partial_{r}\bigl(h_{lm}^{B1}(t,r)\,B(r)\bigr)}{2} - \partial_{t}h_{lm}^{Bt}(t,r) =  -16\pi G s_{lm}^{B2}(t,r)A(r),  
\end{equation}
\begin{align}\label{eq:Bt_equation}
&\left[\partial_r + 2\frac{\Sigma'(r)}{\Sigma(r)} -\frac{1}{2}\left(\frac{A'(r)}{A(r)} -\frac{B'(r)}{B(r)}\right) \right] \partial_{t}h_{lm}^{B1}(t,r) - \left[\partial^2_{r} -\frac{1}{2}\left(\frac{A'(r)}{A(r)} -\frac{B'(r)}{B(r)}\right)\partial_{r} \right.\\ \nonumber 
&+ \left. 2\frac{A'(r)\Sigma'(r)}{A(r)\Sigma(r)} +  \frac{\bar R}{B(r)} - \frac{l(l+1)}{B(r)\Sigma(r)^2} \right] h_{lm}^{Bt}(t,r) = 16\pi G \frac{s_{lm}^{Bt}(t,r)}{B(r)}
\end{align}
\begin{align}\label{eq:B1_eom}
&\partial_{t}^{2}h_{ln}^{B1}(t,r) - \left(\partial_r - 2\frac{\Sigma'(r)}{\Sigma(r)}\right)\partial_{t}h_{lm}^{Bt}(t,r) + \left[\frac{A(r)\,l(l+1)}{\Sigma(r)^{2}} - \bar{R}A(r) -A(r)B(r) \frac{\partial_{r}^2(\Sigma(r)^2)}{\Sigma(r)^2} \right. \nonumber \\
&\left. - \partial_{r} (A(r)B(r))\frac{\Sigma'(r)}{\Sigma(r)}\right]h_{lm}^{B1}(t,r) = 16\pi G s_{lm}^{B1}(t,r) A(r).
\end{align}
Substituting Eq.~(\ref{eq:1}) into Eq.~(\ref{eq:B1_eom}) and defining the tortoise coordinate $r^{*}(r)$ and the Regge-Wheeler variable $Q_{lm}(t,r)$:
\begin{equation}
\frac{\partial r^*}{\partial r} = \frac{1}{\sqrt{A(r)B(r)}},
\end{equation}
\begin{equation}
Q_{l m}(t,r) = \frac{\sqrt{A(r)\,B(r)}}{\Sigma(r)} \;h^{B1}_{l m}(t,r),    
\end{equation}
the perturbation equations can be reformulated as 
\begin{equation}\label{eq2}
\left(\frac{\partial^2}{ \partial{r^*}^2} - \frac{\partial^2}{ \partial t^2} - V_{lm}(r(r^*)) \right) {Q}_{lm}(r,t) 
= {S}^{ax}_{lm}(r,t),
\end{equation}
where,
\begin{equation}
S^{\mathrm{ax}}_{lm}(t,r) = -16 \pi 
\frac{\sqrt{A(r)B(r)}}{\Sigma(r)}
\left[
s^{B1}_{lm}(t,r) A(r) 
+ \left( \partial_r - 2 \frac{\Sigma'(r)}{\Sigma(r)} \right) 
s^{B2}_{lm}(t,r) A(r)
\right],    
\end{equation}
\begin{equation}
V_{lm}(r(r^*)) = A(r) \left[\frac{l(l+1)}{\Sigma(r)^2} - \overline{R} 
-\frac{3 B'(r) \Sigma '(r)}{2 \Sigma (r)}-\frac{3 B(r) \Sigma ''(r)}{\Sigma (r)}-\frac{3 B(r) A'(r) \Sigma '(r)}{2A(r) \Sigma (r)}\right].   
\end{equation}
Finally, performing a Fourier transform with respect to the time, we obtain
\begin{equation}\label{FourierTransform}
\left(\frac{\partial^2}{ \partial{r^*}^2} + w^2 - V_{lm}(r(r^*)) \right) \tilde{Q}_{lm}(r,w) 
= \tilde{S}^{ax}_{lm}(r,w).    
\end{equation}
Perturbing now at first order the matter sector given by
\begin{equation}
T_{\mu\nu} = \big(\rho(r) + p_{t}(r)\big) u_\mu u_\nu 
+ \big(p_{r}(r) - p_{t}(r)\big) x_\mu x_\nu 
+ p_{t}(r) g_{\mu\nu},
\end{equation}
we obtain,
\begin{align}
\delta T_{\mu\nu} &= \big(\bar{\rho}(r) + \bar{p}_{t}(r)\big) (\delta u_{\mu} \bar{u}_{\nu} + \bar{u}_{\mu} \delta u_{\nu}) + \big(\bar{p}_{r}(r) - \bar{p}_{t}(r)\big) (\delta x_{\mu} \bar{x}_{\nu} + \bar{x}_{\mu} \delta \bar{x}_{\nu}) \nonumber \\
&+ \bar{p}_{t}(r) h_{\mu\nu}
+ \big(\bar{g}_{\mu\nu} + \bar{u}_{\mu} \bar{u}_{\nu} - \bar{x}_{\mu} \bar{x}_{\nu}\big)\, \delta p_{t} + \bar{u}_{\mu} \bar{u}_{\nu} \delta \rho
    + \bar{x}_{\mu} \bar{x}_{\nu} \delta p_{r},
\end{align}
with $u_\mu$ a unitary time-like four-vector, $x_\mu$ a unitary space-like four-vector and  $\bar{T}^{\mu}{}_{\nu} =\text{diag}(-\bar{\rho},\bar{p}_r,\bar{p}_t,\bar{p}_t)$. Expanding the axial sector of $u_\mu$ and $x_\mu$ in vector spherical harmonics, and $\rho(r)$, $p_r(r)$, and $p_t(r)$ in scalar spherical harmonics, comparing with (\ref{mperturbation}), performing a Fourier transform  and considering a scalar test field as a matter perturbation, which couples exclusively to the polar sector, leaving the axial sector unaffected, we can rewrite Eq.~(\ref{FourierTransform}) in the form (see \cite{Duran-Cabaces:2025sly} for details)
\begin{equation}\label{eq3}
\left(\frac{\partial^2}{ \partial{r^*}^2} + w^2 - V^{\mathrm{RW}}_{lm}(r(r^*))\right)\Psi_{lm}(r,w)
= 0,  
\end{equation}
and,
\begin{equation}
V_{lm}^{RW}(r(r^*)) = A(r) \left[\frac{l(l+1)-2}{\Sigma(r)^2}-\frac{B'(r) \Sigma '(r)}{2 \Sigma (r)}-\frac{B(r) \Sigma ''(r)}{\Sigma (r)}+\frac{2 B(r) \Sigma '(r)^2}{\Sigma (r)^2}-\frac{B(r) A'(r) \Sigma '(r)}{2 A(r) \Sigma (r)}\right],
\end{equation}where \(V^{RW}_{lm}\) is the Regge–Wheeler potential.

\subsection{TIME DOMAIN}\label{secIV}
To solve Eq.~(\ref{eq3}), we adopt the numerical method developed in Ref.~\cite{Gundlach:1993tp}, defining the null coordinates
$v \equiv t+r^*$ and $u \equiv t -r^*$, which allows us to rewrite in the form
\begin{equation}
\left( 4 \frac{\partial^2}{\partial u \partial v} + V_{lm}^{RW}(u,v) \right) \Psi(u,v) = 0.
\end{equation}
Applying the finite difference method, one obtains
\begin{equation}
\Psi_N = \Psi_E + \Psi_W - \Psi_S - \frac{h^2}{8} V_{lm}^{RW}(S)(\Psi_W + \Psi_E) + O(h^4),
\end{equation}
with $S = (u,v)$, $W = (u + h,v)$, $E = (u, v+h)$ and $N = (u+h,v+h)$, where $h$ denotes the step size between two neighboring grid points. Given initial data on the null surface $v = v_0$, which are kept constant along $u = u_0$, the time evolution of the field can be computed iteratively. We will consider a Gaussian wave packet as the initial profile, 
\begin{equation}
\Psi (0,v) = A \text{exp}[-(v-v_c)^2/\sigma].   
\end{equation}
We will observe the field at $r^* = 10r_h$, with $h=0.1$, $A=1, v_c = 20$ and $\sigma = 9$. To find the dominant QNM that appears in the ringdown phase when the above wave packet interacts with the potential barrier, we can perform a fitting of the numerical integration data considering a linear expansion of the form
\begin{equation}
\Psi = \sum_{i=1}^n A_i e^{Im[w_i]t} \cos (Re[w_i]t + c_i),     
\end{equation}
which allows us to find the coefficients $A_i, w_i$ (real and imaginary parts) and $c_i$. 

\section{SHADOW}\label{secV}
Considering null-particle motion in the spacetime described by Eq. (\ref{eq1}), and assuming without loss of generality that the motion is confined to the equatorial plane, \(\theta=\pi/2\), we obtain
\begin{equation}\label{eq4}
\frac{A(r)}{B(r)} \left(\frac{dr}{d \bar{\lambda}} \right)^2 = \frac{1}{b^2} - V_{Sh}(r);\;\;V_{sh}(r) = \frac{A(r)}{\Sigma^2(r)},   
\end{equation}
where $\bar{\lambda}$ is the affine parameter, $b \equiv \frac{L}{E}$ is the impact parameter and $V_{Sh}(r)$ is the shadow potential. The critical impact parameter $b_c$ is defined as the value of $b$ for which the effective potential exhibits a maximum, and therefore satisfies the following conditions
\begin{equation}
b_c^2 = \frac{1}{V_{Sh}(r_{ps})}; \quad 
V'_{Sh}(r)\big|_{r=r_{ps}} = 0; \quad 
V''_{Sh}(r)\big|_{r=r_{ps}} < 0,    
\end{equation}
where $r_{ps}$ denotes the radius of the \emph{photon sphere}. These conditions imply
\begin{equation}
b_c = \frac{\Sigma(r_{ps})}{\sqrt{A(r_{ps})}}.
\end{equation}
Also, it is  convenient to rewrite (\ref{eq4}) as the variation of the azimuthal angle with respect to the radial coordinate
\begin{equation}
\frac{d\phi}{dr} = \mp \frac{b}{\Sigma^2(r)} \sqrt{\frac{A(r)/B(r)}{1-\frac{b^2A(r)}{\Sigma^2(r)}}}.    
\end{equation}
where $\mp$ represents ingoing (outgoing) geodesics. The photons with impact parameters $b \gtrsim b_{c}$ experience strong gravitational deflection and may execute several half-turns around the black hole before escaping. Labeling by $n$ the number of intersections with the equatorial plane, $n=1$ corresponds to the direct emission from the disk, while $n=2$ describes photons that cross the disk twice and generate the \emph{lensing ring}. Higher-order images with $n=3,4,\ldots$ are associated with photons that perform multiple half-turns around the black hole before reaching the observer. As a result, the observed image is dominated by a bright direct ring ($n=1$), accompanied by a weaker lensing ring ($n=2$) and a sequence of higher-order photon-ring images ($n \geq 3$).

On the other hand, to model the contribution of the accretion disk, we will consider an optically and infinitesimally thin geometrical shape, which leads to the total observed intensity \cite{Olmo:2023lil}
\begin{equation}
I_{\text{ob}}(r) = \int d\nu_{o} I_{\nu_{o}} 
= \int (g\, d\nu_{e}) \,(g^{3} I_{\nu_{e}}) 
= g^{4} I(r),
\end{equation}
where $\nu_e$ and $\nu_o$ are the photon’s frequency in the emission and observer’s frames, respectively, while $I_{\nu_e}$ 
and $I_{\nu_o}$ are the corresponding intensities, with $g = \frac{\nu_{0}}{\nu_e}$. Considering the subsequent intersections with the accretion disk, we found
\begin{equation}
I^{ob}_{total} = \sum_{n=1}^{i} A^{2}(r)\, I(r).
\end{equation}
Finally, with respect to the choice of emission profiles, we consider the unbounded Johnson distribution
\begin{equation}
I(r;\gamma_J,\mu,\sigma_J) =
\frac{\exp\!\left(-\tfrac{1}{2}\left(\gamma_J + \operatorname{arcsinh}\!\left(\tfrac{r-\mu}{\sigma_J}\right)\right)^{2}\right)}
{\sqrt{(r-\mu)^{2} + \sigma_J^{2}}}.    
\end{equation}
with $\gamma_J = -2, \mu = r_{ISCO}$ and $\sigma_J = \frac{M}{4}$ (GML3 \cite{daSilva:2023jxa}). The parameter $\gamma$ governs the growth rate of the intensity profile from infinity down to its maximum; $\mu$ introduces a translation that shifts the profile to a desired position, in our case for the Innermost Stable Circular Orbit (ISCO) of the compact object and $\sigma$ regulates the dilation of the profile. In what follows, we investigate the formation of shadows within the same range of parameters previously analyzed for QNMs, in order to assess whether this effect also extends to photon propagation.

\section{RESULTS}\label{secVI}
\begin{figure}[t]
\begin{center}
\begin{tabular}{ccc}
\includegraphics[height=5.0cm]{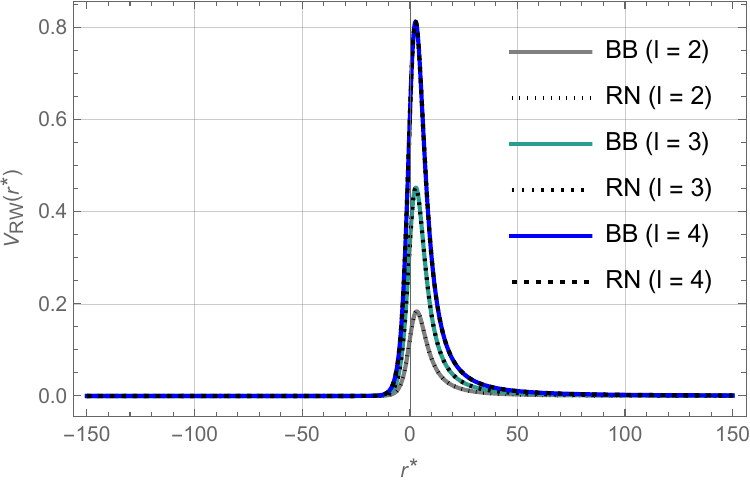} \includegraphics[height=5.0cm]{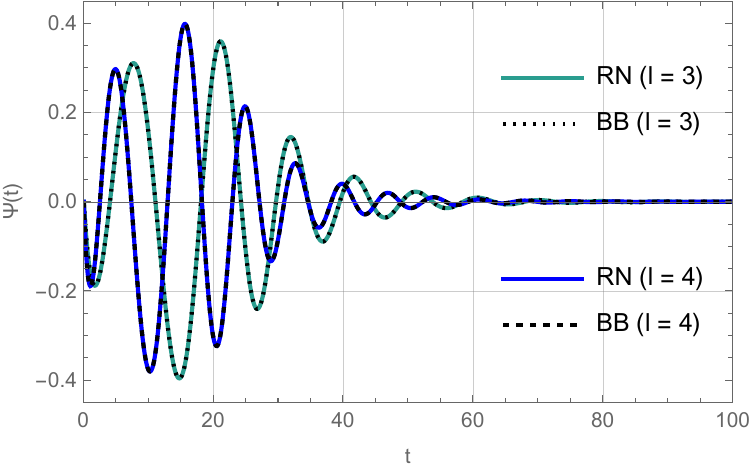}\\ 
(1a) \hspace{7.5 cm}(1b)
\end{tabular}
\end{center}
\caption{The Regge-Wheeler potential (1a) and time evolution of the axial gravitational perturbation (1b) for the symmetric black bounce of Eq. (\ref{symetric}) with $l= 3, l=4, a = M = \omega = 1$ and $\rho_0= 0.5$. The observation point is located at $r^* = 100$. \label{Figure1}}
\end{figure}

\begin{figure}[t]
\begin{center}
\begin{tabular}{ccc}
\includegraphics[height=4cm]{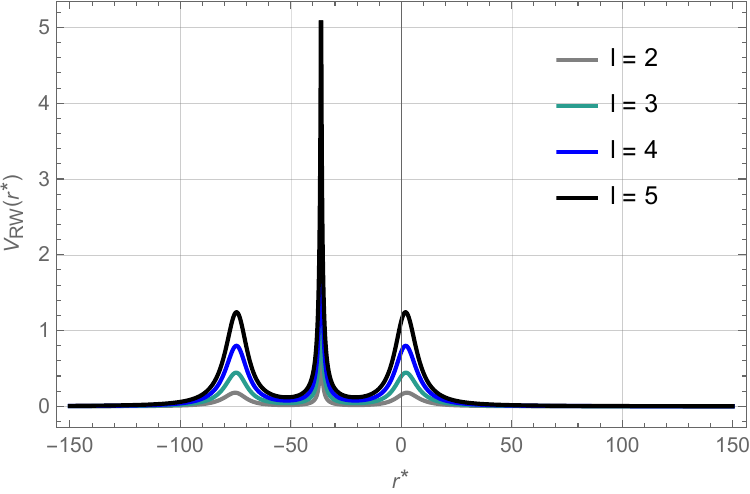} \includegraphics[height=4cm]{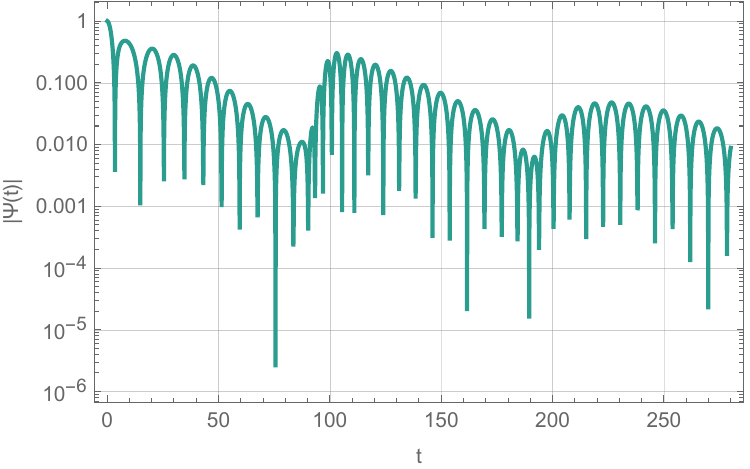}\\ 
(2a) \hspace{5.5 cm}(2b)\\
\includegraphics[height=4cm]{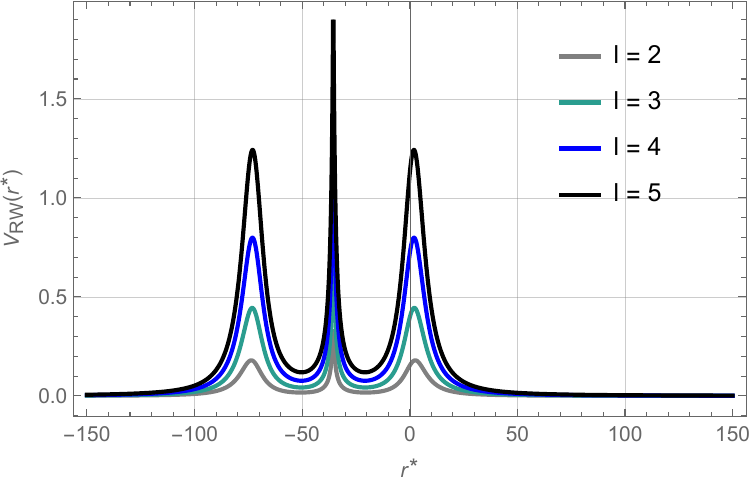} \includegraphics[height=4cm]{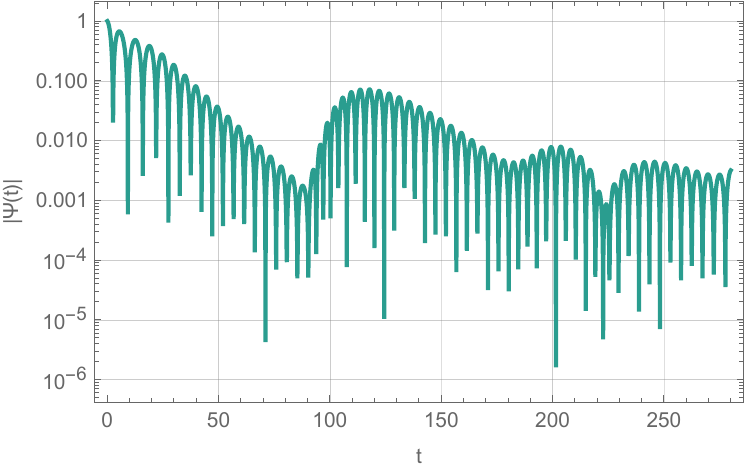}\\ 
(2c) \hspace{5.5 cm}(2d)\\
\includegraphics[height=4cm]{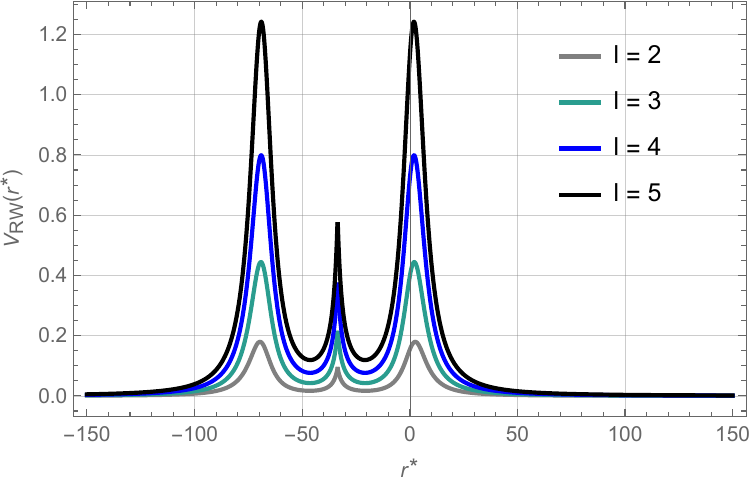} \includegraphics[height=4cm]{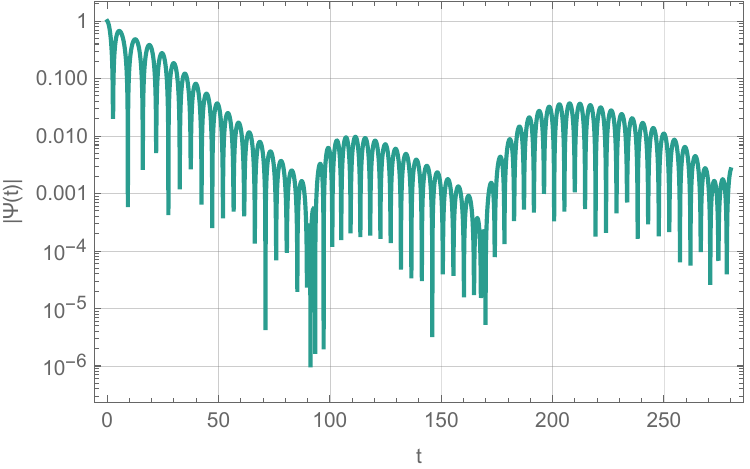}\\
(2e) \hspace{5.5 cm}(2f)\\
\includegraphics[height=4cm]{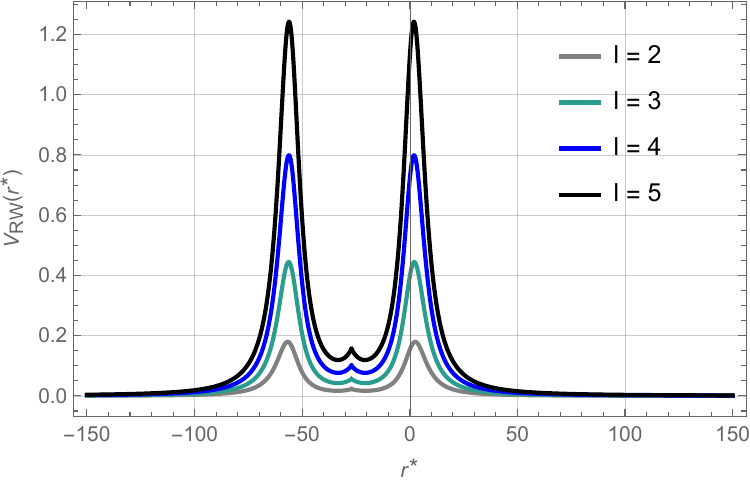} \includegraphics[height=4cm]{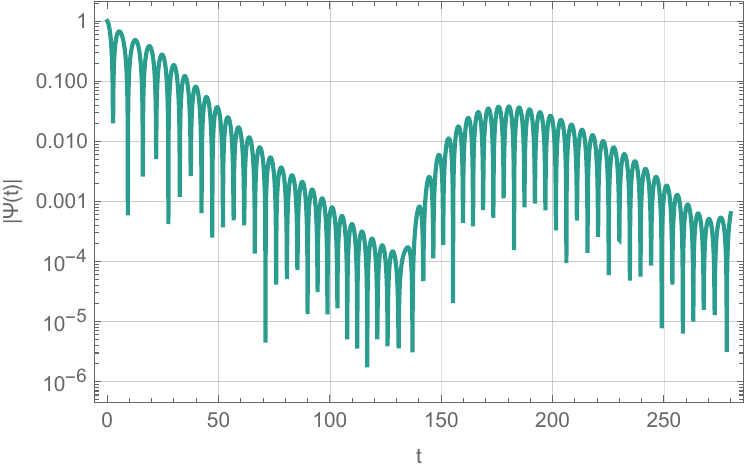}\\
(2g) \hspace{5.5 cm}(2h)
\end{tabular}
\end{center}
\caption{The Regge-Wheeler potential (left panels) and time evolution of the axial gravitational perturbation (right panels) for the symmetric solution (\ref{symetric}) with $\omega = 3/2, M = 1, \rho_0 = 65/27, l=3$, $ a = 1$ (2a), $a = 1.1$ (2c), $a = 1.2$ (2e) and $a  = 1.3$ (2g). The existence of echoes with a significant increase of the amplitude after several oscillations is evident.\label{Figure2}}
\end{figure}

\begin{figure}[t]
\begin{center}
\begin{tabular}{ccc}
\includegraphics[height=5.0cm]{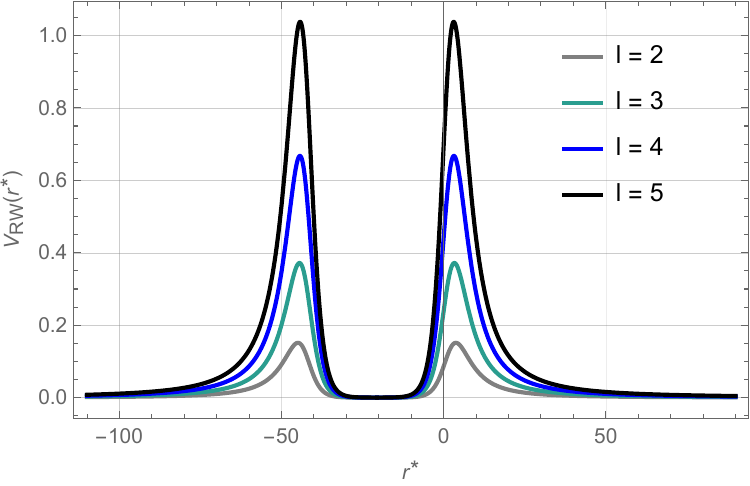} \includegraphics[height=5.0cm]{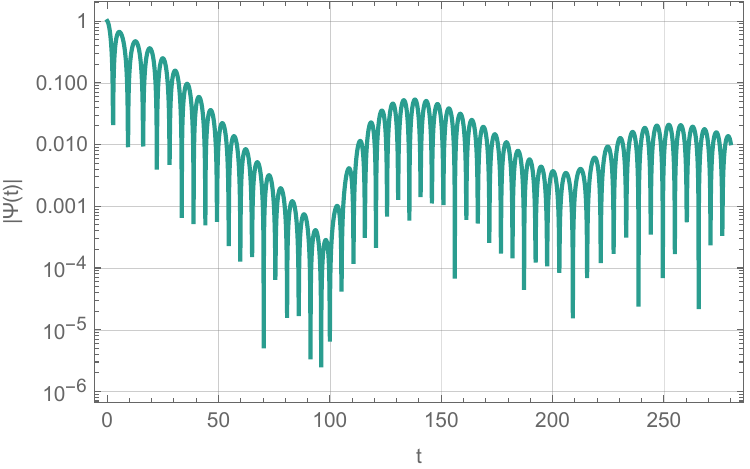}\\ 
(3a) \hspace{7 cm}(3b)\\
\includegraphics[height=5.0cm]{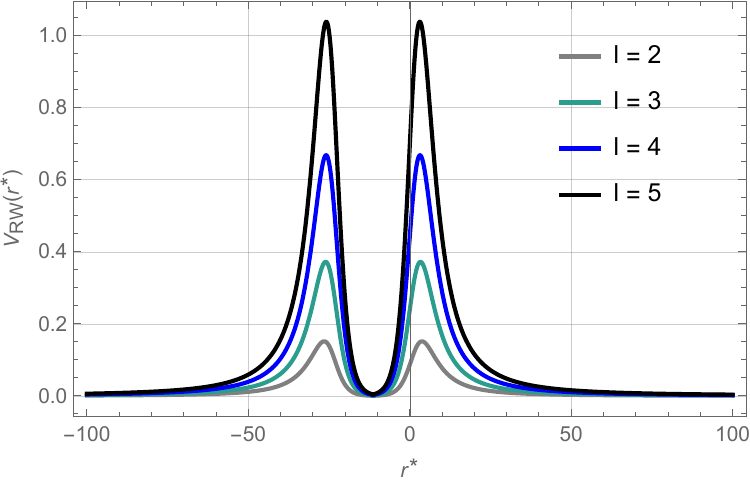} \includegraphics[height=5.0cm]{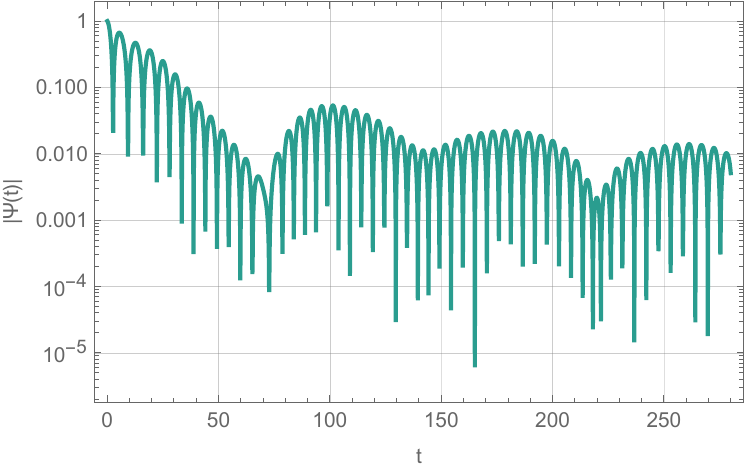}\\ 
(3c) \hspace{7 cm}(3d)\\
\includegraphics[height=5.0cm]{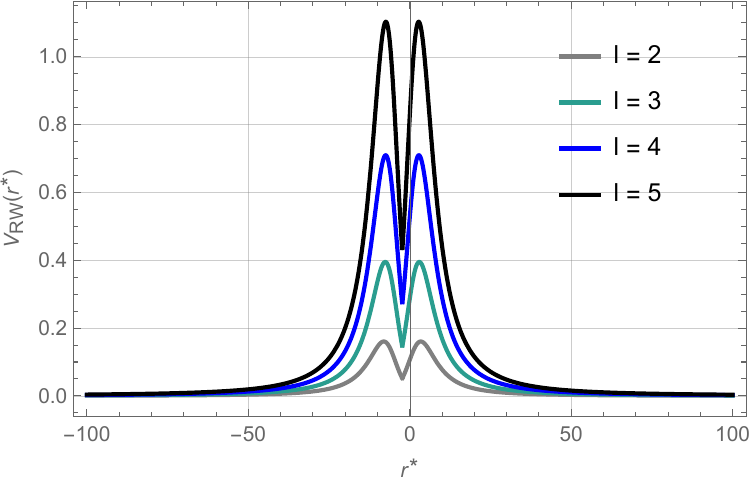} \includegraphics[height=5.0cm]{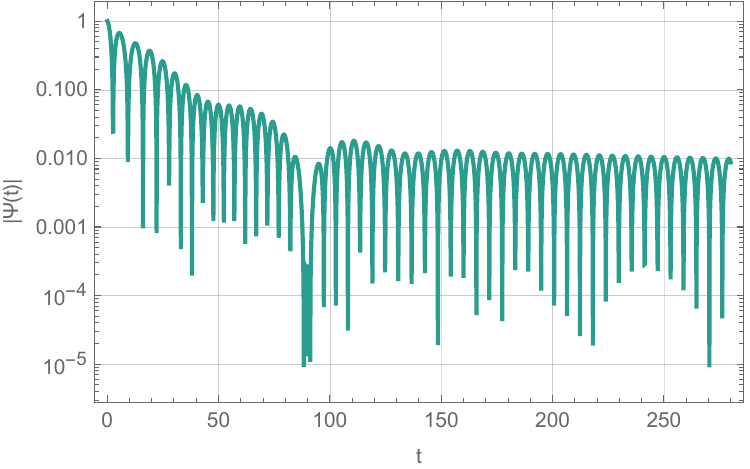}\\
(3e) \hspace{7 cm}(3f)\\
\end{tabular}
\end{center}
\caption{The Regge-Wheeler potential (left panels) and time evolution of the axial gravitational perturbation (right panels) for the symmetric solution (\ref{symetric}) with $l = 3, M = 1, \omega = 3/2, a = 2$, $\rho_0 = 10^{-4}$ (3a), $\rho_0 = 10^{-2}$ (3c) and $\rho_0 = 1$ (3e). The existence of echoes is also evident in these cases.\label{Figure3}}
\end{figure}

\begin{figure}[!h]
\begin{center}
\begin{tabular}{ccc}
\includegraphics[height=5cm]{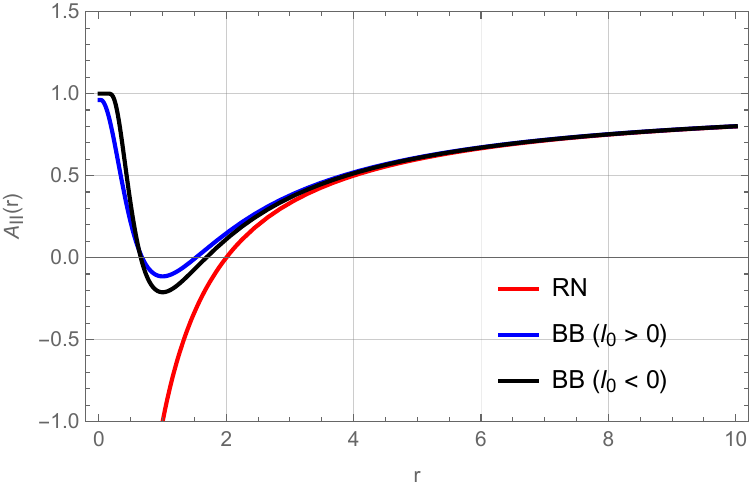} \includegraphics[height=5cm]{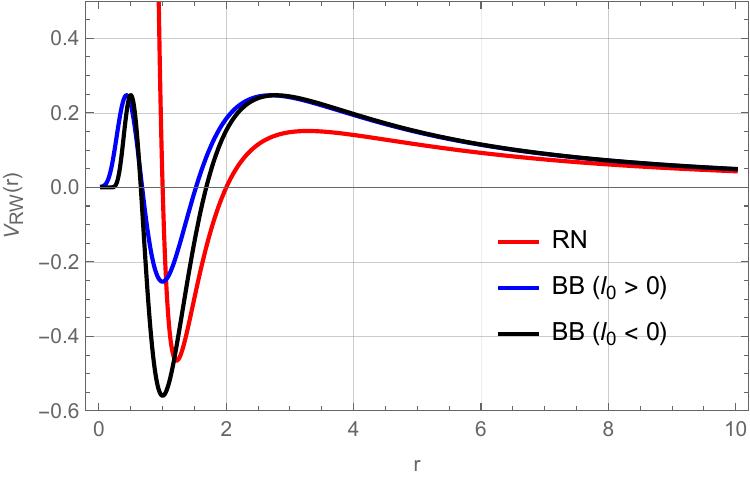}\\ 
(4a) \hspace{7 cm}(4b)\\
\includegraphics[height=5cm]{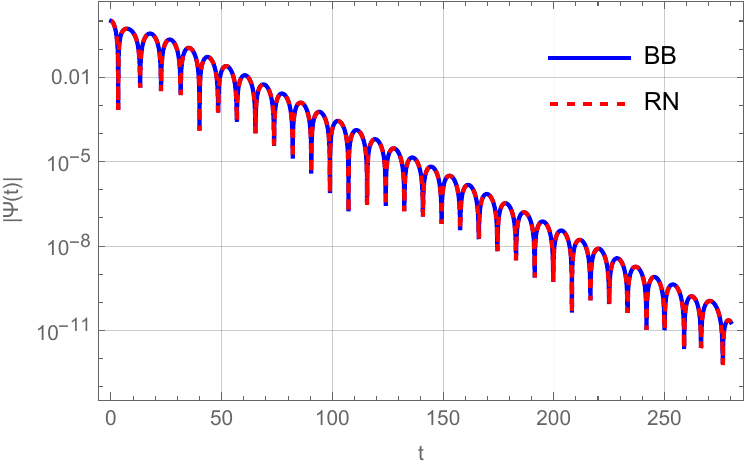}
\includegraphics[height=5cm]{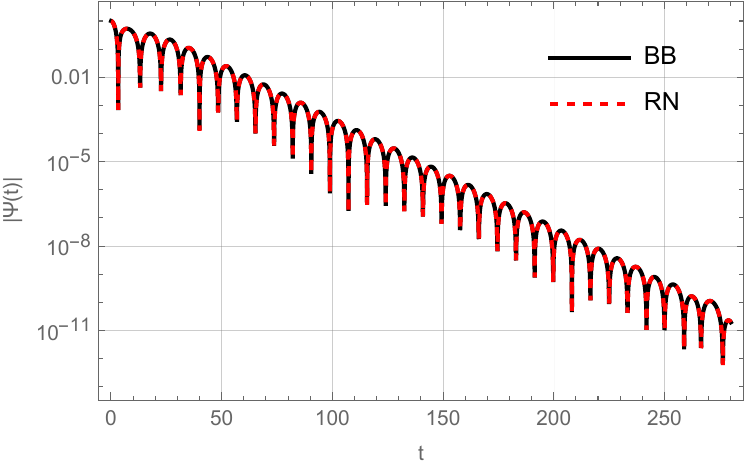} \\
(4c) \hspace{7 cm}(4d)\\
\end{tabular}
\end{center}
\caption{Graphical representation of $A_{II}(r)$ (4a), the Regge-Wheeler potential for angular momentum $l=2$ (4b) and time evolution of the axial gravitational perturbation for the asymmetric solution (\ref{asymetric}) for $l_0 = - 1/5$ (4c) and $l_0 = 1/5$ (4d) with  $\tilde{\rho}_0 = 0, \tilde{M} = r_0 = \omega = \alpha = 1$.\label{Figure4}}
\end{figure}

\begin{figure}[t]
\begin{center}
\begin{tabular}{ccc}
\includegraphics[height=5.0cm]{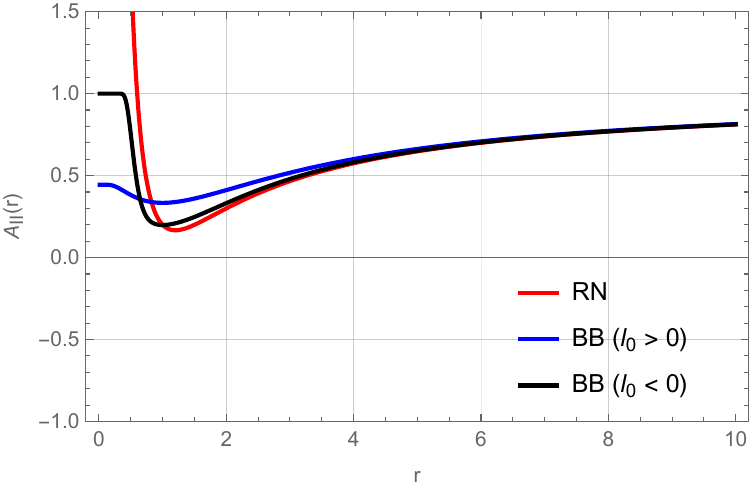} \includegraphics[height=5.0cm]{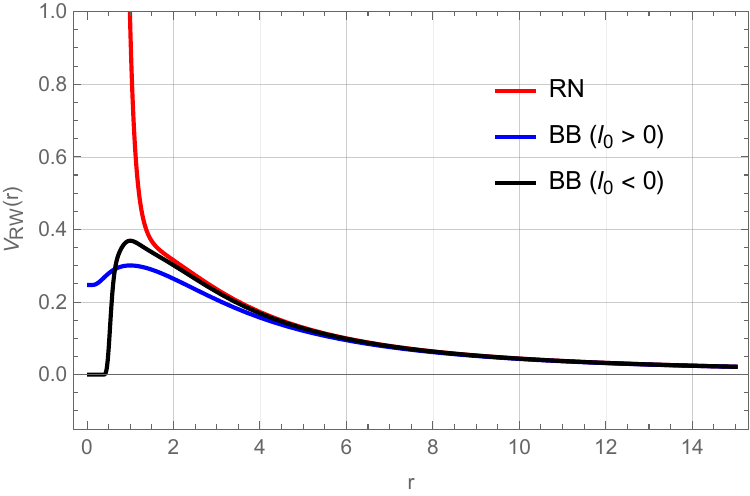}\\ 
(5a) \hspace{7 cm}(5b)\\
\includegraphics[height=5.0cm]{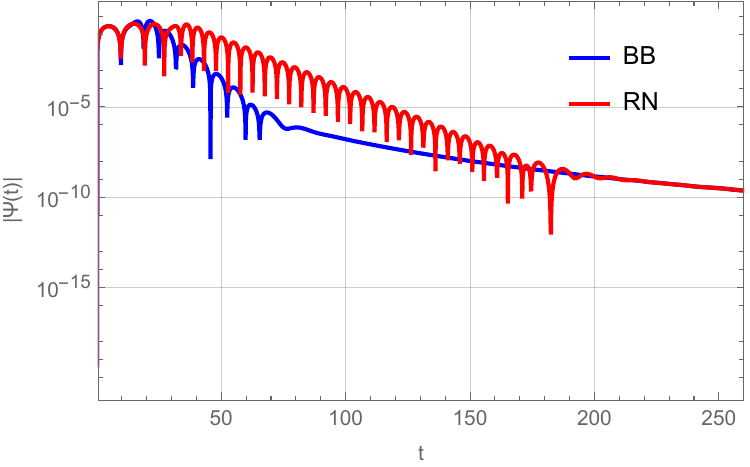} \includegraphics[height=5.0cm]{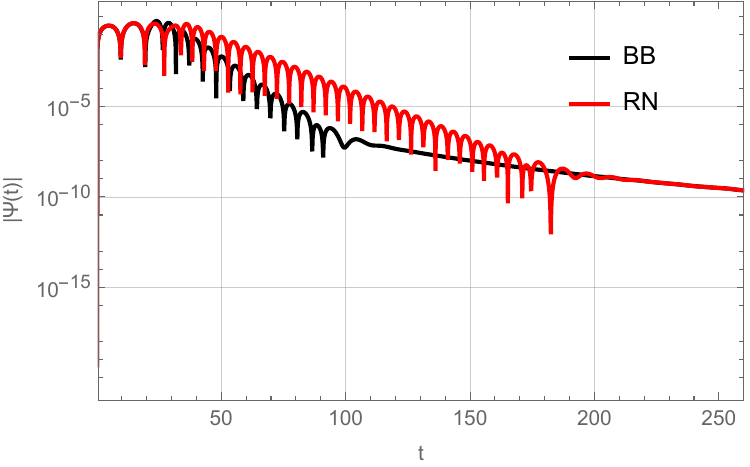}\\
(5c) \hspace{7 cm}(5d)
\end{tabular}
\end{center}
\caption{Graphical representation of $A_{II}(r)$ (5a), the Regge-Wheeler potential  for angular momentum $l=2$ (5b) and time evolution of the axial gravitational perturbation for the asymmetric solution (\ref{asymetric}) with $\tilde{M} = r_0 = \omega = \alpha = 1, \tilde{\rho_0} = {1.2}/{\tilde{\Sigma}_0^{2 \omega +2}}, l_0 = 1$ (5c) and $l_0= -1$ (5d). The observation point is located at $r^* = 20$. \label{Figure5}}
\end{figure}

\begin{figure}[t]
\includegraphics[height=5.1cm]{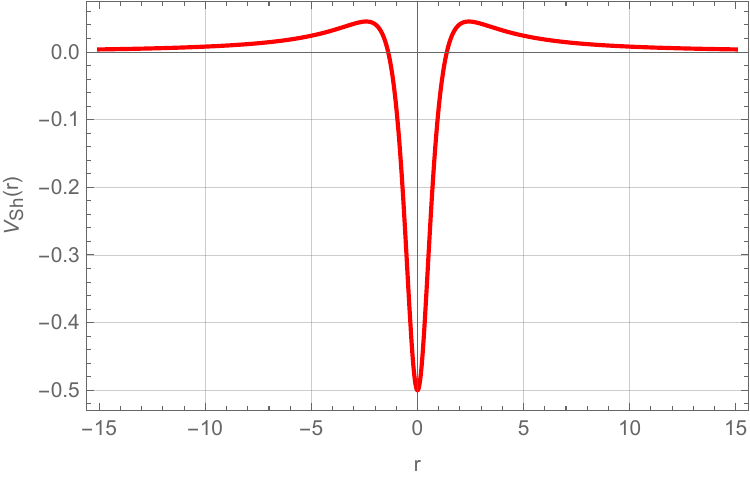}
\includegraphics[height=5.1cm]{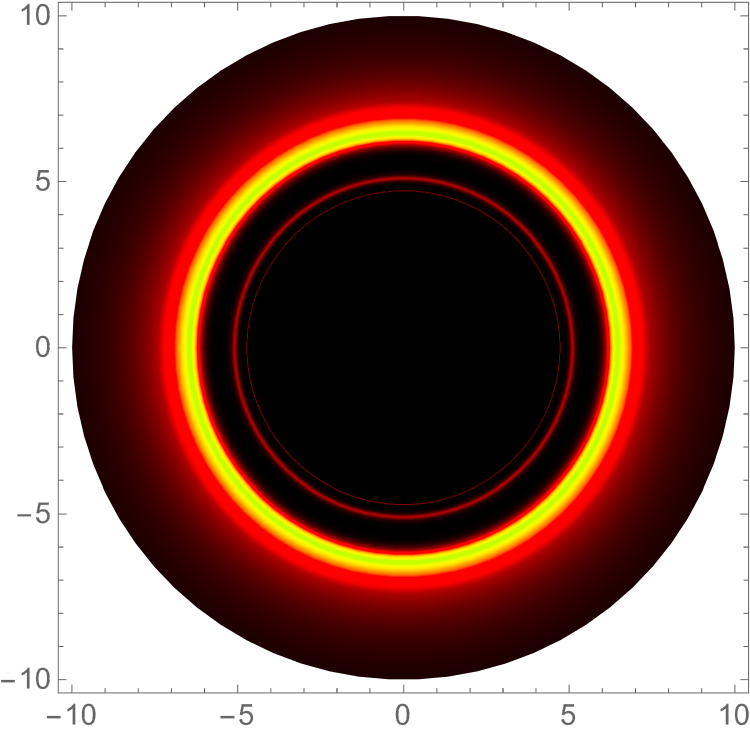}
\includegraphics[height=5.1cm]{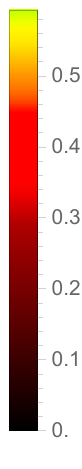}\\
(6a) \hspace{6 cm}(6b)\\
\includegraphics[height=5.1cm]{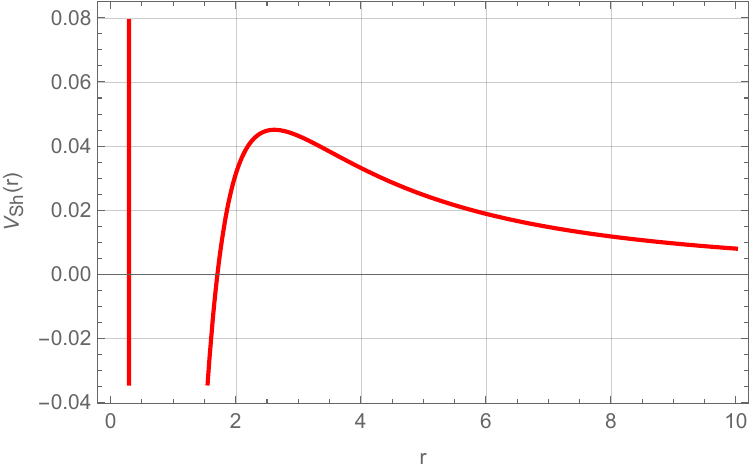}
\includegraphics[height=5.1cm]{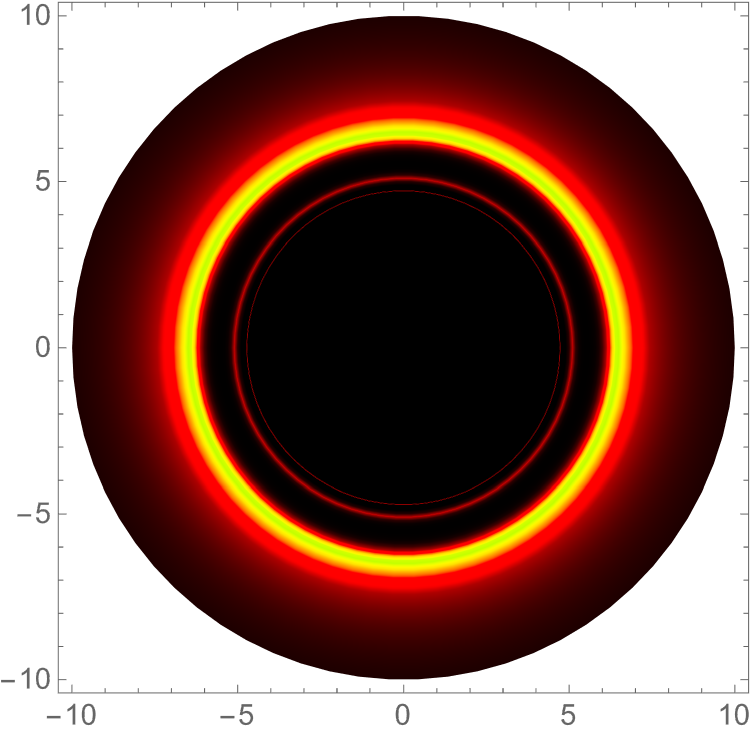}
\includegraphics[height=5.1cm]{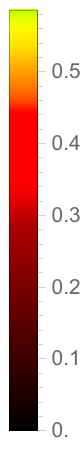}\\
(6c) \hspace{6 cm}(6d)\\
\caption{The shadow potential (left panels) and the optical images (right panels) for the symmetric solution (\ref{symetric}) (6a) and RN (6c) with $a = M = \omega = 1$ and $\rho_0= 0.5$.\label{Figure6}}
\end{figure}

\begin{figure}[t]
\includegraphics[height=5.1cm]{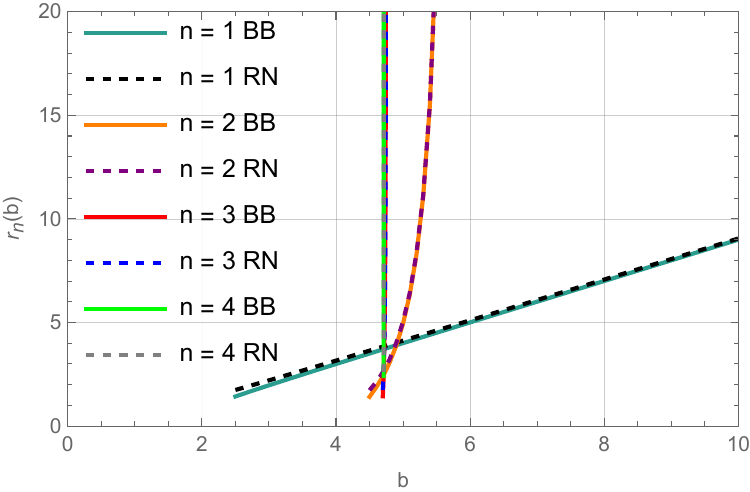}
\caption{Transfer function $r=r_n(b)$ for the symmetric solution (\ref{symetric}) compared with RN for $a = M = \omega = 1$ and $\rho_0= 0.5$.\label{Figure7}}
\end{figure}

\begin{figure}[t]
\begin{center}
\begin{tabular}{ccc}
\includegraphics[height=4cm]{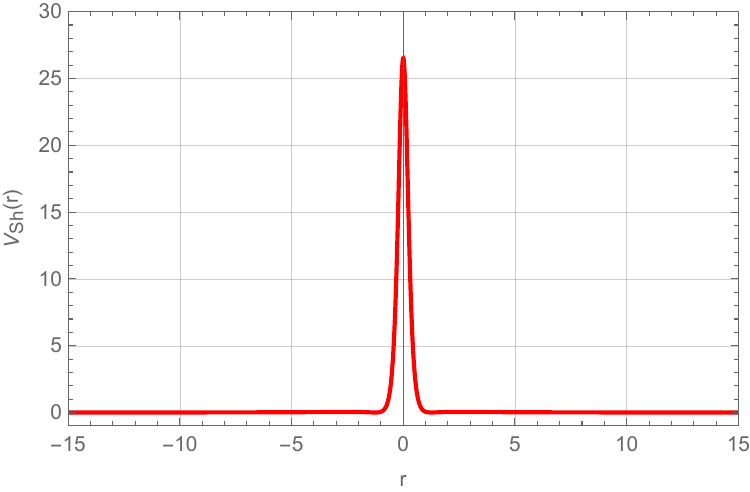} \includegraphics[height=4cm]{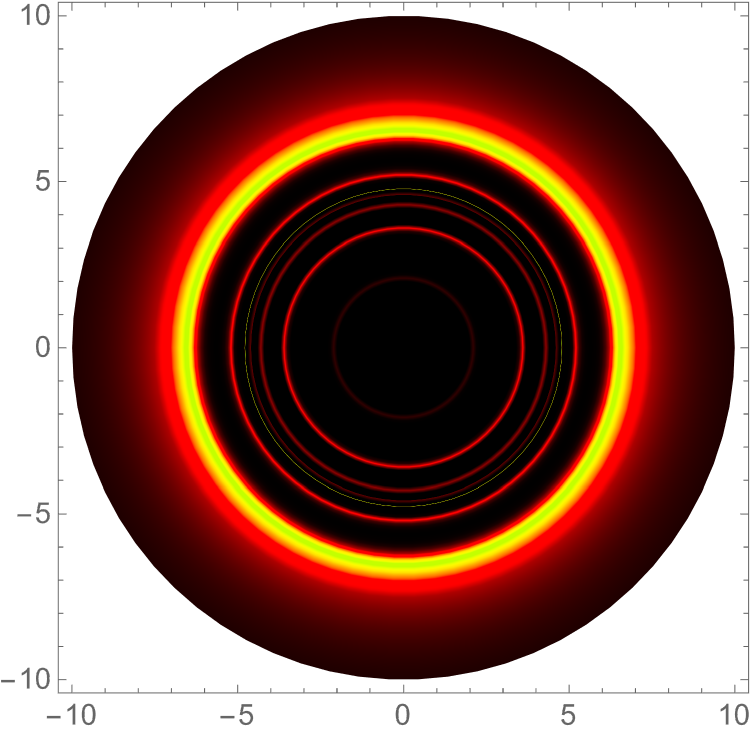} 
\includegraphics[height=4cm]{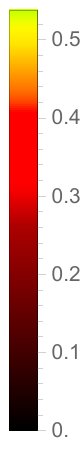}\\ 
(8a) \hspace{4.5 cm}(8b)\\
\includegraphics[height=4cm]{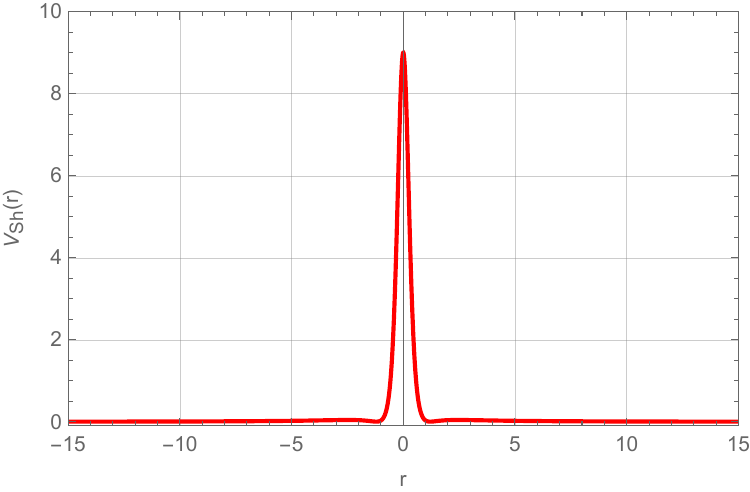} \includegraphics[height=4cm]{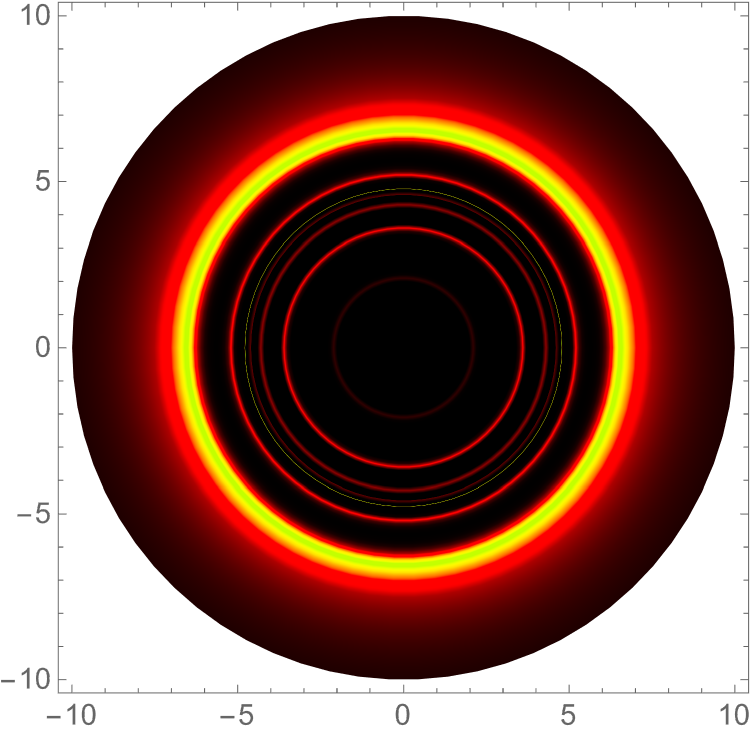} 
\includegraphics[height=4cm]{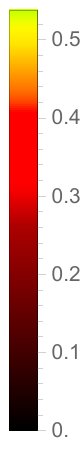}\\ 
(8c) \hspace{4.5 cm}(8d)\\
\includegraphics[height=4cm]{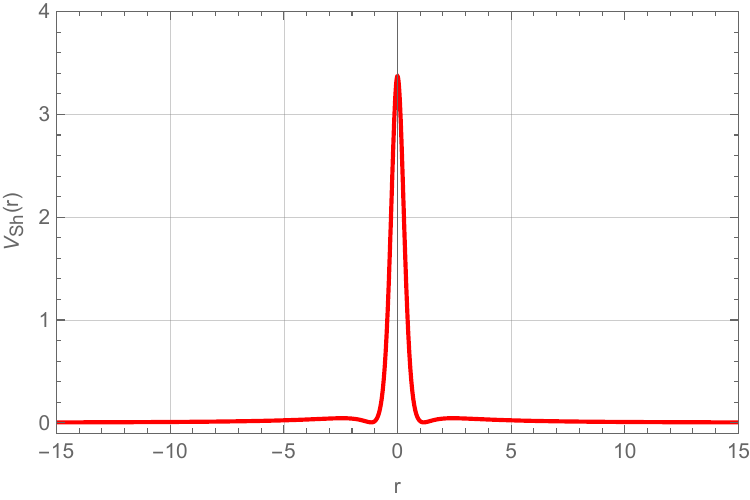} \includegraphics[height=4cm]{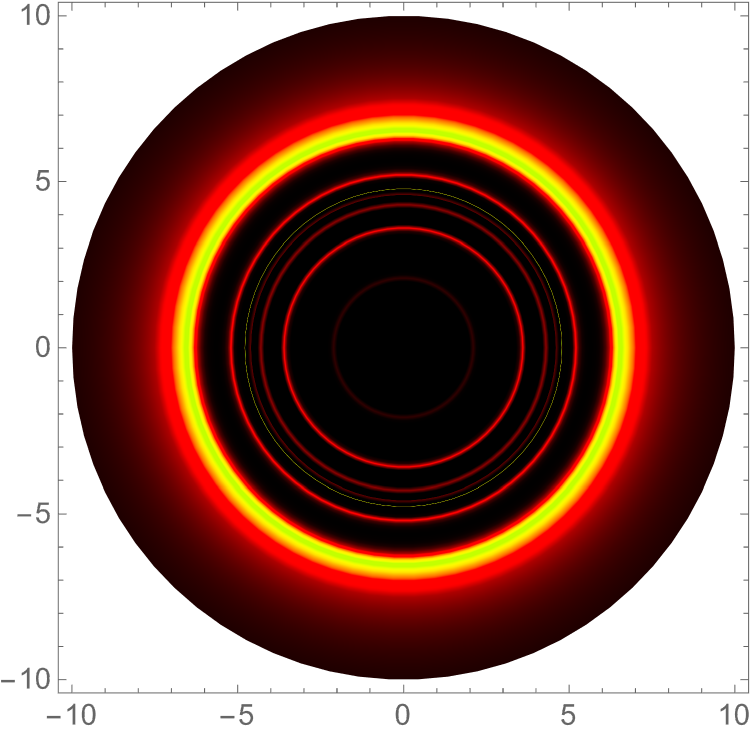}
\includegraphics[height=4cm]{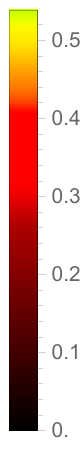}\\ 
(8e) \hspace{4.5 cm}(8f)\\
\includegraphics[height=4cm]{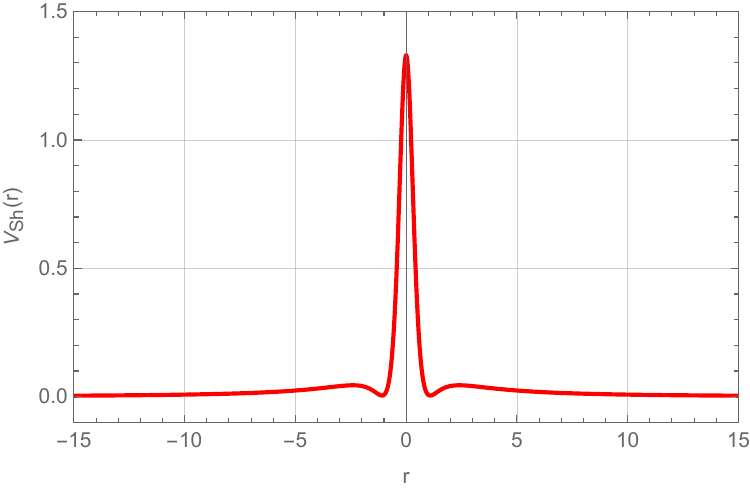} \includegraphics[height=4cm]{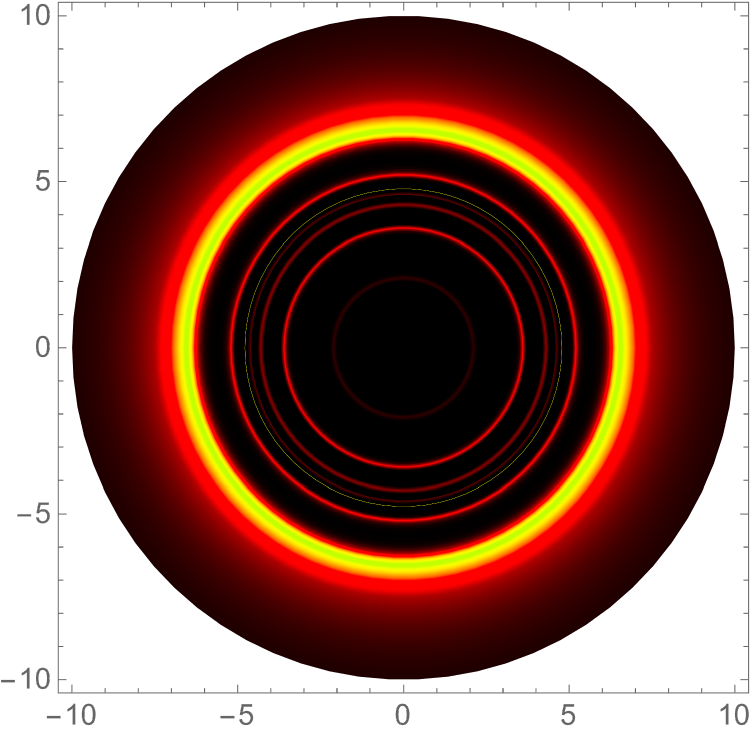}
\includegraphics[height=4cm]{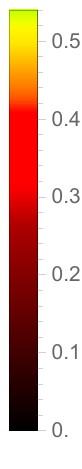}\\ 
(8g) \hspace{4.5 cm}(8h)
\end{tabular}
\end{center}
\caption{The shadow potential (left panels) and optical image (right panels) for the symmetric solution (\ref{symetric}) with $\omega = 3/2, M = 1, \rho_0 = 65/27$, $ a = 0.5$ (8a), $a = 0.6$ (8c), $a = 0.7$ (8e) and $a  = 0.8$ (8g).\label{Figure8}}
\end{figure}

\begin{figure}[t]
\begin{center}
\begin{tabular}{ccc}
\includegraphics[height=4cm]{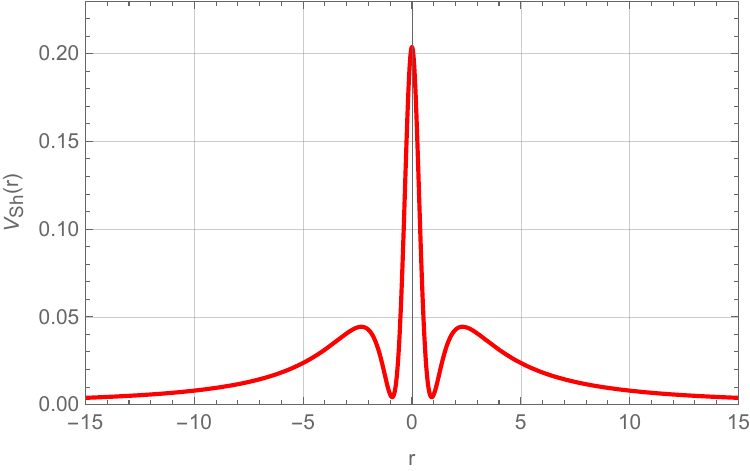} \includegraphics[height=4cm]{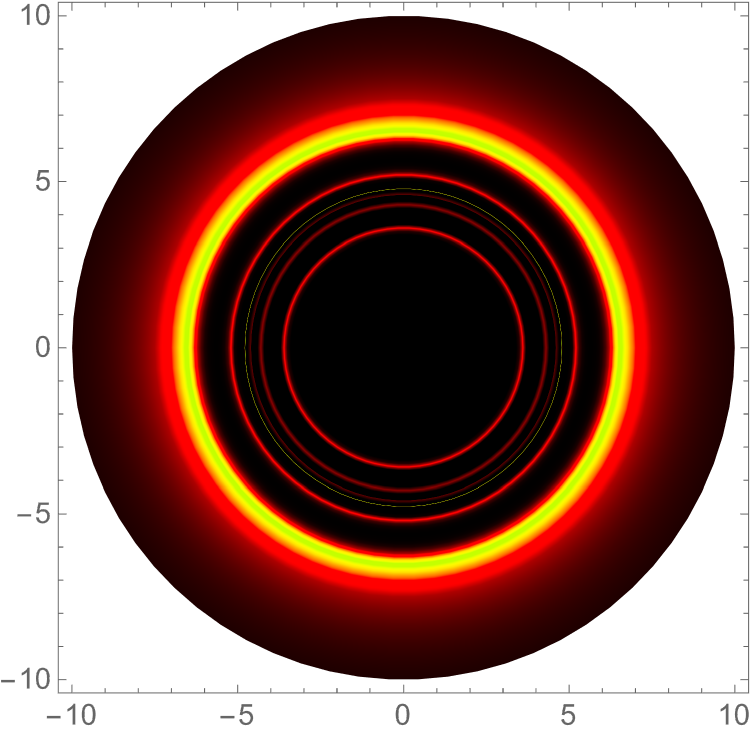} 
\includegraphics[height=4cm]{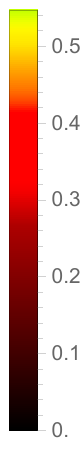}\\ 
(9a) \hspace{4.5 cm}(9b)\\
\includegraphics[height=4cm]{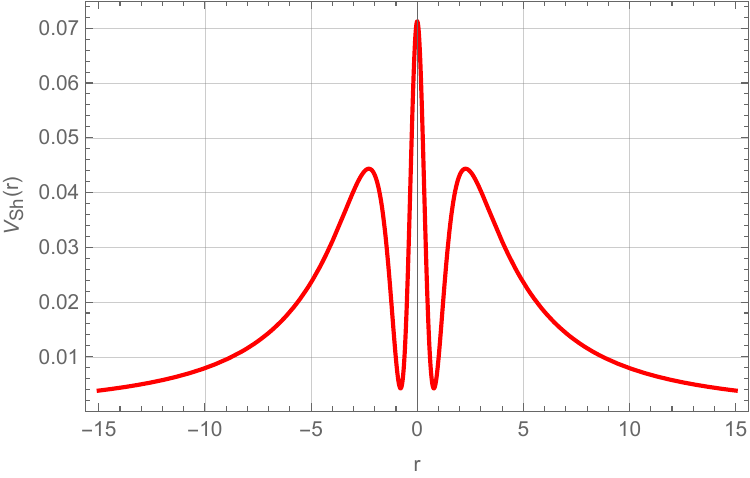} \includegraphics[height=4cm]{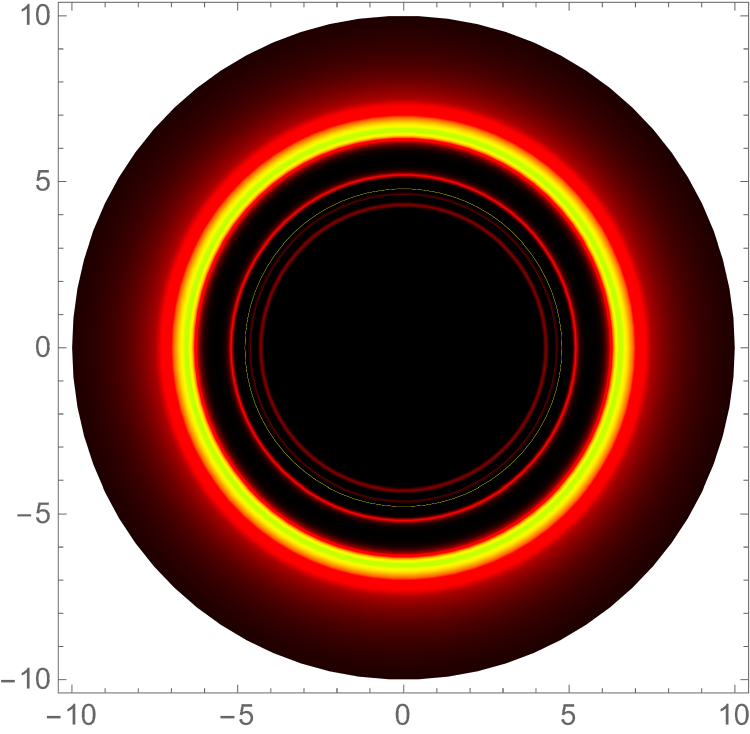} 
\includegraphics[height=4cm]{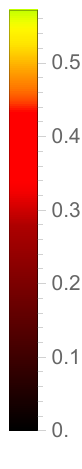}\\ 
(9c) \hspace{4.5 cm}(9d)\\
\includegraphics[height=4cm]{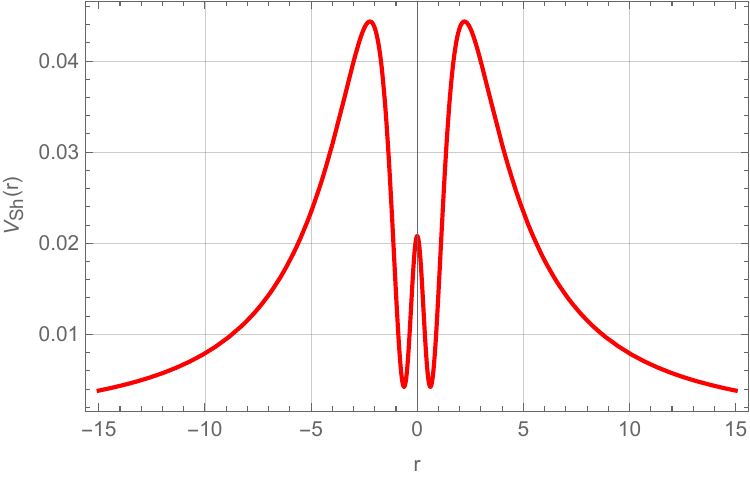} \includegraphics[height=4cm]{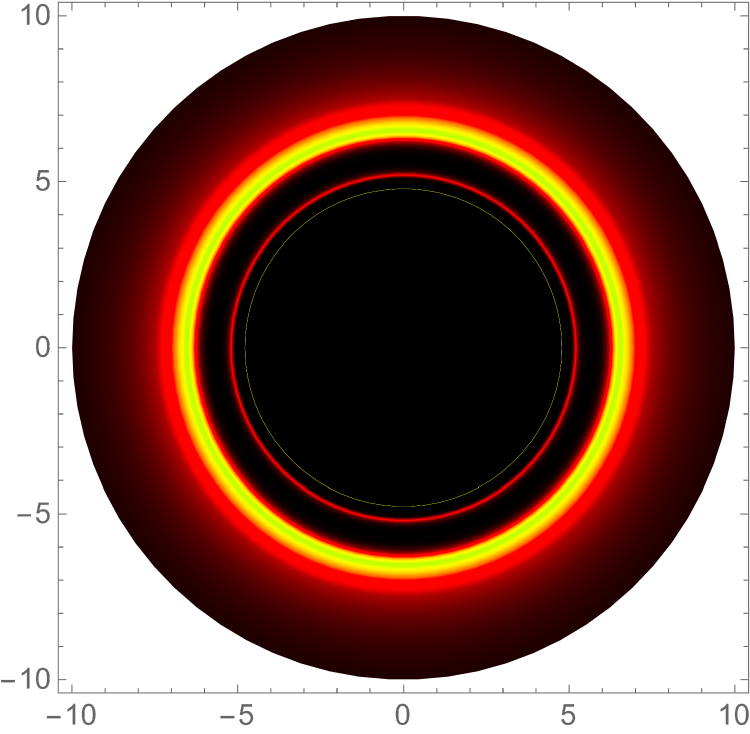}
\includegraphics[height=4cm]{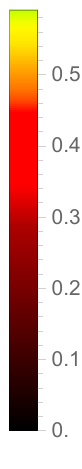}\\ 
(9e) \hspace{4.5 cm}(9f)\\
\includegraphics[height=4cm]{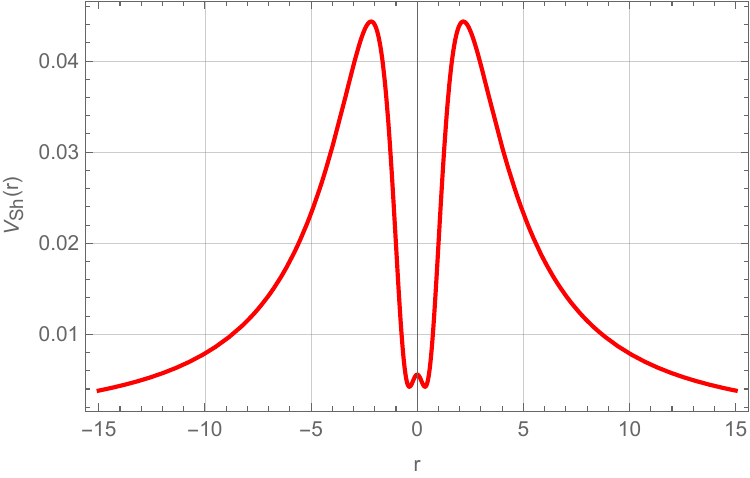} \includegraphics[height=4cm]{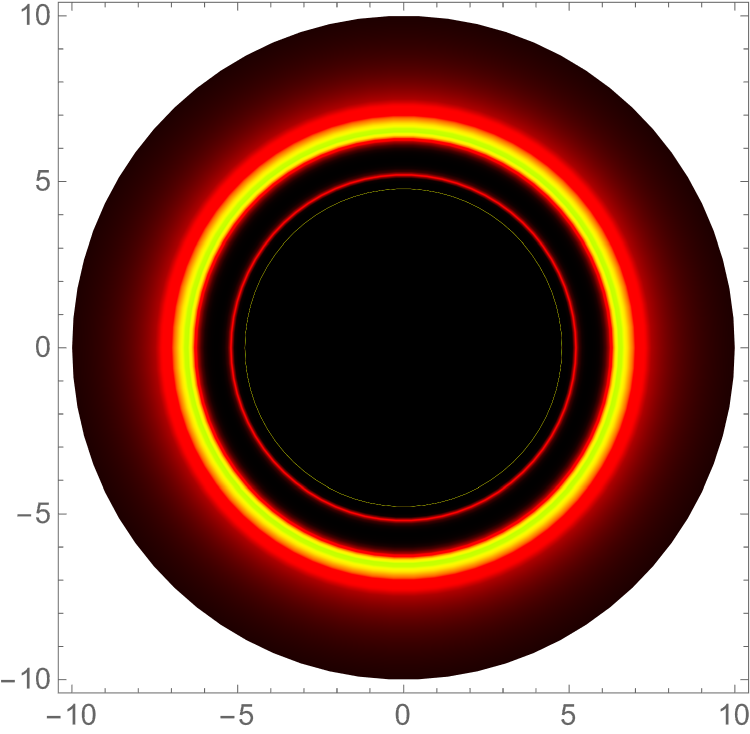}
\includegraphics[height=4cm]{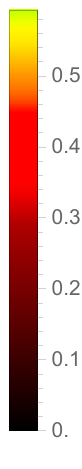}\\ 
(9g) \hspace{4.5 cm}(9h)
\end{tabular}
\end{center}
\caption{The shadow potential (left panels) and optical image (right panels) for the symmetric solution (\ref{symetric}) with $\omega = 3/2, M = 1, \rho_0 = 65/27$, $ a = 1$ (9a), $a = 1.1$ (9c), $a = 1.2$ (9e) and $a  = 1.3$ (9g).\label{Figure9}}
\end{figure}

\begin{figure}[t]
\begin{center}
\begin{tabular}{ccc}
\includegraphics[height=4cm]{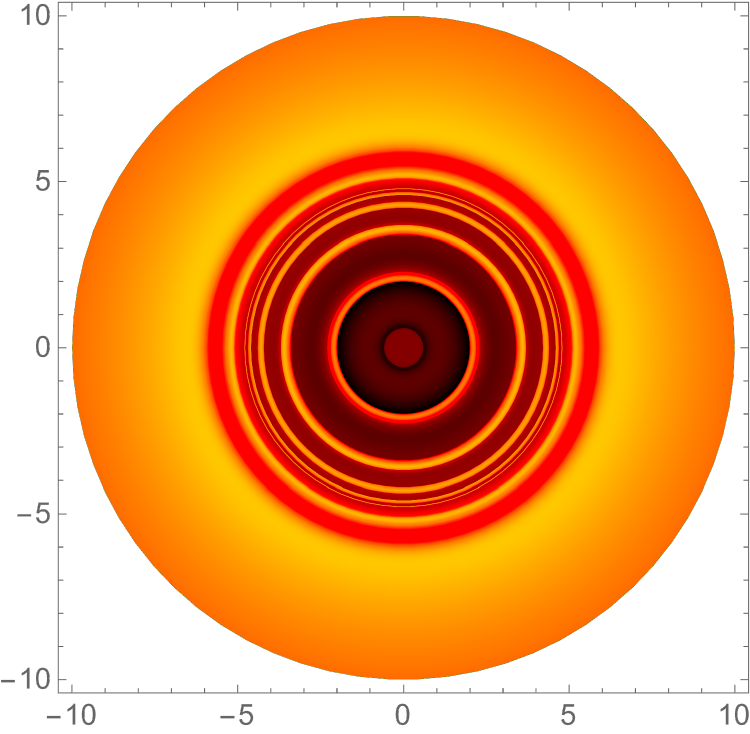} \includegraphics[height=4cm]{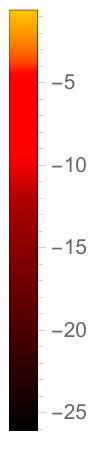}
\includegraphics[height=4cm]{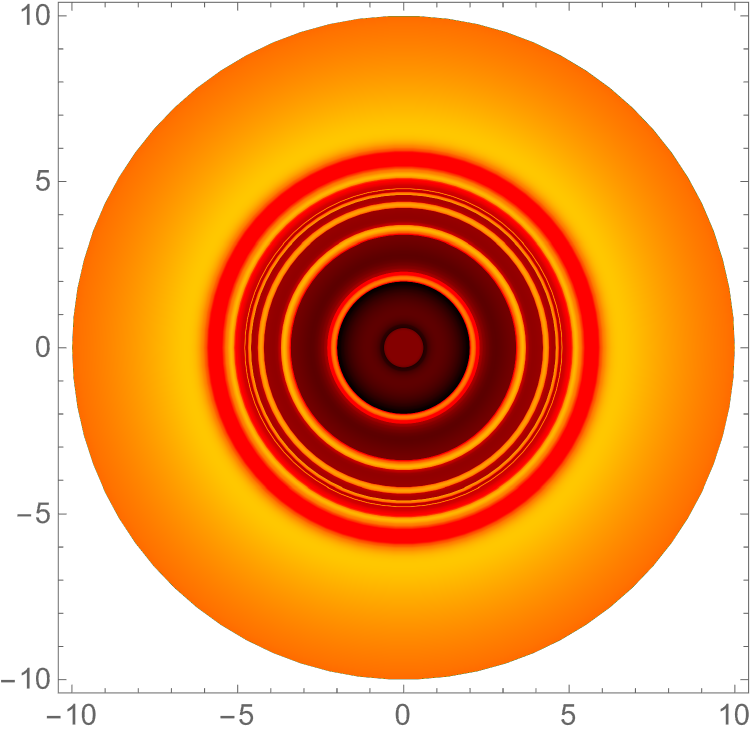} 
\includegraphics[height=4cm]{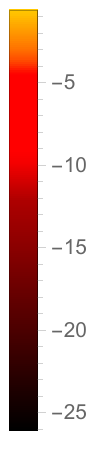}
\\
(10a) \hspace{4 cm}(10b)\\
\includegraphics[height=4cm]{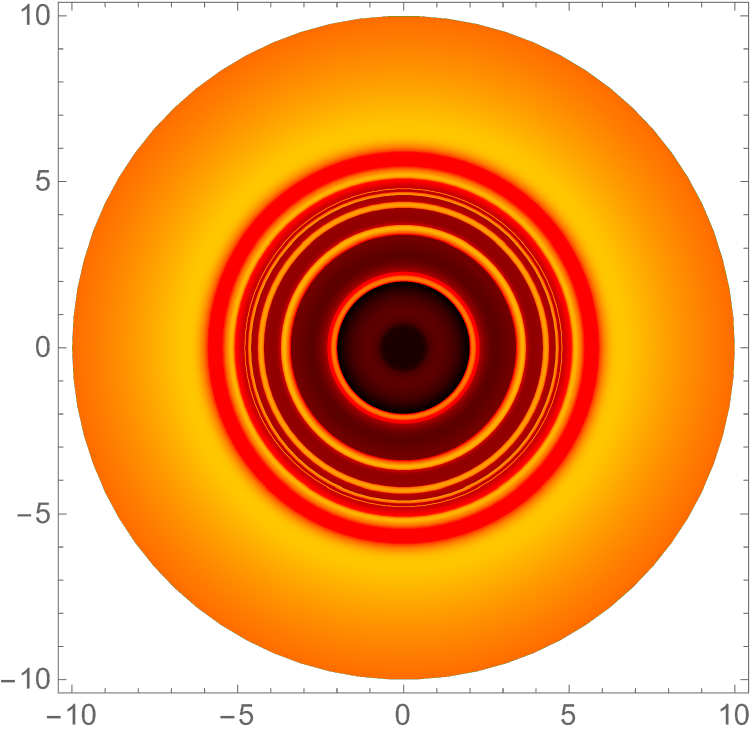}
\includegraphics[height=4cm]{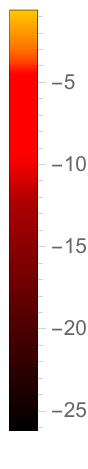}
\includegraphics[height=4cm]{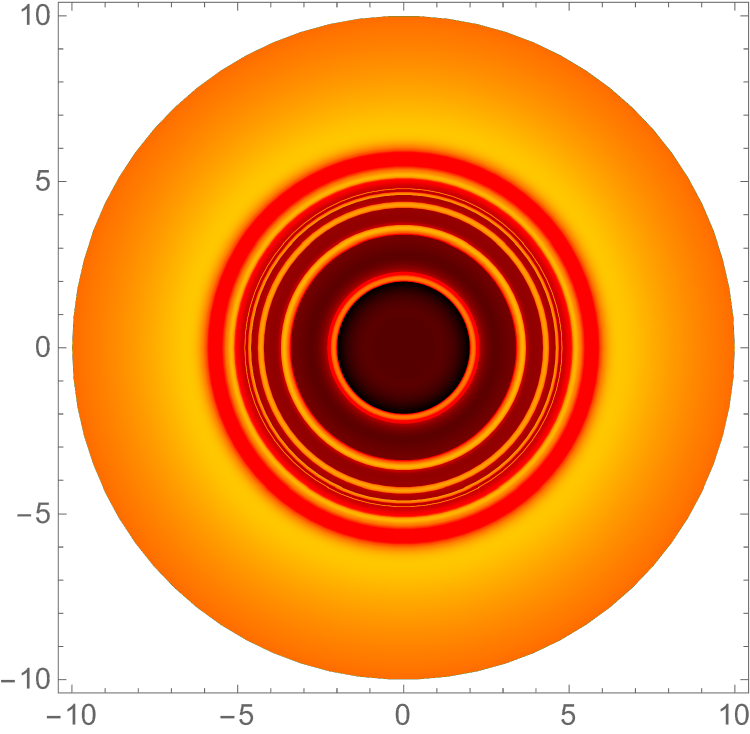}
\includegraphics[height=4cm]{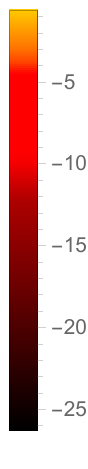}\\
(10c) \hspace{4 cm}(10d)\\
\includegraphics[height=4cm]{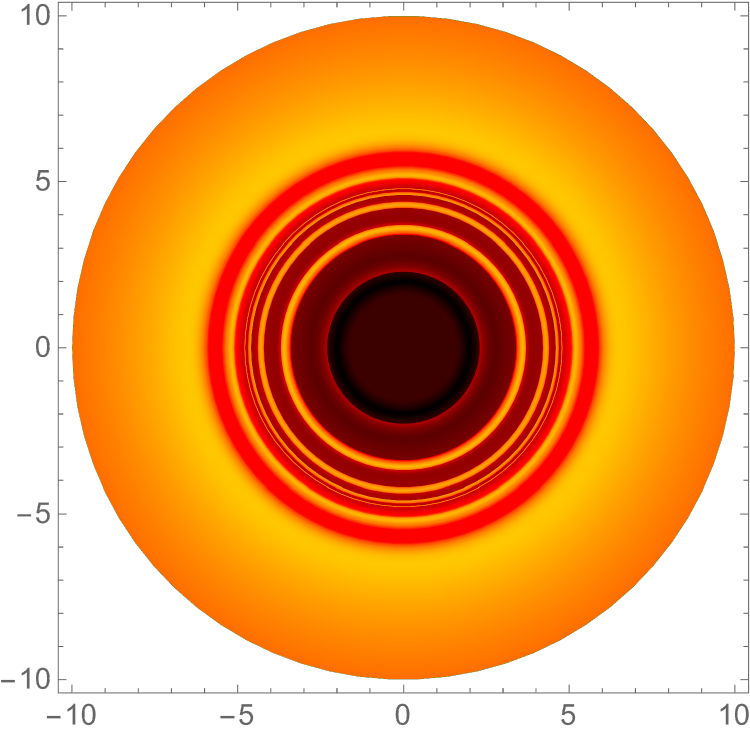}
\includegraphics[height=4cm]{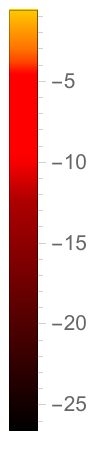}
\includegraphics[height=4cm]{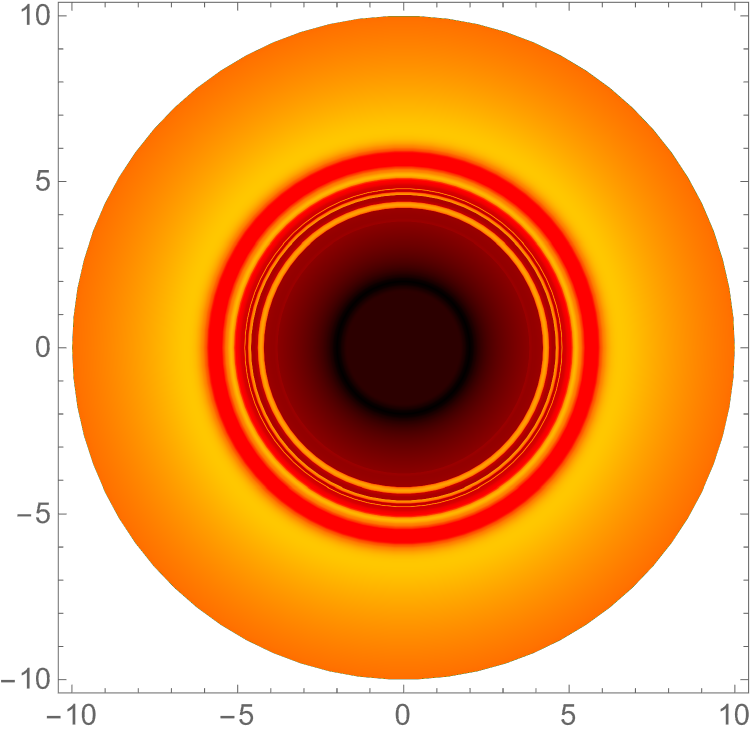}
\includegraphics[height=4cm]{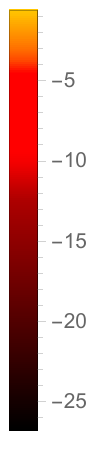}\\
(10e) \hspace{4 cm}(10f) \\
\includegraphics[height=4cm]{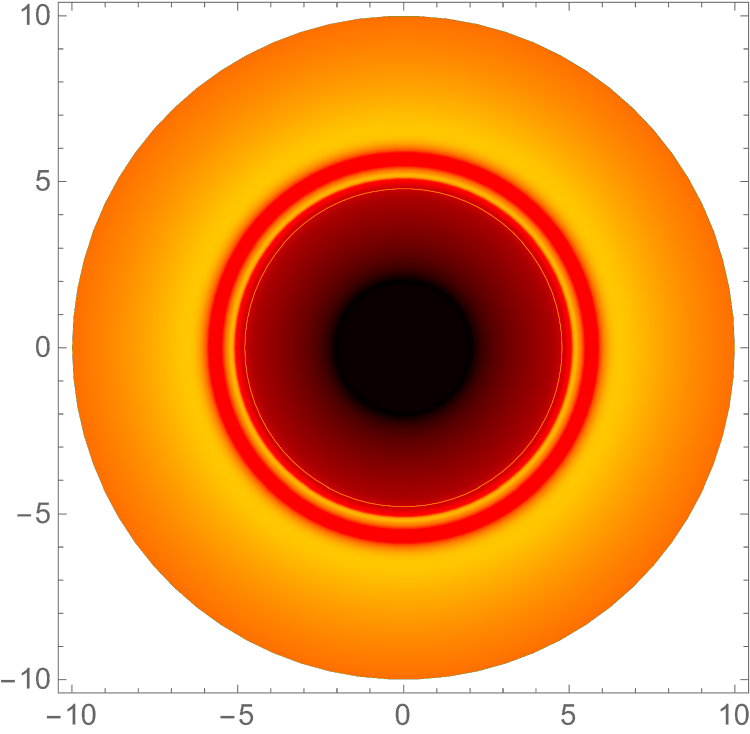}
\includegraphics[height=4cm]{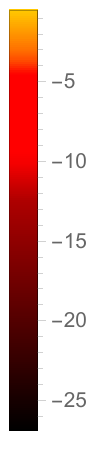}
\includegraphics[height=4cm]{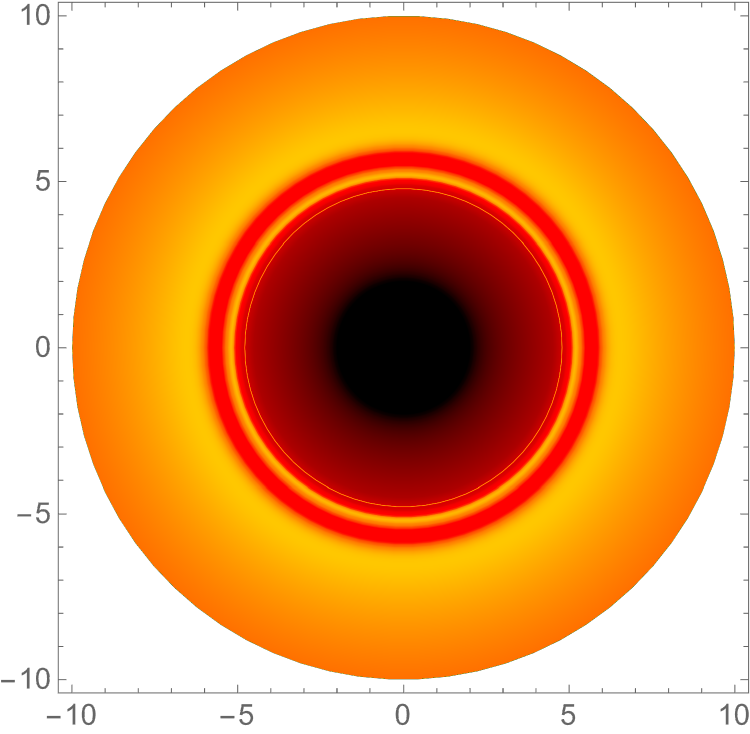}
\includegraphics[height=4cm]{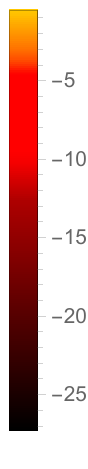}\\
(10g) \hspace{4 cm}(10h) \\
\end{tabular}
\end{center}
\caption{The logarithmic-scale optical image for the symmetric solution (\ref{symetric}), with $\omega = 3/2, M = 1, \rho_0 = 65/27$, $a = 0.5$ (10a), $a = 0.6$ (10b), $a = 0.7$ (10c), $a = 0.8$, (10d), $a = 1.0$ (10e), $a = 1.1$ (10f),  $a = 1.2$ (10g),  $a = 1.3$ (10h).\label{Figure10}}
\end{figure}

\begin{figure}[t]
\begin{center}
\begin{tabular}{ccc}
\includegraphics[height=4.5cm]{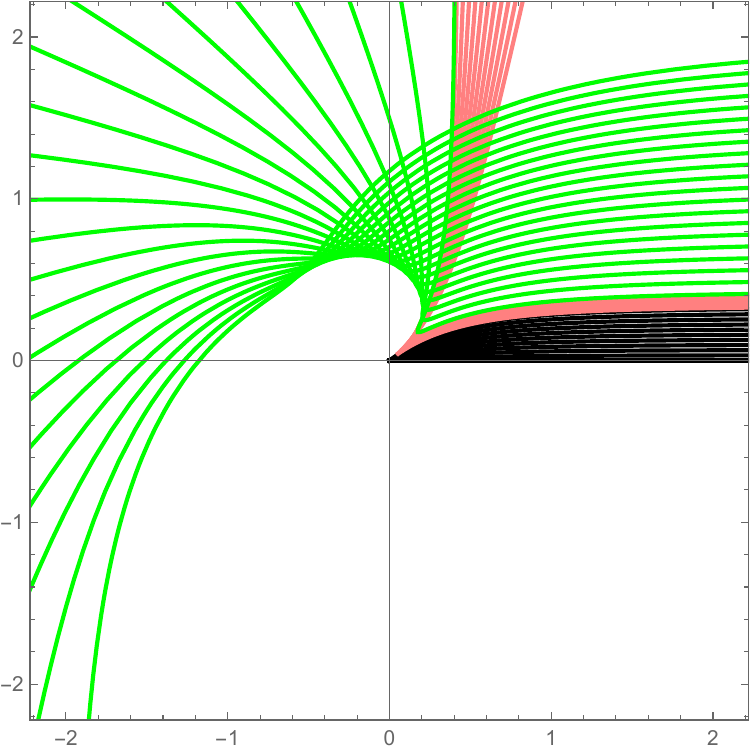} \includegraphics[height=4.5cm]{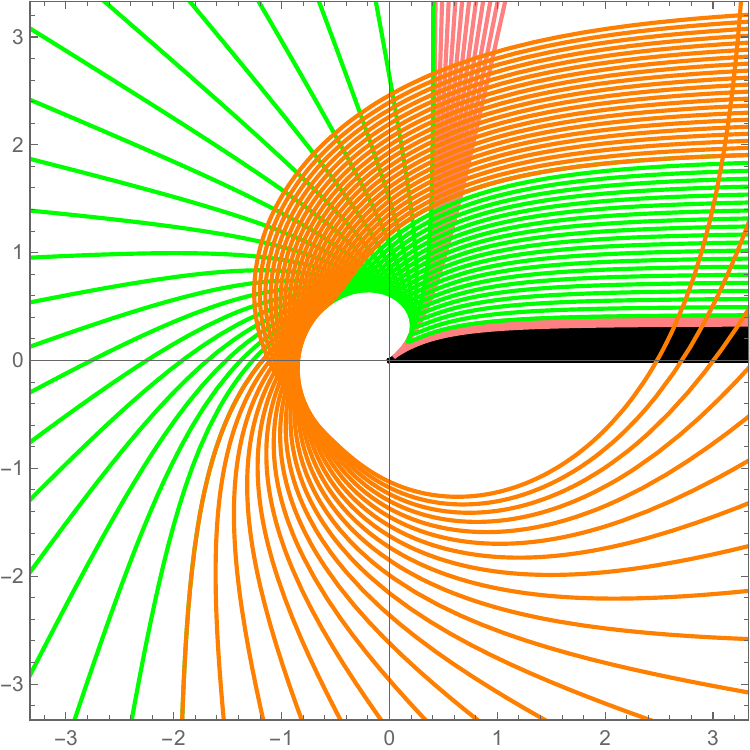}
\includegraphics[height=4.5cm]{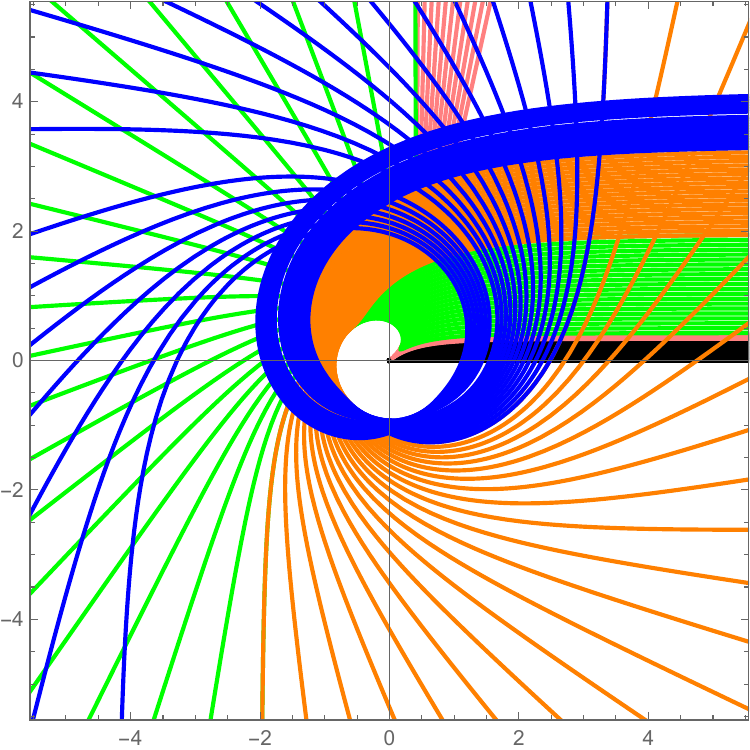} \\
(11a) \hspace{4.5 cm}(11b) \hspace{4.5 cm}(11c)\\
\includegraphics[height=4.5cm]{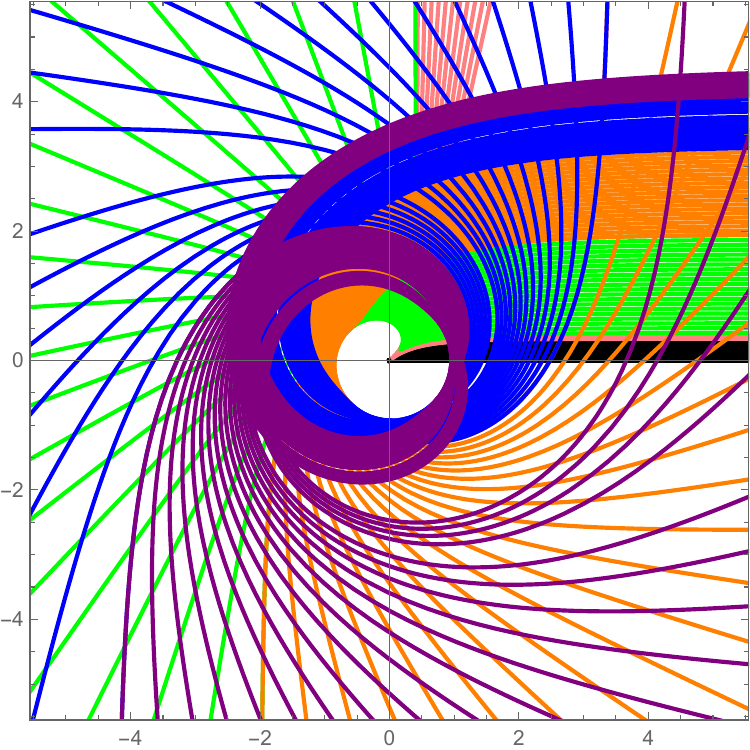}
\includegraphics[height=4.5cm]{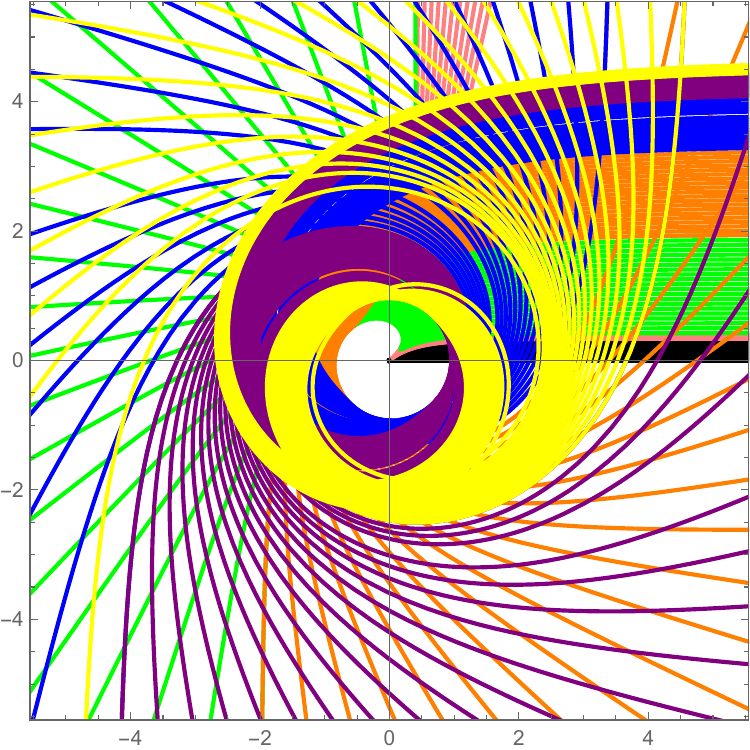}
\includegraphics[height=4.5cm]{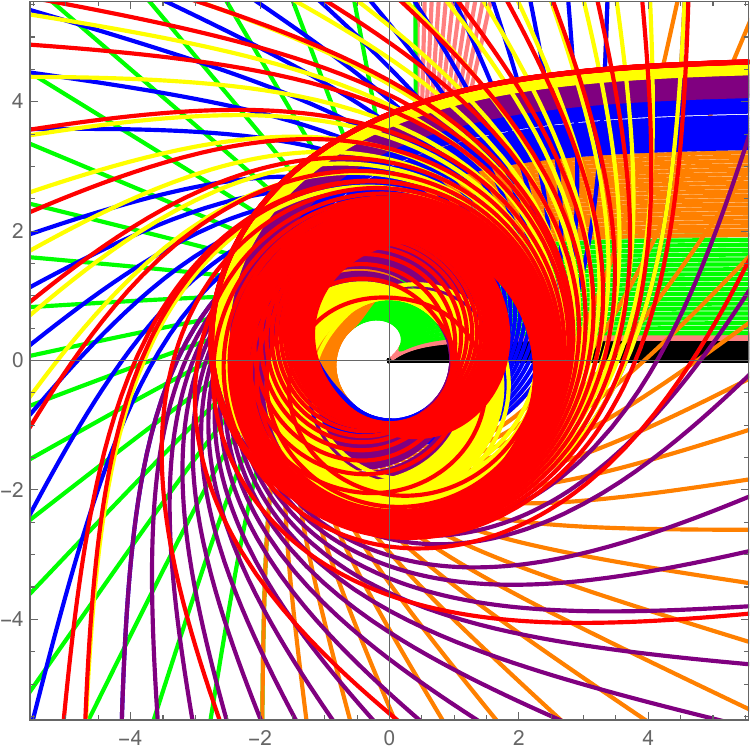}\\
(11d) \hspace{4.5 cm}(11e) \hspace{4.5 cm}(11f)\\
\includegraphics[height=4.5cm]{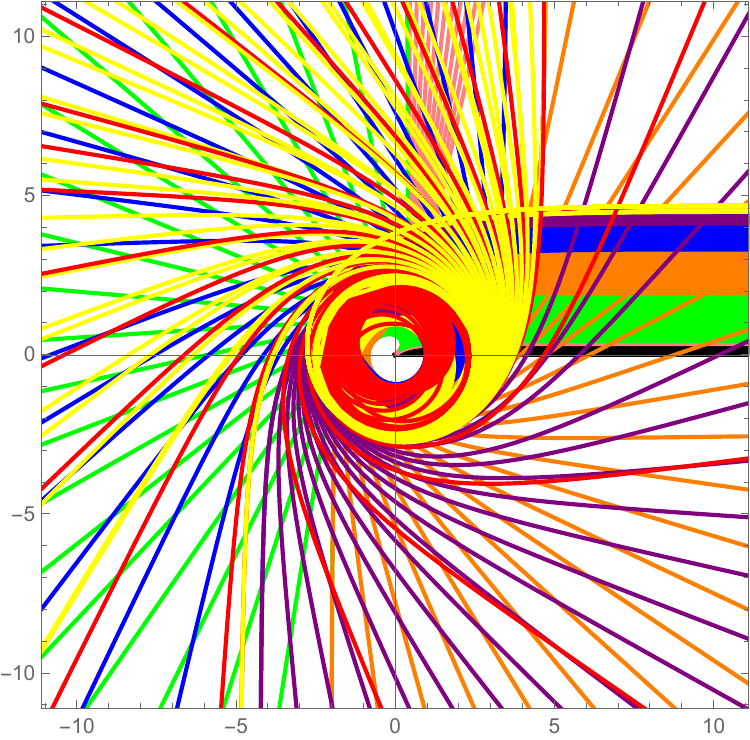}
\includegraphics[height=4.5cm]{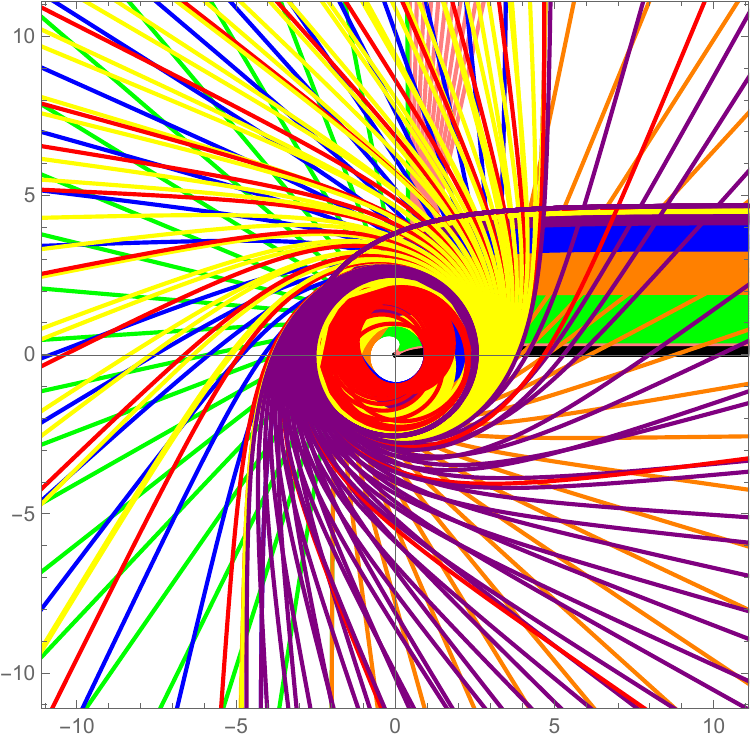}
\includegraphics[height=4.5cm]{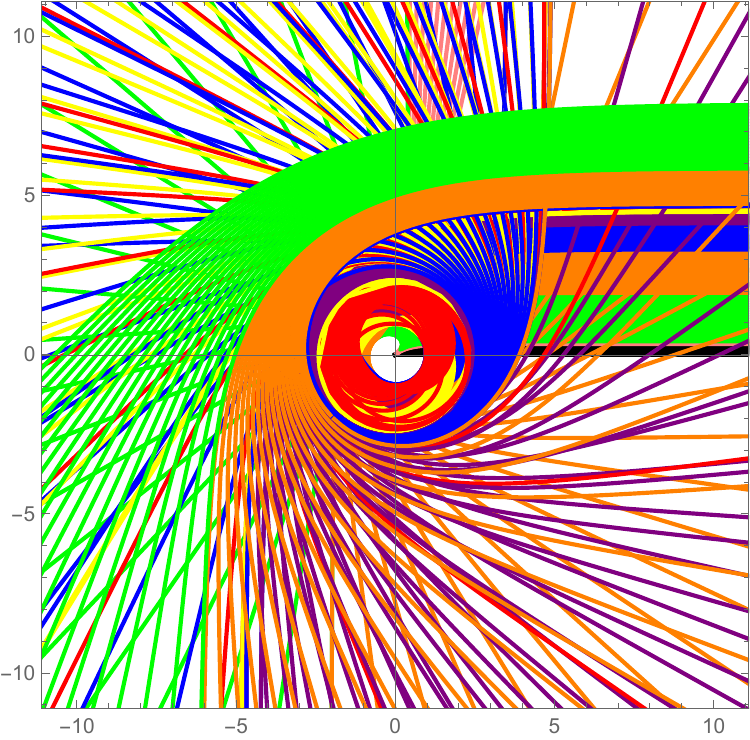}\\
(11g) \hspace{4.5 cm}(11h) \hspace{4.5 cm}(11i)\\
\end{tabular}
\end{center}
\caption{Ray tracing for the symmetric solution (\ref{symetric}), with $\omega = 3/2, M = 1, \rho_0 = 65/27$, $a = 0.6$ and $b = (0,1.97)$ (11a), $b = (0,3.35)$ (11b), $ b = (0,4.19)$ (11c), $ b = (0,4.55)$ (11d), $b = (0,4.69)$ (11e), $b = (0,4.7490)$ (11f), $ b = (0,4.7494)$ (11g), $b = (0,4.75)$ (11h), $ b = (0,8)$ (11i). 
\label{Figure11}}
\end{figure}

\begin{figure}[t]
\begin{center}
\begin{tabular}{ccc}
\includegraphics[height=5cm]{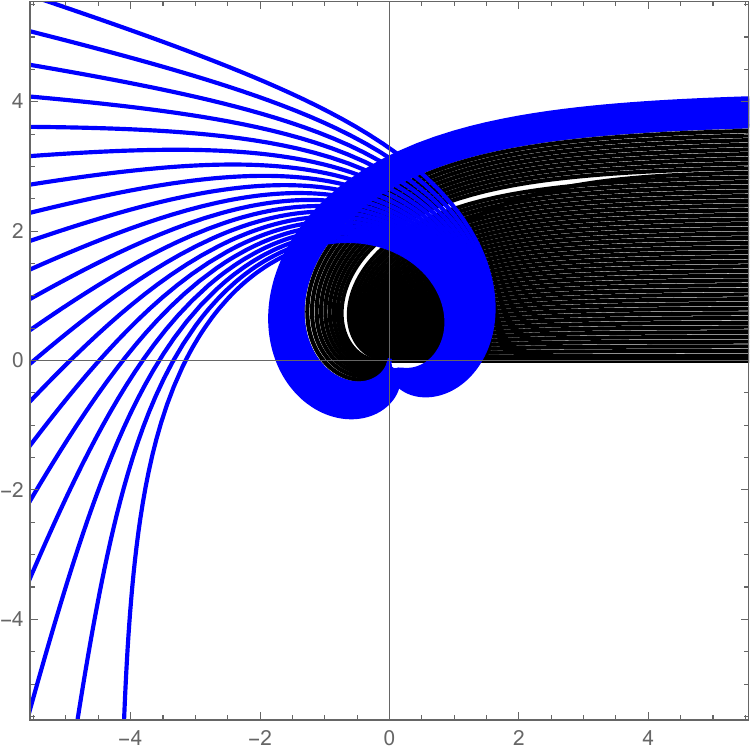} \includegraphics[height=5cm]{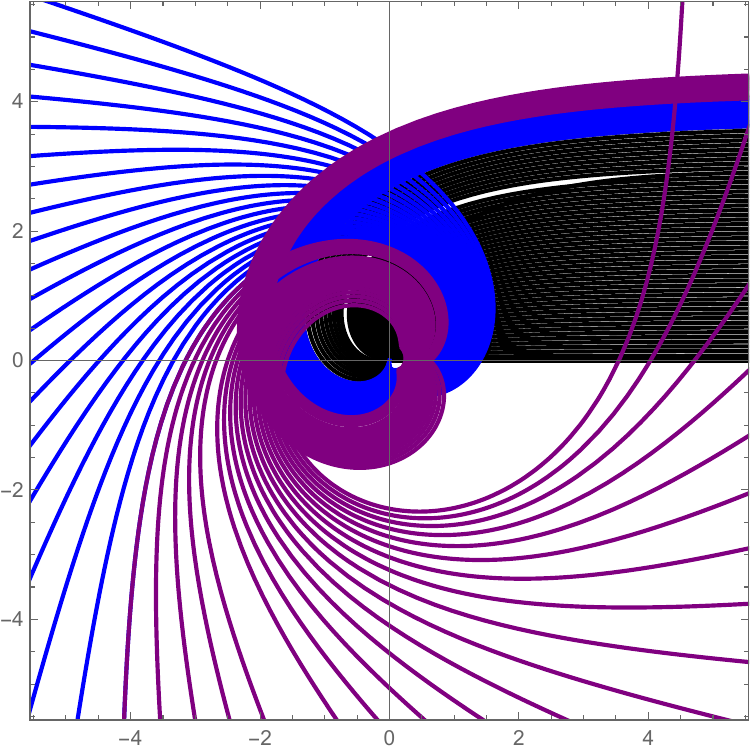}
\includegraphics[height=5cm]{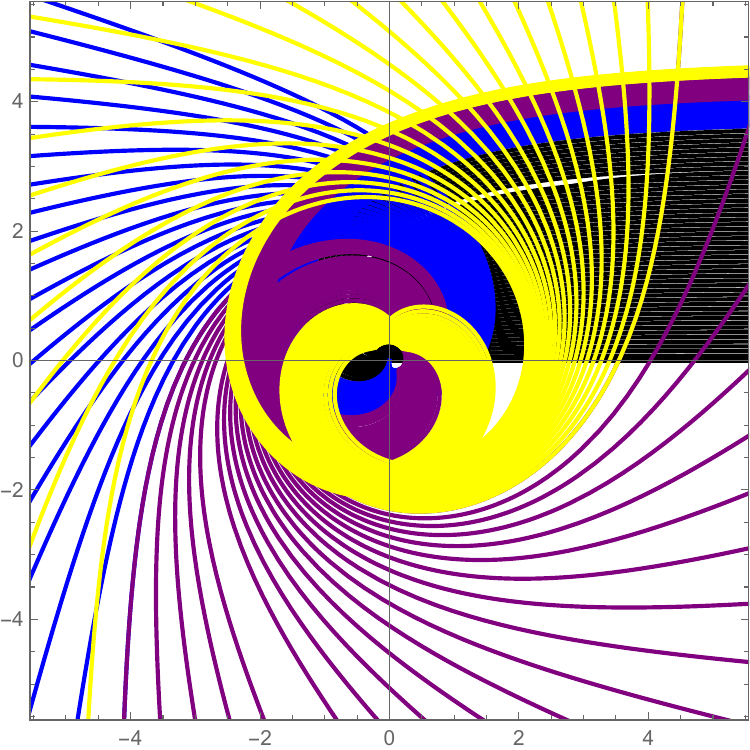} \\
(12a) \hspace{4.5 cm}(12b) \hspace{4.5 cm}(12c)\\
\includegraphics[height=5cm]{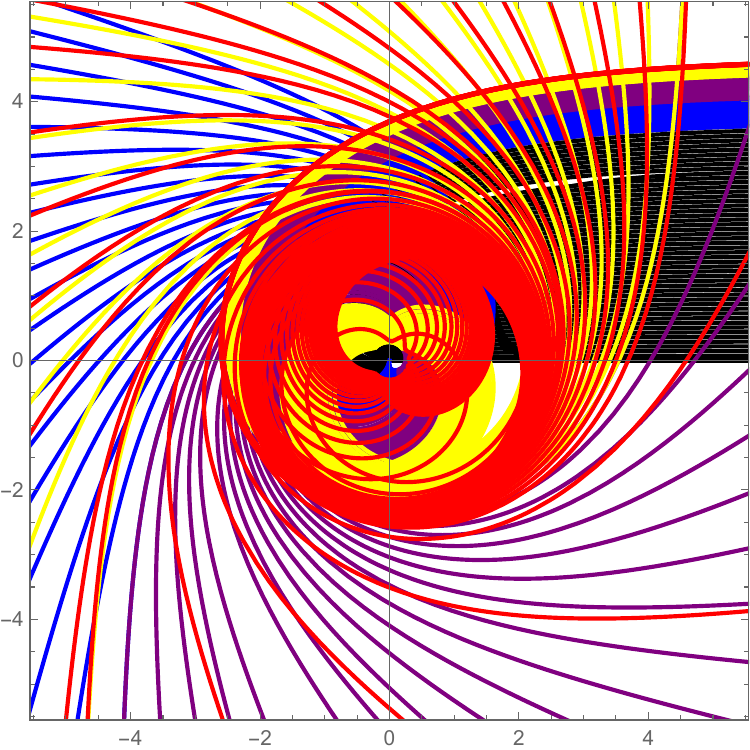}
\includegraphics[height=5cm]{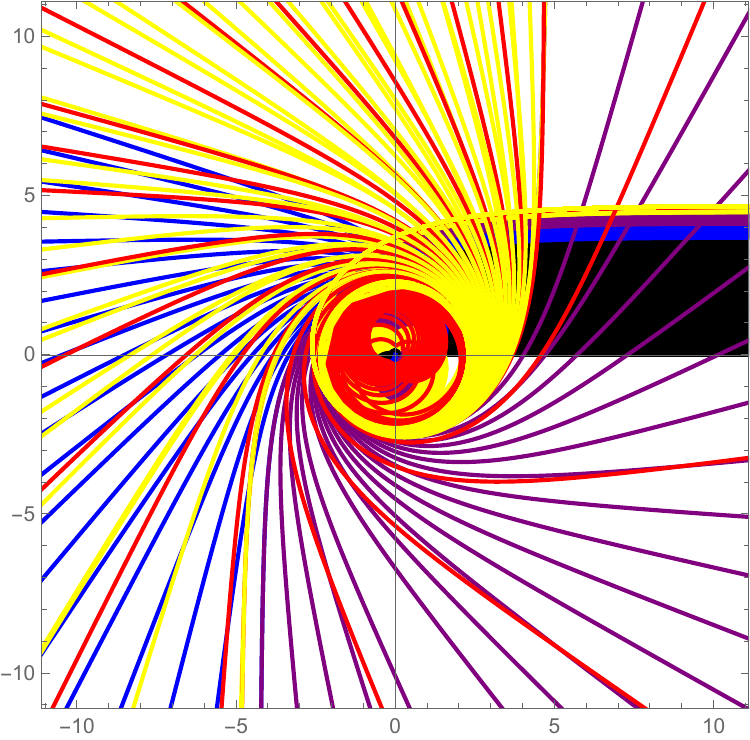}
\includegraphics[height=5cm]{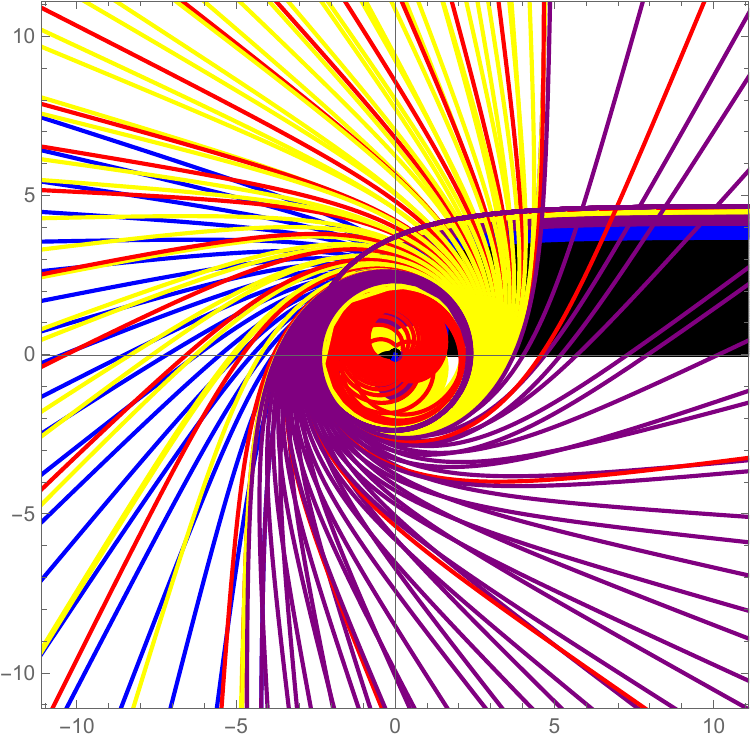}\\
(12d) \hspace{4.5 cm}(12e) \hspace{4.5 cm}(12f)\\
\includegraphics[height=5cm]{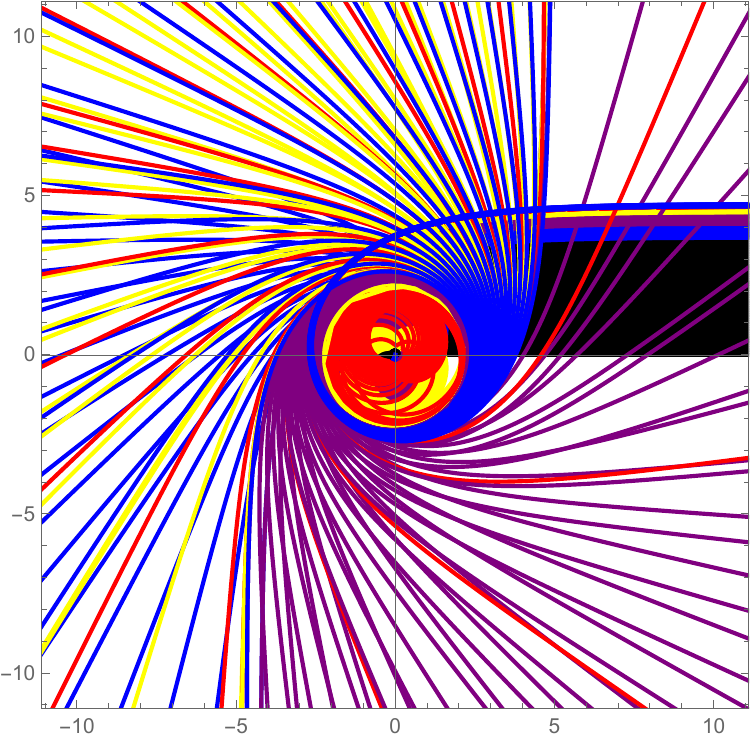}
\includegraphics[height=5cm]{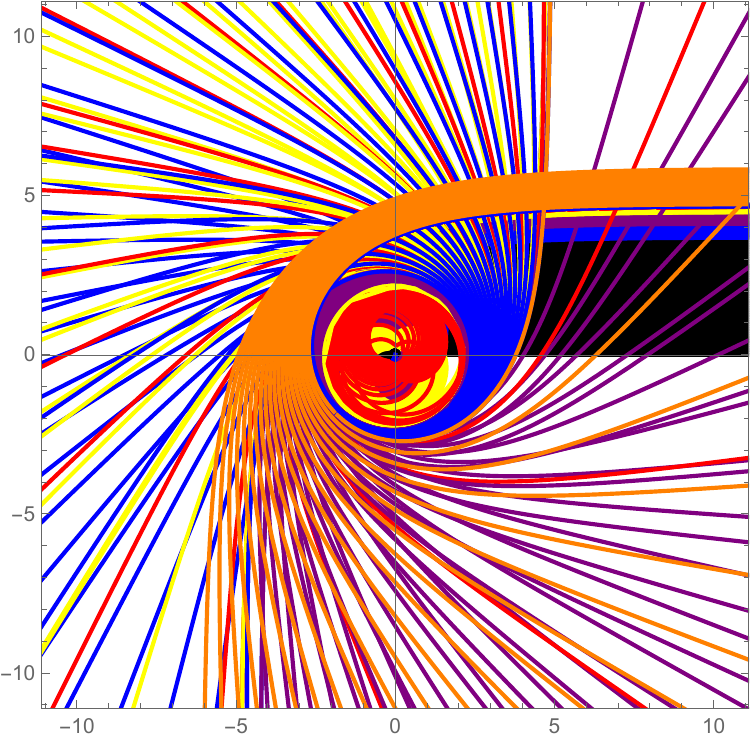}
\includegraphics[height=5cm]{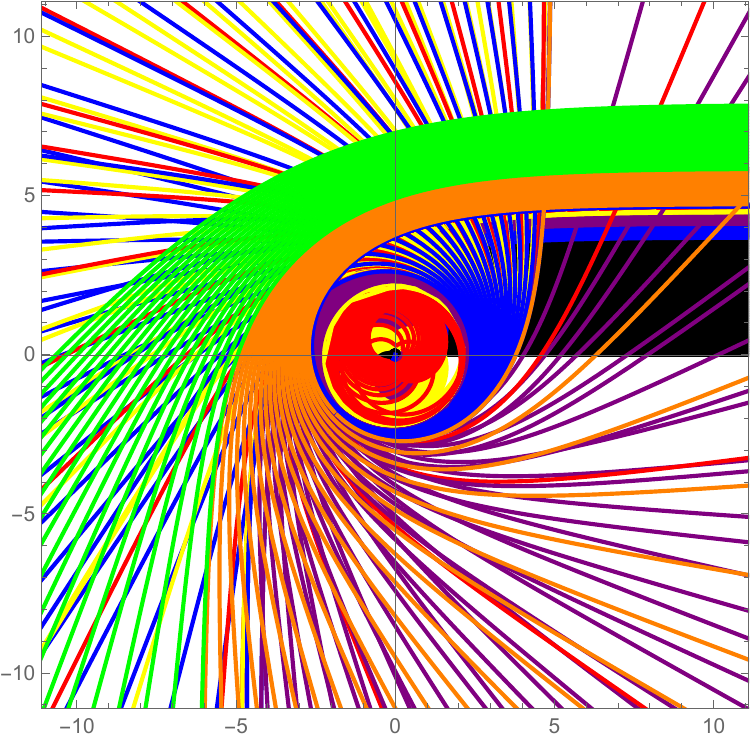}\\
(12g) \hspace{4.5 cm}(12h) \hspace{4.5 cm}(12i)\\
\end{tabular}
\end{center}
\caption{Ray tracing for the symmetric solution (\ref{symetric}), with $\omega = 3/2, M = 1, \rho_0 = 65/27$, $a = 1.1$ and $b = (0,4.19)$ (12a), $b = (0,4.55)$ (12b), $ b = (0,4.69)$ (12c), $ b = (0,4.7490)$ (12d), $b = (0,4.7494)$ (12e), $b = (0,4.75)$ (12f), $ b = (0,4.81)$ (12g), $b = (0,5.94)$ (12h), $ b = (0,8)$ (12i).\label{Figure12}}
\end{figure}

\begin{figure}[t]
\begin{center}
\begin{tabular}{ccc}
\includegraphics[height=5cm]{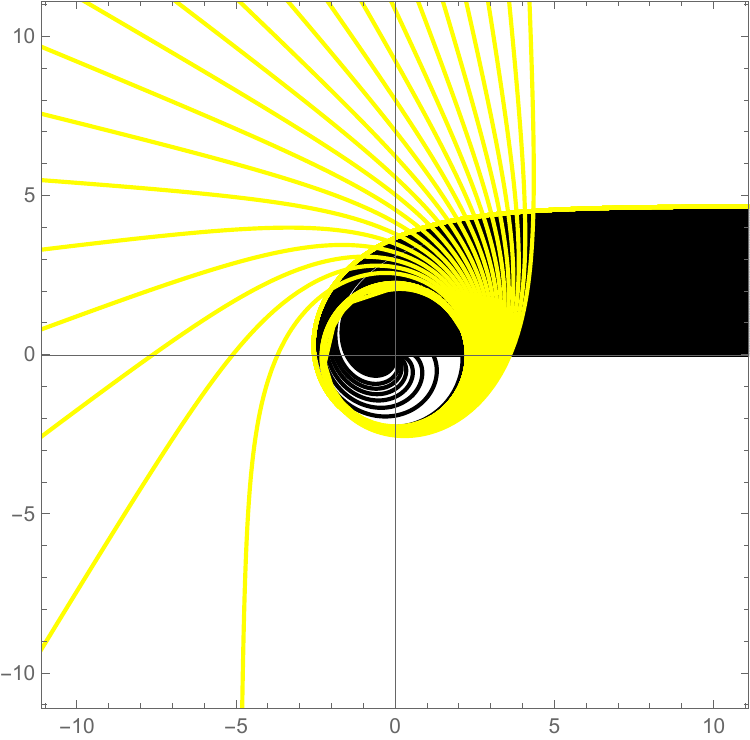} \includegraphics[height=5cm]{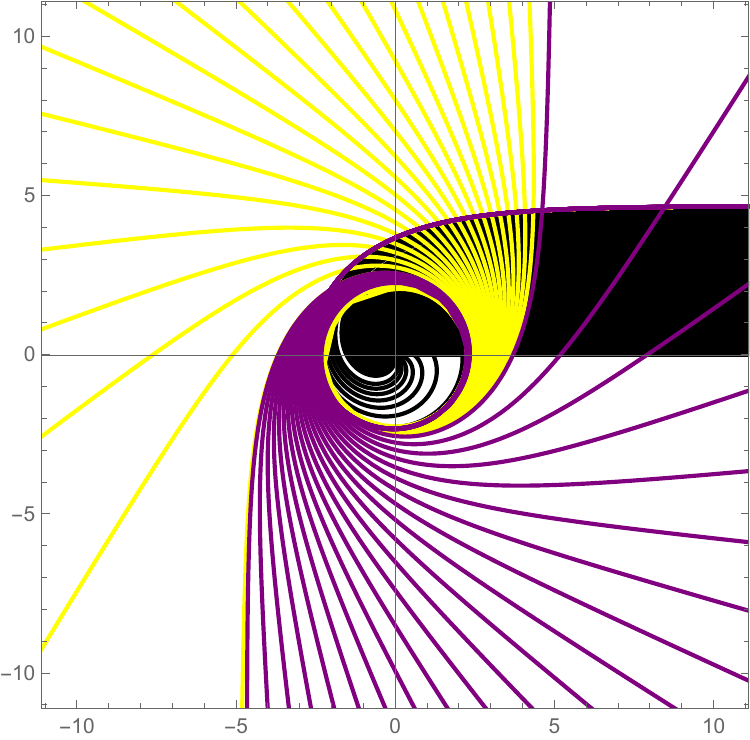}
\includegraphics[height=5cm]{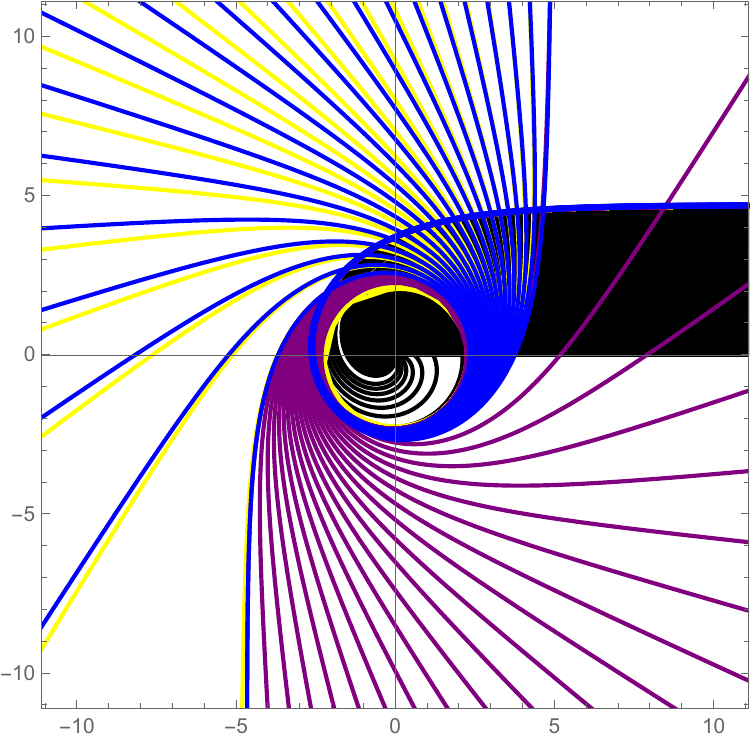} \\
(13a) \hspace{4.5 cm}(13b) \hspace{4.5 cm}(13c)\\
\includegraphics[height=5cm]{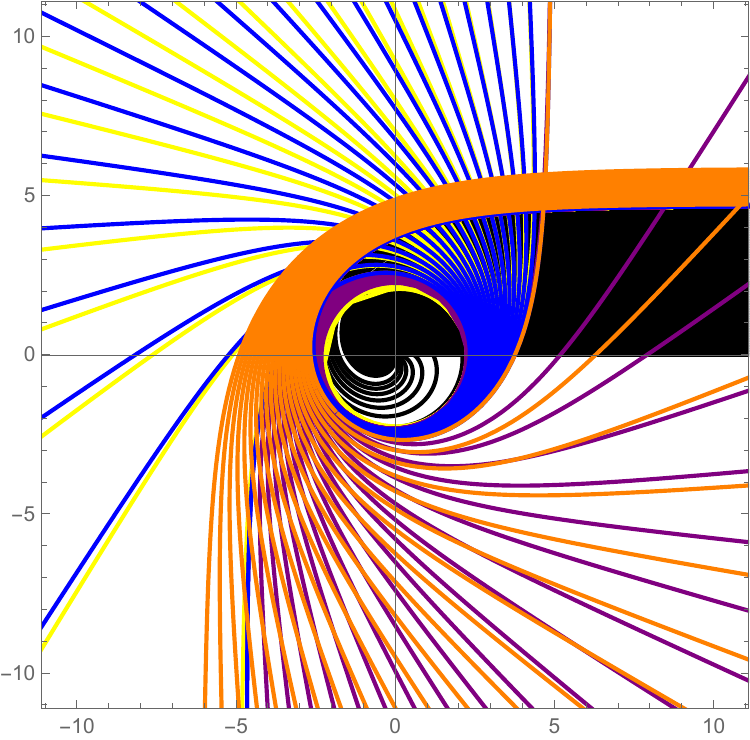}
\includegraphics[height=5cm]{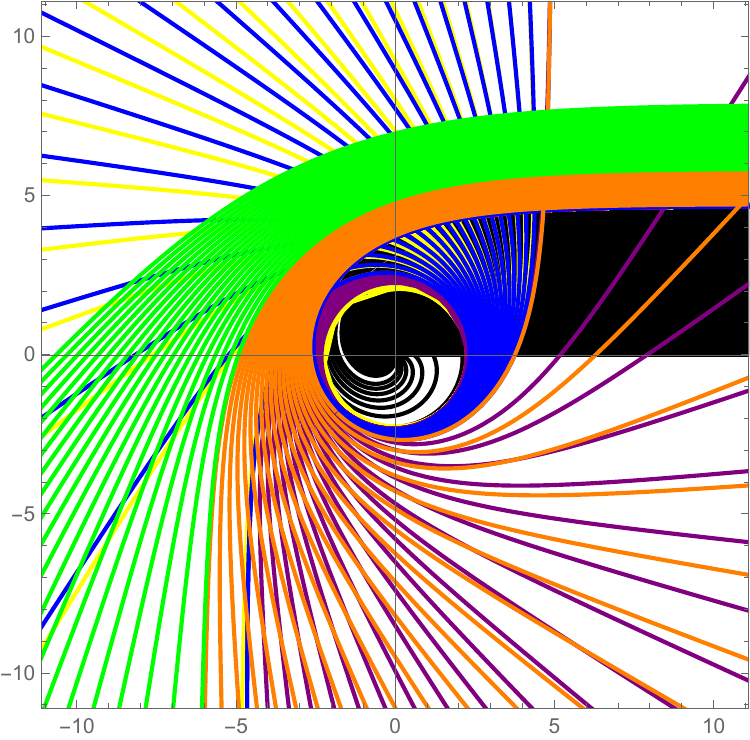}\\
(13d) \hspace{4 cm}(13e)
\end{tabular}
\end{center}
\caption{Ray tracing for the symmetric solution (\ref{symetric}), with $\omega = 3/2, M = 1, \rho_0 = 65/27$, $a = 1.2$ and $b = (0,4.74)$ (13a), $b = (0,4.75)$ (13b), $ b = (0,4.81)$ (13c), $ b = (0,5.94)$ (13d), $b = (0,8)$ (13e).\label{Figure13}}
\end{figure}

\begin{figure}[t]
\begin{center}
\begin{tabular}{ccc}
\includegraphics[height=5cm]{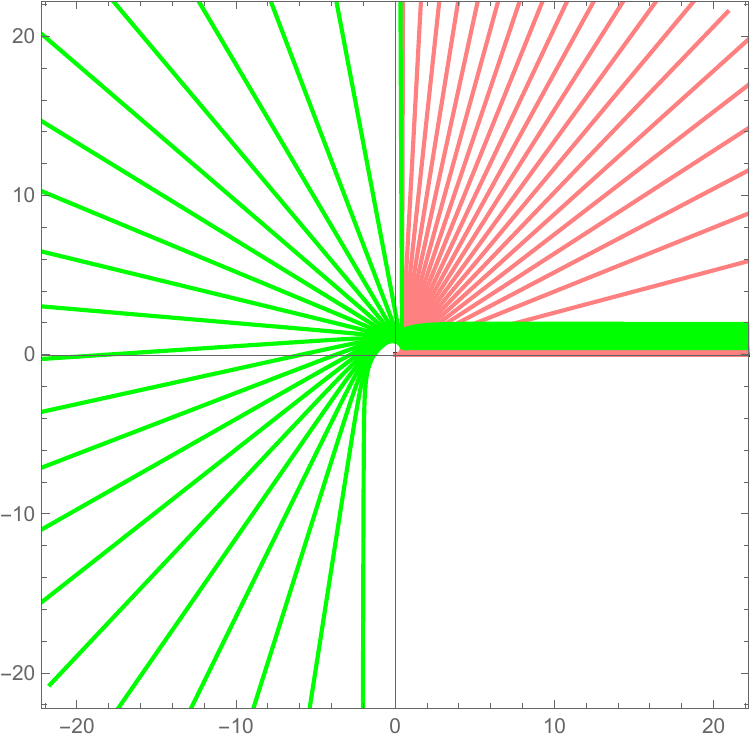} \includegraphics[height=5cm]{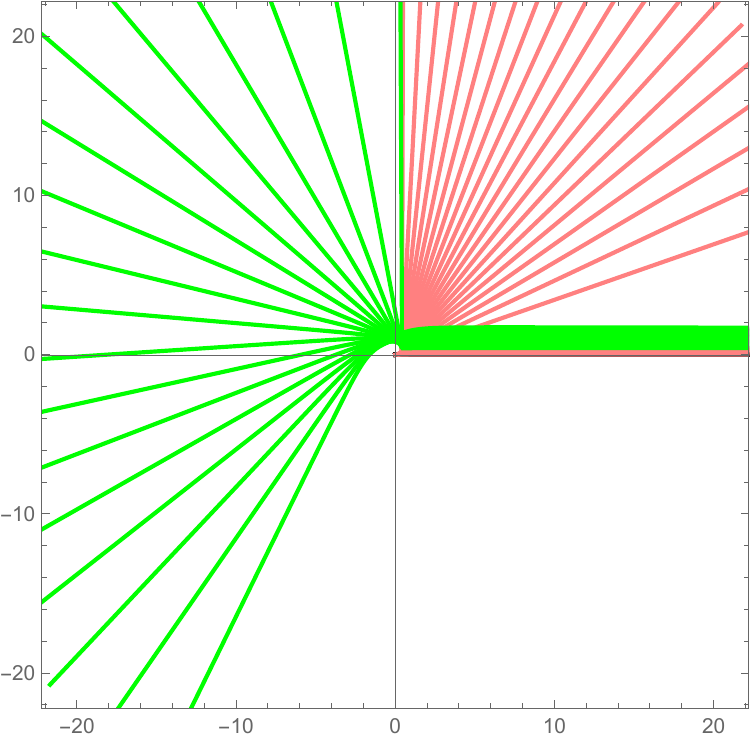}
\includegraphics[height=5cm]{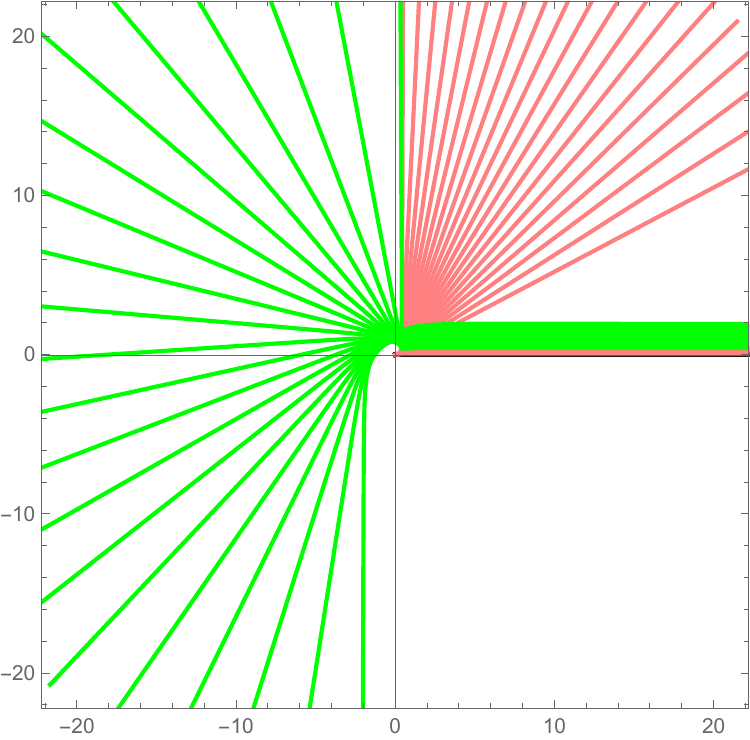} \\
(14a) \hspace{4.5 cm}(14b) \hspace{4.5 cm}(14c)\\
\includegraphics[height=5cm]{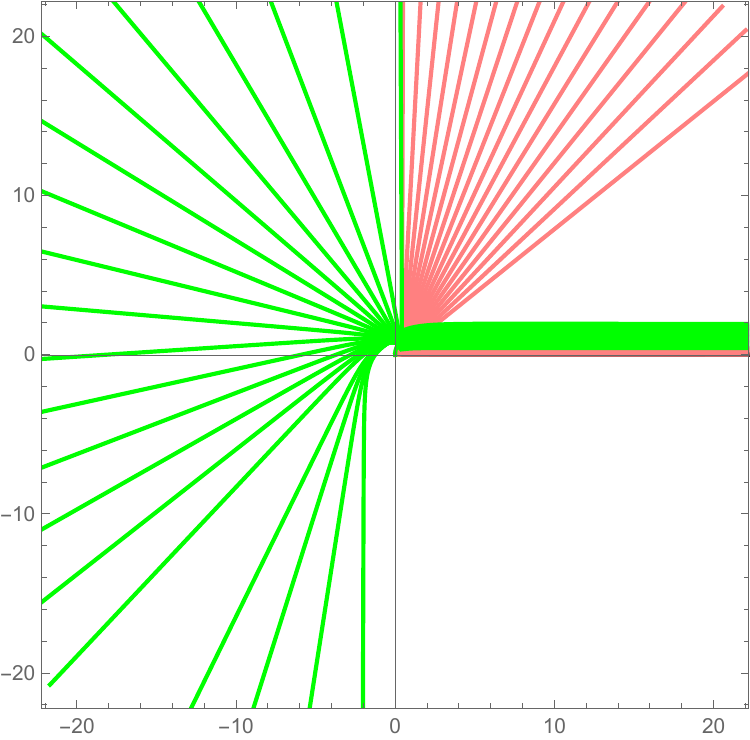}
\includegraphics[height=5cm]{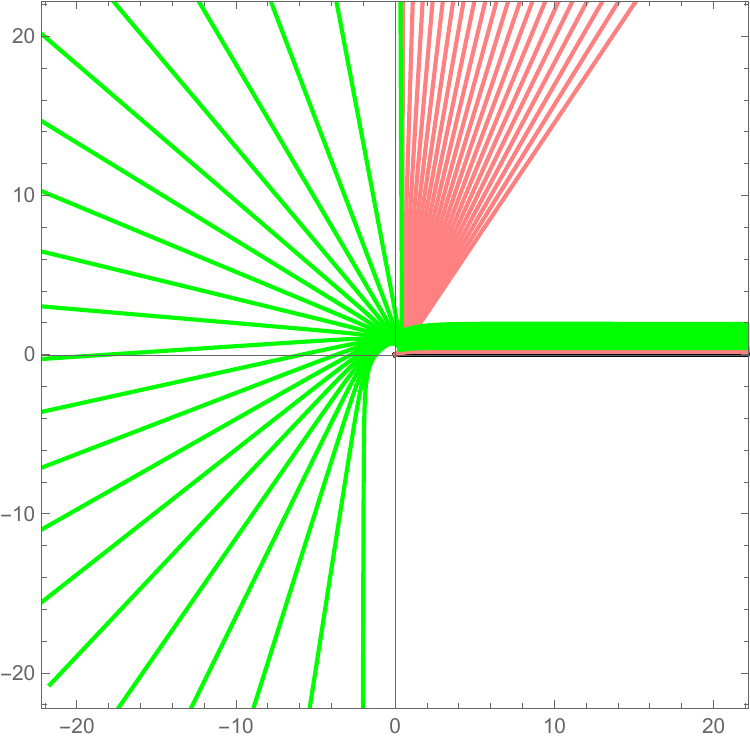}
\includegraphics[height=5cm]{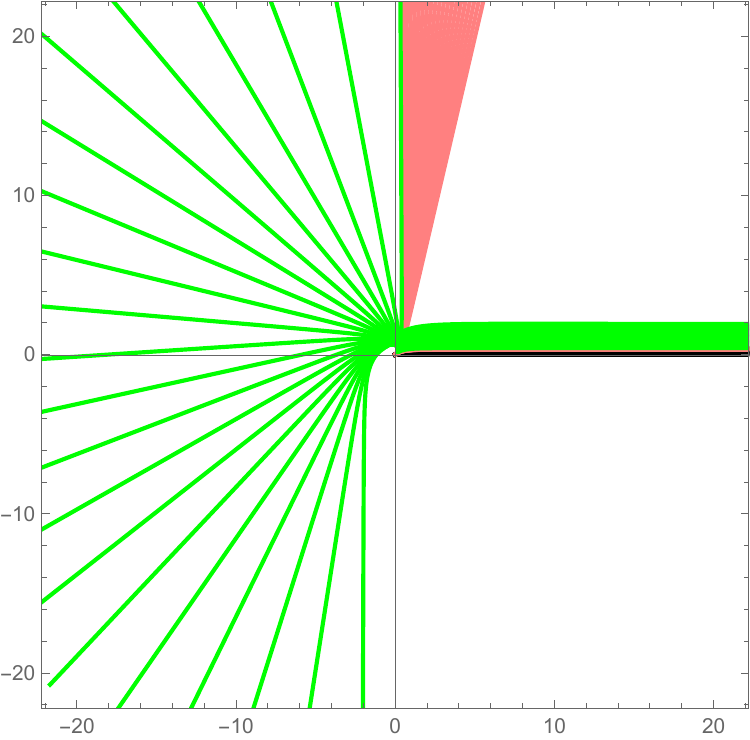}\\
(14d) \hspace{4 cm}(14e)  \hspace{4.5 cm}(14f)\\
\end{tabular}
\end{center}
\caption{Ray tracing for the symmetric solution (\ref{symetric}), with $\omega = 3/2, M = 1, \rho_0 = 65/27, b = (0,1.97)$, $a = 0.1$, $\theta_{open} \approx 73$°  (14a), $a = 0.2$, $\theta_{open} \approx 70$° (14b), $a = 0.3$,  $\theta_{open} \approx 62$° (14c), $a = 0.4$,  $\theta_{open} \approx 51$° (14d), $a = 0.5$, $\theta_{open} \approx 34$° (14e), $a = 0.6$, $\theta_{open} \approx 13$° (14f). Pink lines represent light rays with small impact parameters (the smallest are black, barely visible here but check Fig. (11.a)). Such rays are completely reflected by the potential barrier and never reach the equatorial plane (vertical axis). The largest angular dispersion in each case is given by $\theta_{open}$. This actually means that any light ray with a sufficiently small impact parameter coming from infinity towards the central object forming an angle $0<\theta<\theta_{open}$  will be collimated around the horizontal axis (North pole) of the object.  
Higher impact parameters are represented by green lines and are deflected forward crossing the equatorial plane once, forming a structure which can be seen with more clarity in Fig. (11.a). 
\label{Figure14}}
\end{figure}

\begin{figure}[t]
\begin{center}
\begin{tabular}{ccc}
\includegraphics[height=5cm]{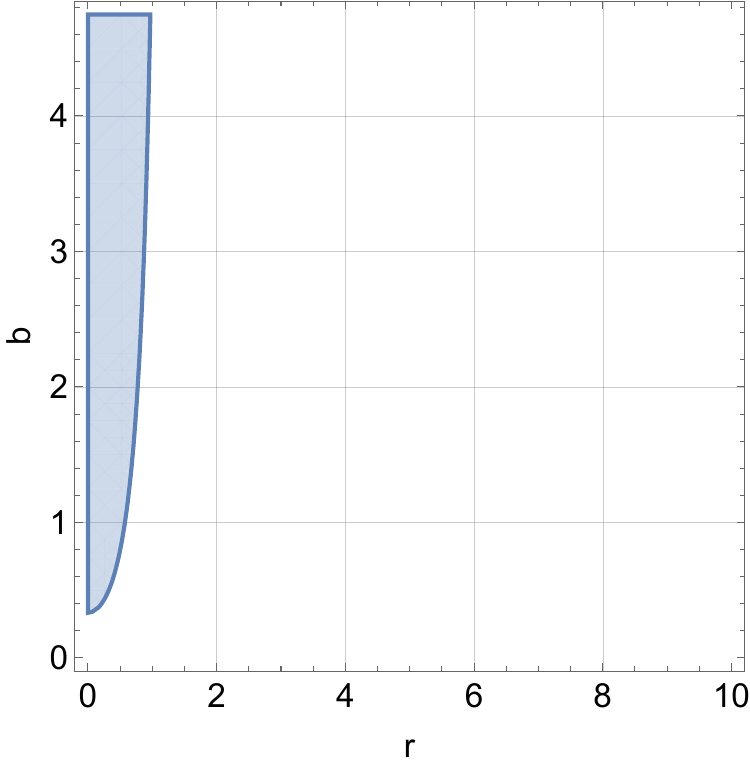} \includegraphics[height=5cm]{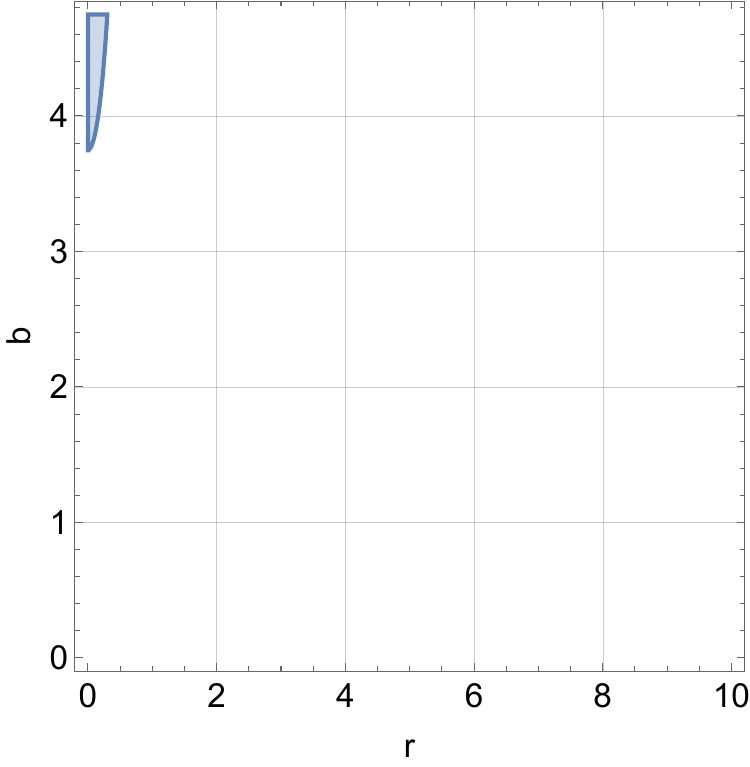}
\includegraphics[height=5cm]{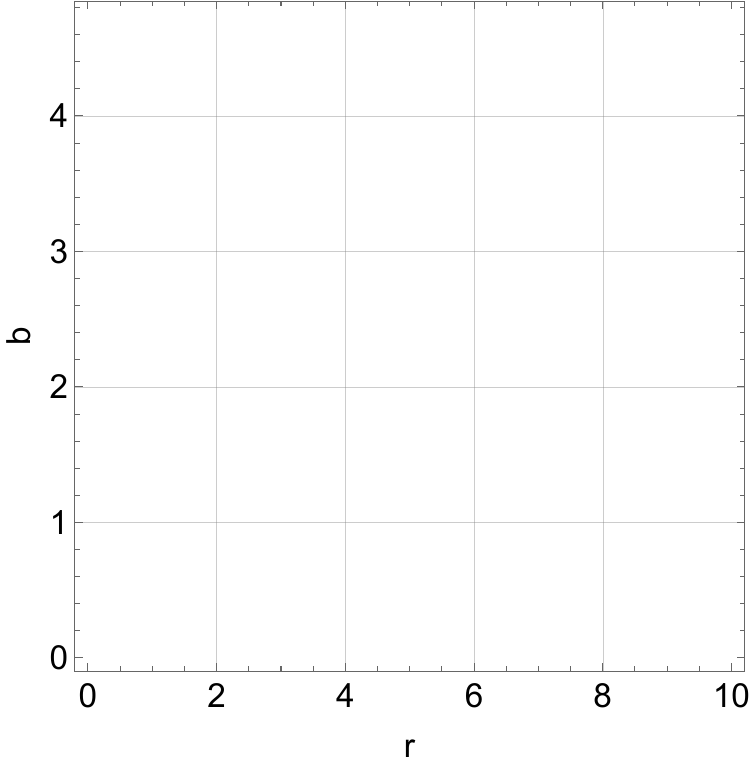} \\
(15a) \hspace{4.5 cm}(15b) \hspace{4.5 cm}(15c)\\
\end{tabular}
\end{center}
\caption{Region forbidden to geodesic motion for the symmetric solution (\ref{symetric}), with $\omega = 3/2, M = 1, \rho_0 = 65/27$, $a = 0.6$ (15a), $a = 1.1$ (15b), $a = 1.2$ (15c).\label{Figure15}}
\end{figure}

\begin{figure}[t]
\begin{center}
\begin{tabular}{ccc}
\includegraphics[height=5.0cm]{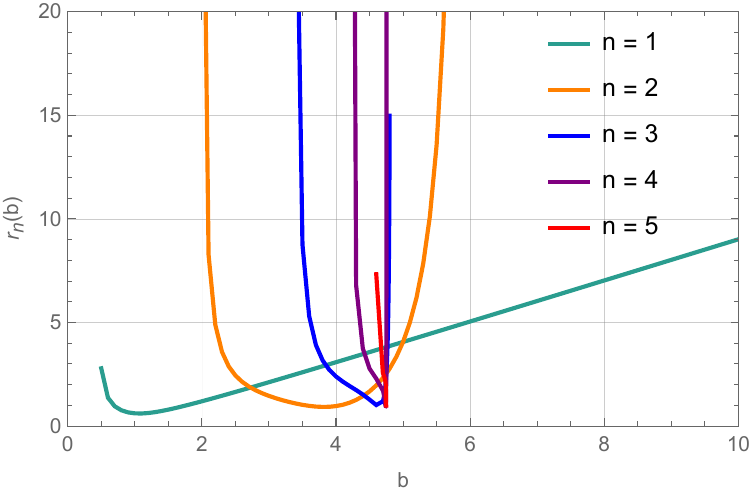}
\includegraphics[height=5.0cm]{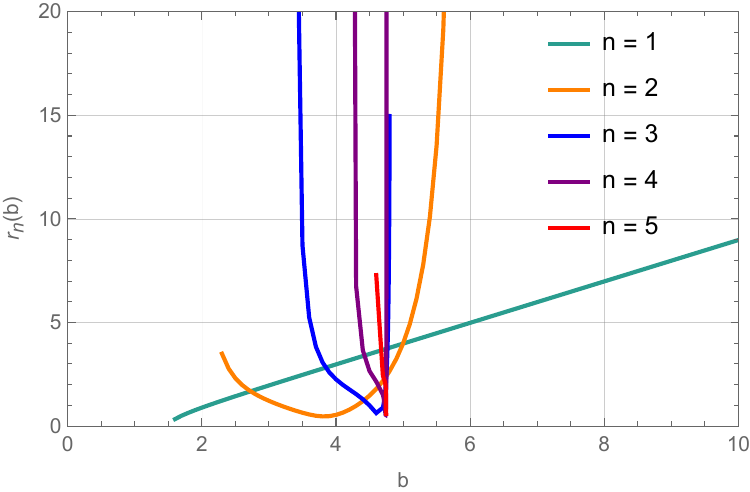} \\ 
(16a) \hspace{7 cm}(16b)\\
\includegraphics[height=5.0cm]{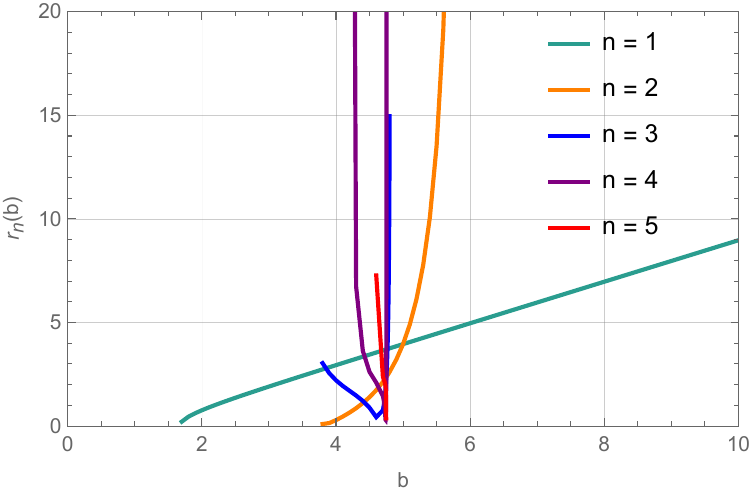}
\includegraphics[height=5.0cm]{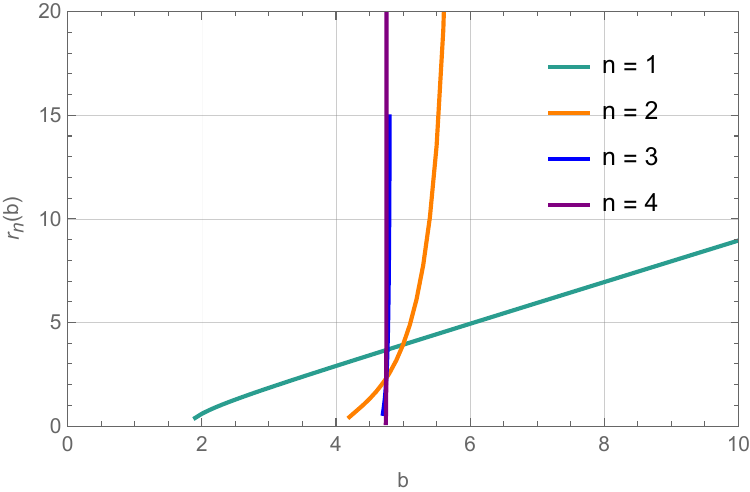} \\ 
(16c) \hspace{7 cm}(16d)\\
\end{tabular}
\end{center}
\caption{Transfer function $r=r_n(b)$ for the symmetric solution (\ref{symetric}) with $\omega = 3/2, M = 1, \rho_0 = 65/27$, $a = 0.6$ (16a), $a = 1$ (16b), $a = 1.1$ (16c), $a  = 1.2$ (16d) .\label{Figure16}}
\end{figure}

\begin{center}
\begin{table}[h!]
\centering
\caption{Relative intensity of the different intersections as a function of the central maximum of the shadow potential, $V_{sh}(r_{01})$, for $\omega = 3/2, m = 1$ and $\rho_0 = 65/27$.}
\begin{tabular}{c||c||c||c||c||c||c}
\hline
\hline
$a$ & $V_{sh}(r_{01})$ & $I_r (\%)\; (n = 1)$ &  $I_r (\%)\; (n = 2)$ & $I_r (\%)\; (n = 3)$ & $I_r (\%)\; (n = 4)$ & $I_r (\%)\; (n = 5)$ \\ 
\hline
\hline
0.6 & 9 & 86.90 & 5.91 & 4.35 & 2.14 & 0.69 \\
\hline
\hline
1.0 & 0.20 & 87.60 & 5.12 & 4.43 & 2.14 & 0.69 \\
\hline
\hline
1.1 &0.07 & 91.30 & 5.33 & 0.40 & 2.23 & 0.71\\
\hline
\hline
1.2 & 0.02 & 94.07 & 5.48 & 0.41 & 0.04 & 0.01\\
\hline
\hline
\end{tabular}
\label{TableI}
\end{table}
\end{center}

\begin{figure}[t]
\begin{center}
\begin{tabular}{ccc}
\includegraphics[height=4.0cm]{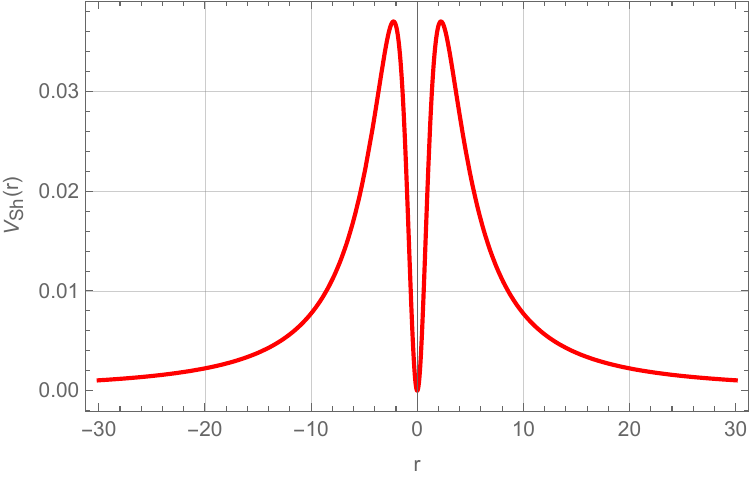} \includegraphics[height=4.0cm]{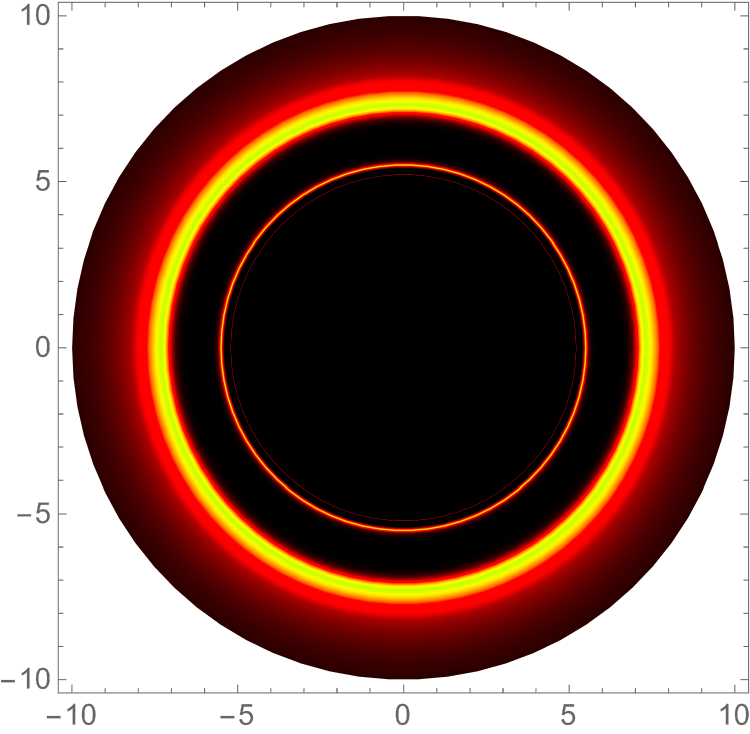} \includegraphics[height=4.0cm]{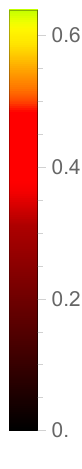}\\ 
(17a) \hspace{4.5 cm}(17b)\\
\includegraphics[height=4.0cm]{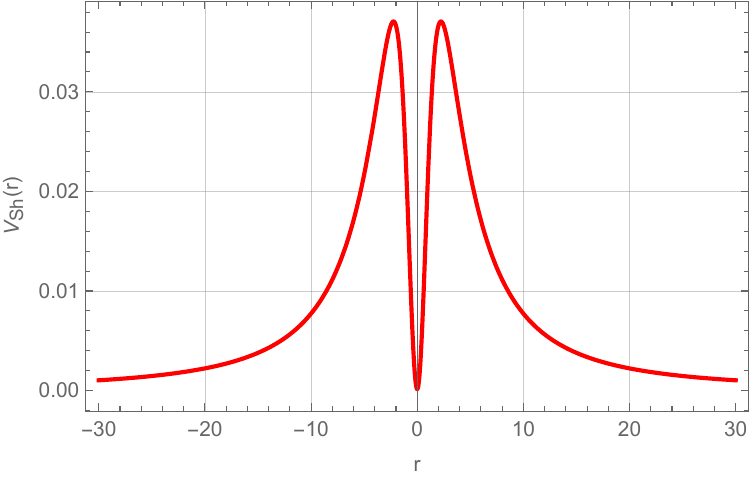} \includegraphics[height=4.0cm]{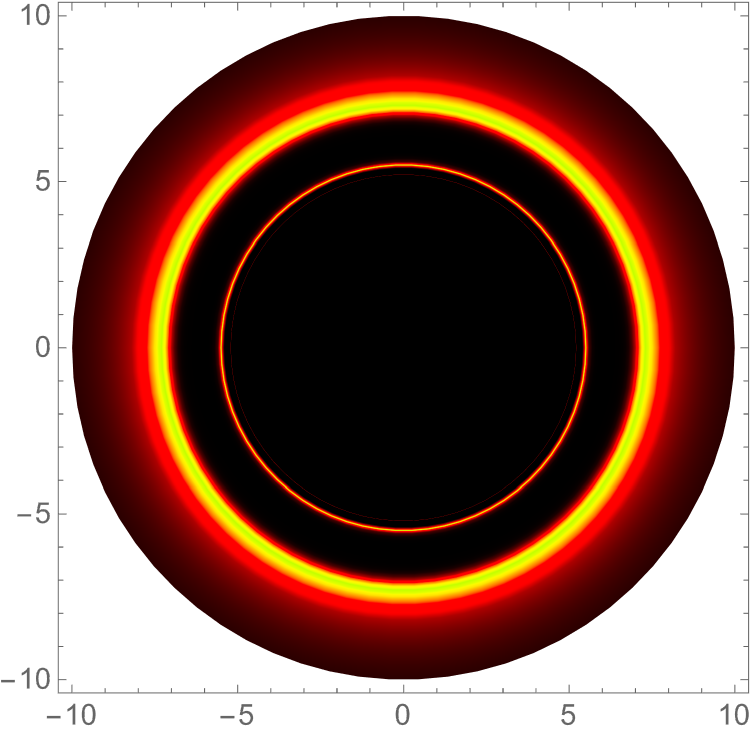} \includegraphics[height=4.0cm]{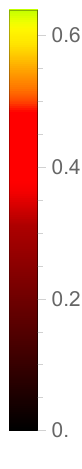}\\  
(17c) \hspace{4.5 cm}(17d)\\
\includegraphics[height=4.0cm]{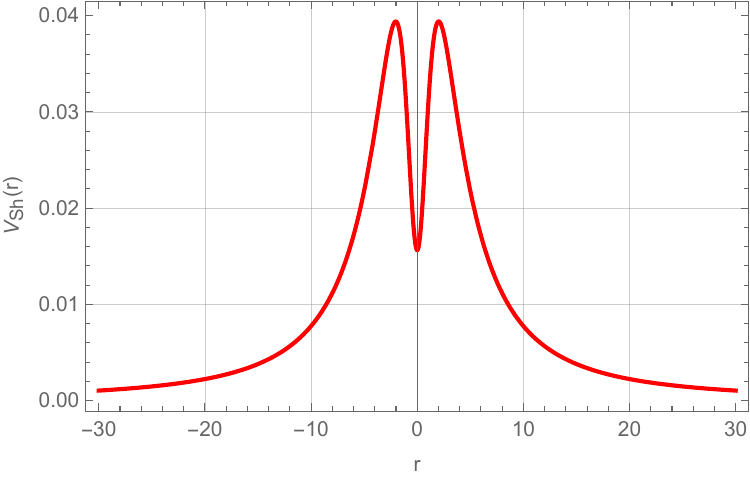} \includegraphics[height=4.0cm]{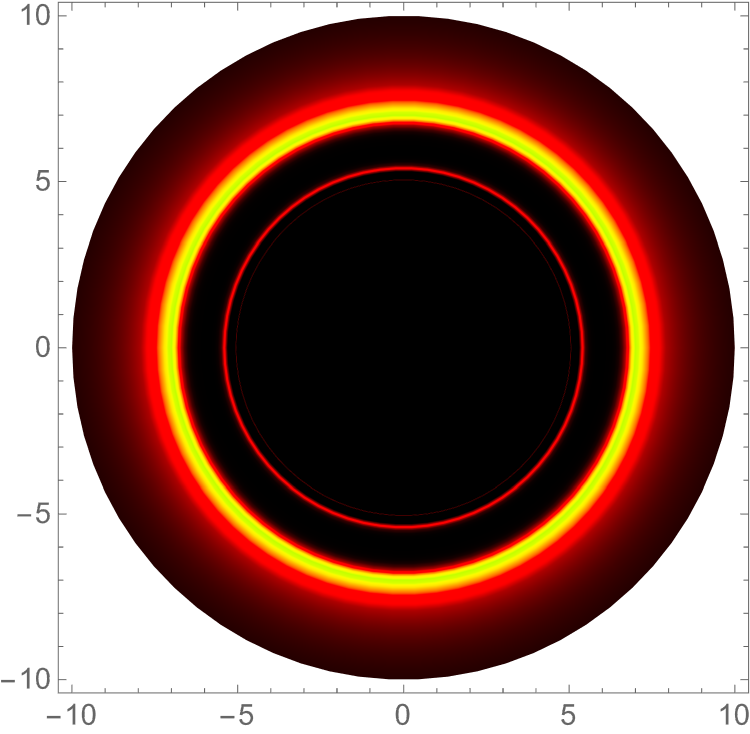} \includegraphics[height=4.0cm]{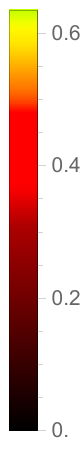}\\
(17e) \hspace{4.5 cm}(17f)\\
\includegraphics[height=4.0cm]{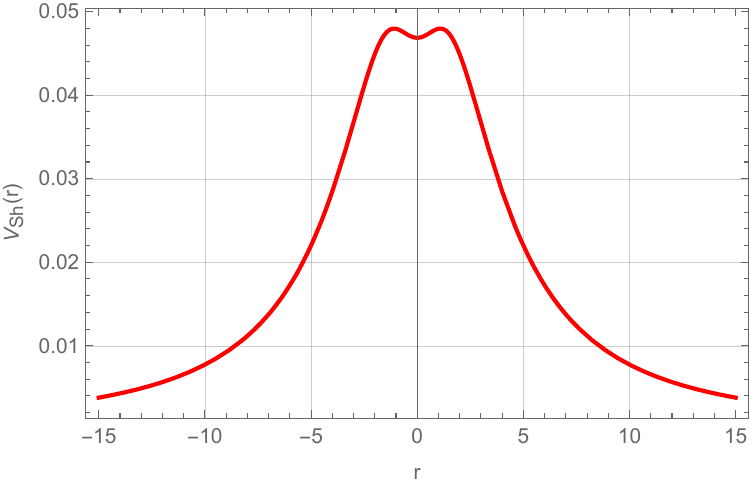} \includegraphics[height=4.0cm]{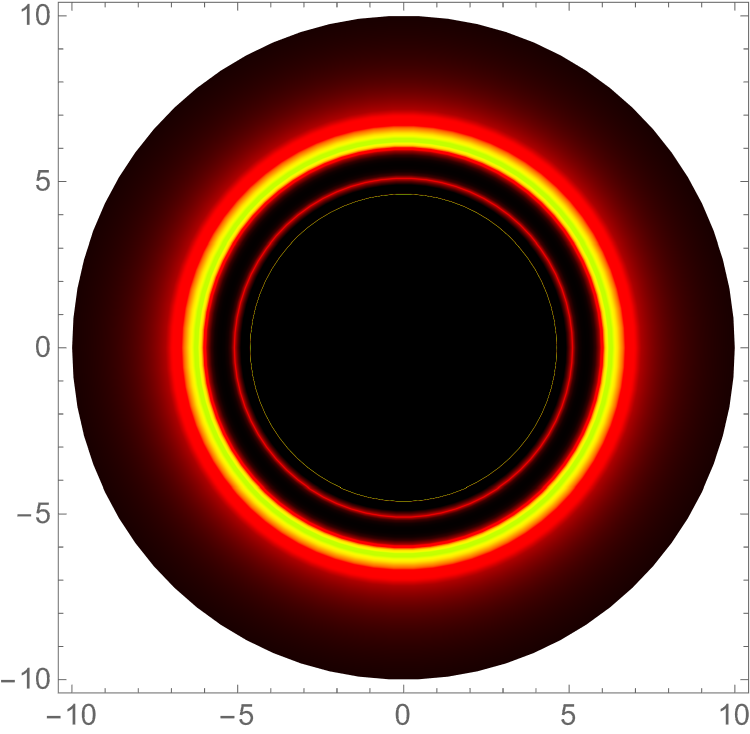} \includegraphics[height=4.0cm]{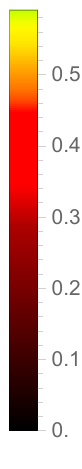}\\
(17g) \hspace{4.5 cm}(17h)\\
\end{tabular}
\end{center}
\caption{The shadow potential (left panels) and optical image (right panels) for the symmetric solution (\ref{symetric}) with $M = 1, \omega = 3/2, a = 2$, $\rho_0 = 10^{-4}$ (17a), $\rho_0 = 10^{-2}$ (17c), $\rho_0 = 1$ (17e) and $\rho_0 = 3$ (17g).\label{Figure17}}
\end{figure}

\begin{figure}[t]
\includegraphics[height=5.1cm]{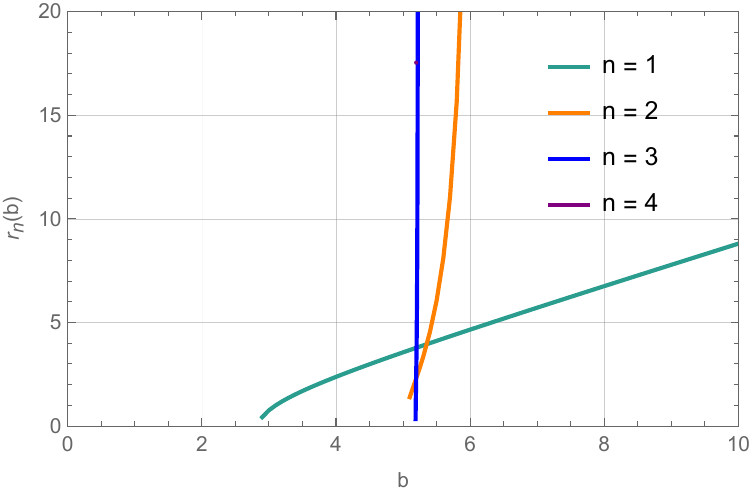}
\includegraphics[height=5.1cm]{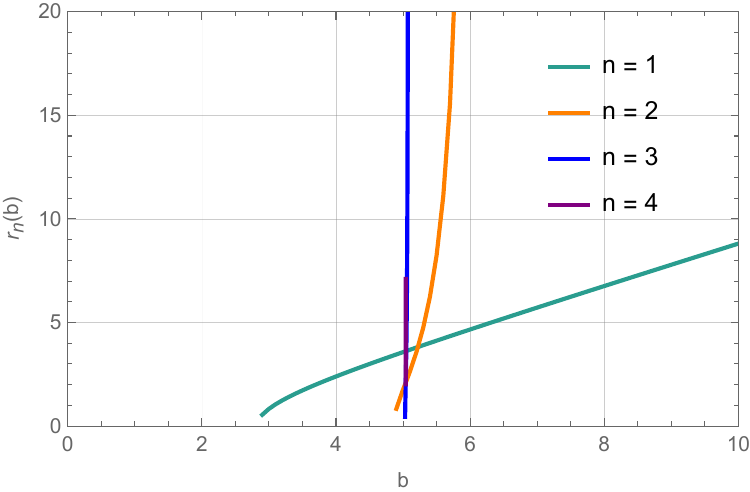}\\
(18a) \hspace{6 cm}(18b)
\caption{Transfer function $r=r_n(b)$ for the symmetric solution (\ref{symetric}) with $M = 1, \omega = 3/2, a = 2$, $\rho_0 = 10^{-2}$ (18a) and $\rho_0 = 1$ (18b).\label{Figure18}}
\end{figure}

\begin{figure}[t]
\includegraphics[height=5.1cm]{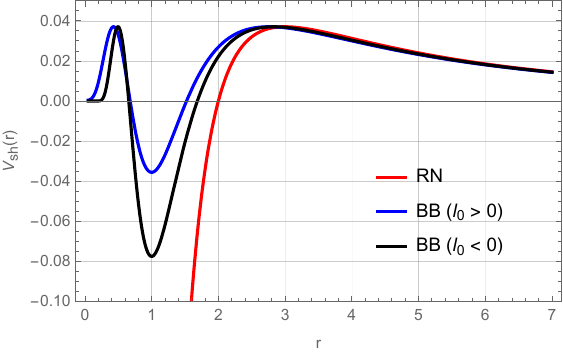}
\includegraphics[height=5.1cm]{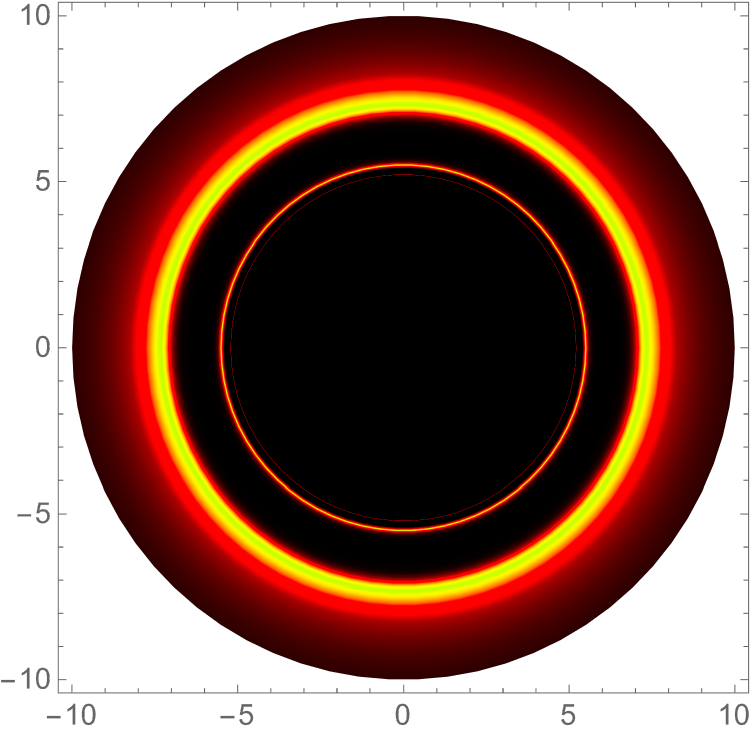}
\includegraphics[height=5.1cm]{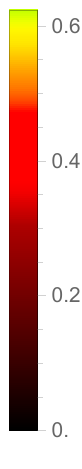}\\
(19a) \hspace{6 cm}(19b)\\
\includegraphics[height=5.1cm]{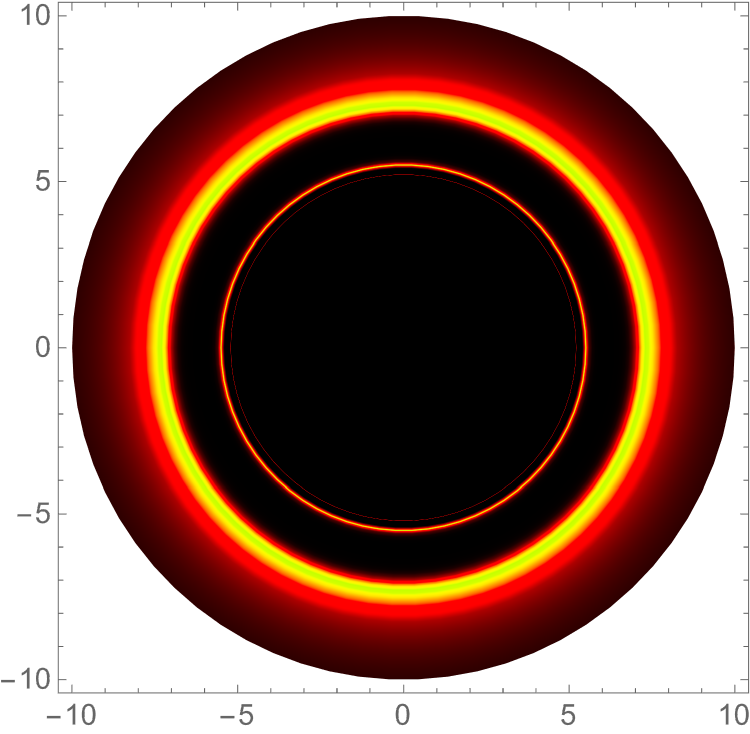}
\includegraphics[height=5.1cm]{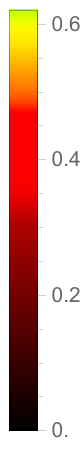}
\includegraphics[height=5.1cm]{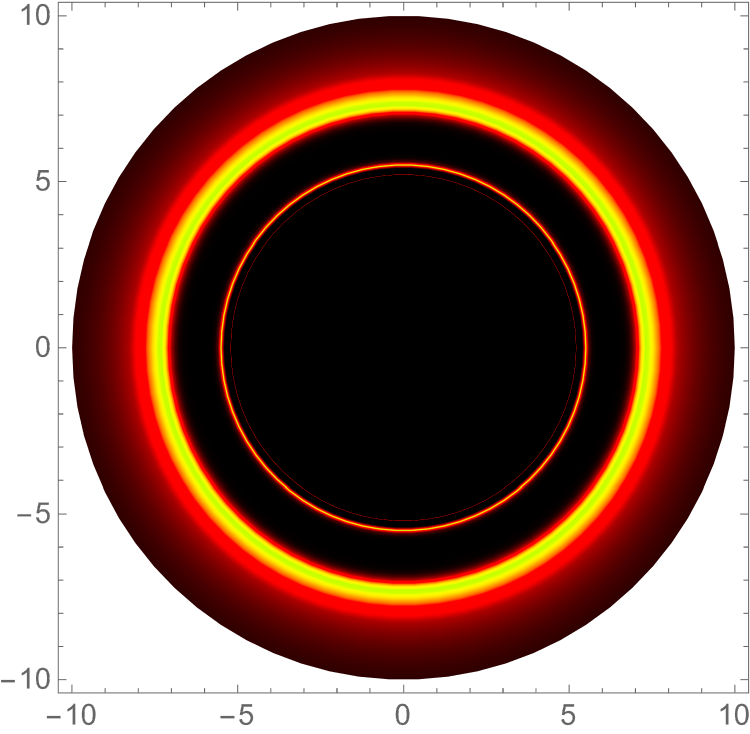}
\includegraphics[height=5.1cm]{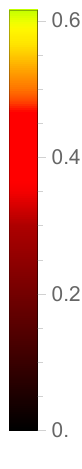}\\
(19c) \hspace{6 cm}(19d)\\
\caption{The shadow potential (19a) and the optical image for the asymmetric solution (\ref{asymetric}) for $l_0 = 1/5$ (19b), $l_0 = -1/5$ (19c) and RN (19d) with  $\tilde{\rho}_0 = 0, \tilde{M} = r_0 = \omega = \alpha = 1$.\label{Figure19}}
\end{figure}

\begin{figure}[t]
\includegraphics[height=5.1cm]{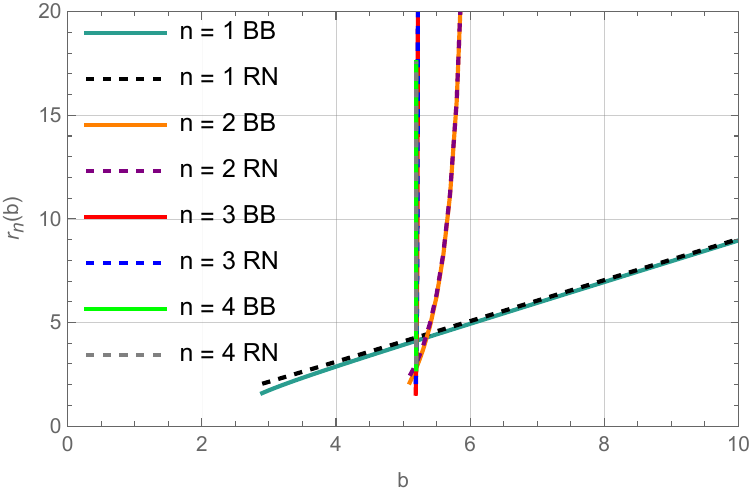}
\includegraphics[height=5.1cm]{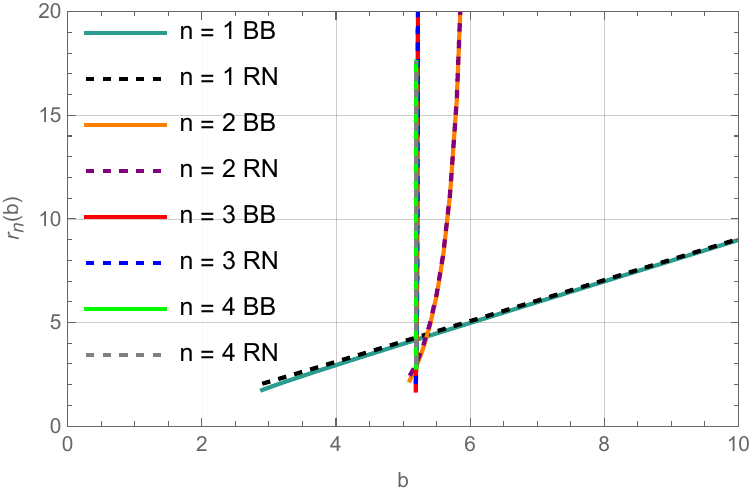}\\
(20a) \hspace{6 cm}(20b)\\
\caption{Transfer function $r=r_n(b)$ for the asymmetric solution (\ref{asymetric}) for $l_0 = 1/5$ (20a) and $l_0 = -1/5$ (20b) compared with RN for $\tilde{\rho}_0 = 0, \tilde{M} = r_0 = \omega = \alpha = 1$.\label{Figure20}}
\end{figure}

\begin{figure}[t]
\includegraphics[height=5.1cm]{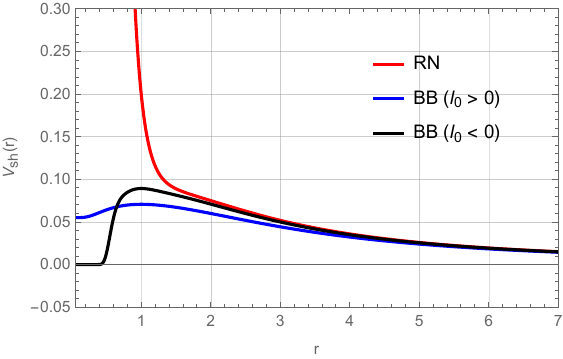}
\includegraphics[height=5.1cm]{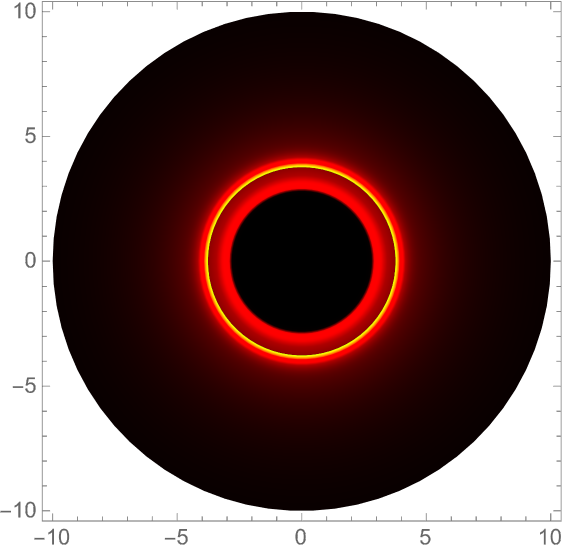}
\includegraphics[height=5.1cm]{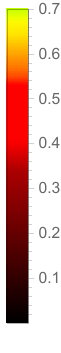}\\
(21a) \hspace{6 cm}(21b)\\
\includegraphics[height=5.1cm]{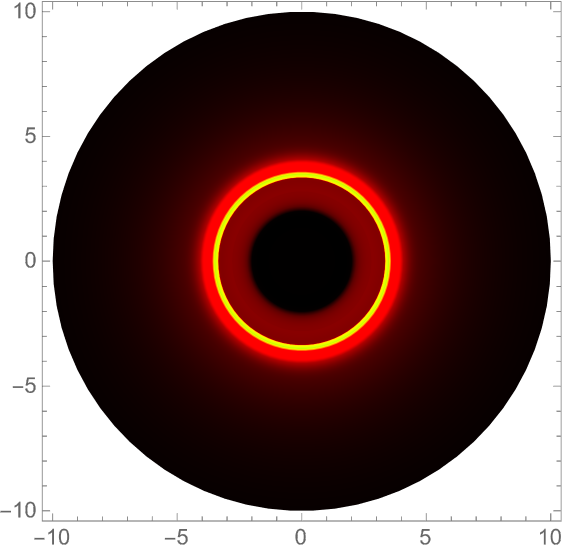}
\includegraphics[height=5.1cm]{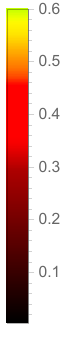}
\includegraphics[height=5.1cm]{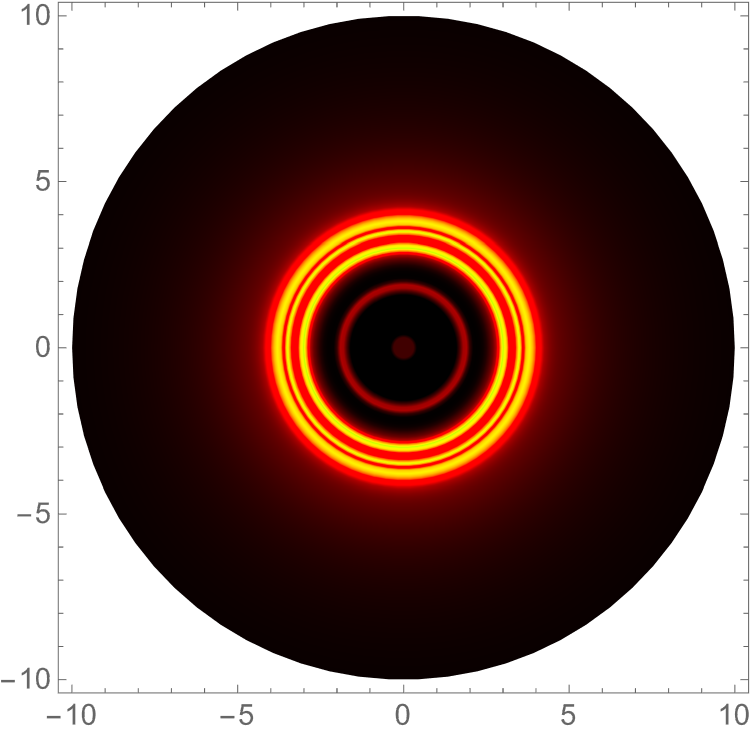}
\includegraphics[height=5.1cm]{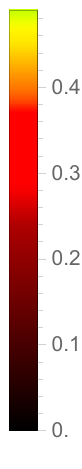}\\
(21c) \hspace{6 cm}(21d)\\
\caption{The shadow potential (21a) and optical image for the asymmetric solution (\ref{asymetric}) for $l_0 = 1$ (21b), $l_0 = -1$ (21c) and RN (21d) with $\tilde{M} = r_0 = \omega = \alpha = l = 1, \tilde{\rho_0} = {1.2}/{\tilde{\Sigma}_0^{2 \omega +2}}$.\label{Figure21}}
\end{figure}

\begin{figure}[t]
\includegraphics[height=5.1cm]{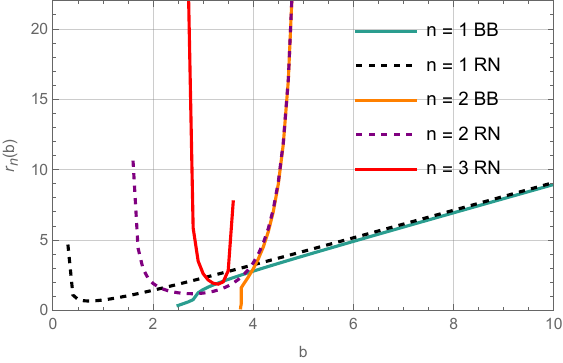}
\includegraphics[height=5.1cm]{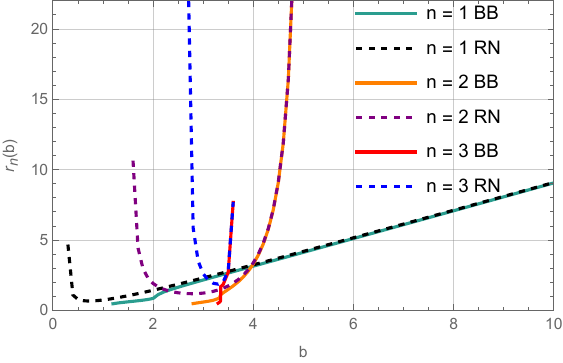}
(22a) \hspace{6 cm}(22b)\\
\caption{Transfer function $r=r_n(b)$ for the asymmetric solution (\ref{asymetric}) for $l_0 = 1$ (22a), $l_0 = -1$ (22b) and RN with $\tilde{M} = r_0 = \omega = \alpha = l = 1, \tilde{\rho_0} = {1.2}/{\tilde{\Sigma}_0^{2 \omega +2}}$.\label{Figure22}}
\end{figure}

\subsection{Quasinormal modes and echoes}
When analyzing the time-domain profiles of axial gravitational perturbations, most of the results previously obtained for scalar perturbations are confirmed. We examined the symmetric configurations with horizons and compared them to the standard Reissner–Nordström case, recovered by setting $\Sigma_I(r)$ equal to the radial coordinate. In the left column of Fig.~\ref{Figure1}, Fig.~(\ref{Figure1}a) shows the effective potentials of the symmetric Kiselev-type black-bounce geometry and the Reissner–Nordström spacetime for different values of the angular momentum, plotted in terms of the corresponding tortoise coordinate. The associated ringdown signals for $l = 3$ and $l = 4$ are displayed in the right column, Fig.~(\ref{Figure1}b). The potential exhibits a single-barrier profile that asymptotically vanishes both near the horizon and at spatial infinity. The resulting time-domain waveforms are almost indistinguishable between the two configurations, a feature that also persists in the asymmetric horizon case, and is directly related to the causal structure encoded in the tortoise coordinate.

For horizonless symmetric solutions, multiple potential barriers may arise, giving rise to the appearance of echoes. Specifically, for small values of the parameter $a$ with constant density, an additional third barrier emerges [see Fig.~\ref{Figure2}], whose amplitude decreases as $a$ increases. This additional peak is responsible for the appearance of intermediate echoes, which also weaken with increasing $a$. When the gravitational perturbation crosses the potential peak, part of it is reflected while the remainder propagates into the inner region. The transmitted component encounters the central barrier and is partially reflected again, giving rise to a sequence of echoes in the time-domain signal. The time delay between echoes, $\Delta t$, corresponds approximately to the round-trip time of a null ray trapped within this effective cavity and can thus be estimated from the separation between the barriers \cite{Magalhaes:2023xya}
\begin{equation}
\Delta t \simeq 2\,\Delta r^{*} ,
\end{equation}
where $\Delta r^{*}$ denotes the separation between the potential peaks in tortoise coordinates. In the first configuration [Fig. (\ref{Figure2}a)], we find $\Delta r^{*} \approx 39$, yielding $\Delta t \approx 78$, which is consistent with the time evolution shown in Fig. (\ref{Figure2}b). A similar behavior is observed in Figs. (\ref{Figure2}c) and (\ref{Figure2}e), where the amplitude of the inner barrier decreases but its position remains basically the same. In Fig. (\ref{Figure2}g), where the central barrier becomes essentially negligible, the separation between the remaining barriers increases to about $\Delta r^{*} \approx 59$, leading to $\Delta t \approx 118$, once again in good agreement with the time domain in Fig. (\ref{Figure2}h).

On the other hand, increasing the density while keeping the parameter $a$ fixed reduces the separation between the barriers [see Fig.~\ref{Figure3}]; as a consequence, the temporal spacing between successive echoes becomes shorter, producing signals that are more closely spaced in time and slightly attenuated. Adopting the same line of analysis as in the previous case, we observe that in Fig.~(\ref{Figure3}a) the separation between the potential barriers is approximately $\Delta  r^{*}  \approx 47.5$, yielding an estimated echo delay of $\Delta t \approx 95$, which is consistent with the time-domain evolution shown in Fig.~(3b). In Fig.~(\ref{Figure3}c), we find $\Delta  r^{*}  \approx 30$, leading to $\Delta t \approx 60$, again in good agreement with the waveform in Fig.~(\ref{Figure3}d). Finally, in Fig.~(\ref{Figure3}e), the cavity becomes more confined, with $\Delta  r^{*}  \approx 11$, corresponding to $\Delta t \approx 22$. In this latter case, the proximity of the potential barriers leads to significant wave interference, rendering the approximation inaccurate, as illustrated in Fig.~(\ref{Figure3}f). These results confirm the expected correlation between the width of the effective cavity and the temporal separation of the echoes.

Focusing now on the asymmetric solution (\ref{asymetric}), the configurations with horizons share essentially the same mode, as shown in Fig.~\ref{Figure4}, since the tortoise coordinate effectively maps $r_{H} \to -\infty$. Horizonless asymmetric configurations, on the other hand, exhibit regularized potentials, in contrast to the Reissner–Nordström case, and do not produce echoes, as illustrated in Fig.~\ref{Figure5}. In these cases, however, the ringdown phase is sensitive to the different geometries.

\subsection{Optical appearance}

To further explore how these behaviors can be related to or understood from the propagation of light, we now analyze null geodesics in the same parameter configurations for these spacetimes. Following the same sequence, Fig. \ref{Figure6} exhibits the shadow associated with a horizon configuration for the symmetric solution, where we recover the standard behavior with a single photon sphere, which remain very close to the classical Reissner-Nordström case, as confirmed by their transfer functions 
in Fig.~\ref{Figure7}, which encode the relation between the radial emission point $r_n$ and the impact parameter $b$. The effect produced by the distinct behavior of the potential near the origin in these cases is masked by the presence of the horizon.

As shown in Figs.~\ref{Figure8} and \ref{Figure9}, for horizonless configurations with fixed $\rho$ and varying $a$, smaller values of $a$ lead to a larger number of photon rings, while this number decreases as $a $ increases, as further illustrated by the logarithmic profiles in Fig.~\ref{Figure10}. These features can be understood by analyzing the corresponding ray-tracing structure. In Fig.~\ref{Figure11}, for $a = 0.6$, some photons reach $r = 0$ (black trajectories). Nearby trajectories are then strongly repelled by the high potential peak near the origin and turn back before intersecting the vertical axis (pink trajectories).  This behavior also extends to nearby trajectories, which are successively deflected so as to intersect the vertical axis once (green trajectories), twice (orange trajectories), three times (blue trajectories), four times (purple trajectories), five times (yellow trajectories), six times (red trajectories), and so on. At this stage, the photons approach the photon sphere associated with the critical impact parameter of the second barrier and undergo a large number of turns. Beyond this regime, one finds the lensing and direct images associated with the second barrier. \\
It is worth noting the unusual effect produced by light rays with increasing impact parameter in Fig.(\ref{Figure11}a). The smallest impact parameters (black lines) can reach the center, but those immediately above (pink) are scattered back, never crossing the vertical axis (where the accretion disk lies). Then green rays experience some repulsion but eventually feel attractive gravity and end up following concave trajectories. As the range of impact parameters is increased (orange lines in Fig.(\ref{Figure11}b)), we see that a small portion of the geometry remains in white color, meaning that no geodesic can reach it. This region is better shaped in the following figures, showing that a snail-like (or pacman) region remains inaccessible to light rays emitted with $b>0$. If the $b<0$ region is taken into account, the snail shape turns into a cashew nut (a typical nut from the Brazilian Northeast). This means that there is a portion of the space-time geometry which is not accessible to null geodesics. 

In Fig.~\ref{Figure12}, for $a = 1.1$, the barrier near the origin is lower, allowing a larger number of photons to reach the origin. As a consequence, the pink, green, and orange images associated with the first barrier are no longer present, as the corresponding trajectories now fall into the central region. Nevertheless, trajectories that intersect the vertical axis three times (blue), four times (purple), five times (yellow), and six times (red) still remain. As before, the photons then approach the photon sphere associated with the critical impact parameter of the second barrier and undergo a large number of turns. Beyond this regime, the direct and lensing images associated with the second barrier emerge.

Finally, in Fig.~\ref{Figure13}, for $a = 1.2$, the barrier near the origin becomes lower than the second barrier and no longer produces the effects identified in the previous cases. The standard behavior is then recovered: photons with impact parameters smaller than the critical value reach the origin, those with impact parameters close to the critical one undergo a large number of turns, and the corresponding lensing and direct images are subsequently formed. This behavior is illustrated by the transfer functions shown in Fig.~\ref{Figure16}. As $a$ increases, the branches associated with higher-order intersections progressively disappear, reflecting the loss of the additional families of null geodesics generated by the inner potential barrier. Consequently, the transfer-function structure becomes progressively simpler and approaches the standard single-barrier behavior.

This transition also explains the relative intensities reported in Table~\ref{TableI}. While the contribution of the direct image ($n=1$) increases from $86.90 \%$ to $94.07\%$, the contribution of higher-order images progressively decreases. The region in which the geodesics are strongly deflected without intersecting the vertical axis (pink trajectories) increases as $a$ decreases, as shown in Fig.~\ref{Figure14}. This behavior is a direct consequence of the increasing influence of the central potential barrier at smaller $a$, which reflects a broader set of trajectories before they can reach the central region. Accordingly, the corresponding opening angle $\theta_{\rm open}$ increases considerably as $a$ decreases, showing that the domain of strongly scattered trajectories becomes progressively larger.

This behavior can also be understood from the regions of forbidden geodesic motion shown in Fig.~\ref{Figure15}. From the radial geodesic equation, such regions occur whenever
\begin{equation}
\frac{1}{b^{2}}-V_{\rm sh}(r)<0,    
\end{equation}
for which radial motion is not allowed. For small $a$, the high central barrier generates a sizeable forbidden region in the $(r,b)$ plane. As $a$ increases, this region progressively shrinks and eventually disappears, consistently with the disappearance of the additional trajectory families associated with the inner barrier.

Fixing $a$ and increasing $\rho_0$, we observe that the two potential barriers move progressively closer until they eventually merge into a single one [see Fig.~(\ref{Figure17}g)], in agreement with the behavior found in the quasinormal-mode analysis. This merging modifies mainly the relative luminosity and contrast of the lensing and photon rings visible in Figs.~(\ref{Figure17}b)--(\ref{Figure17}h), rather than generating the sequence of additional rings observed when varying $a$. As the two maxima approach each other, the families of strongly deflected trajectories associated with them become less distinguishable, leading to a redistribution of the intensity among the different image orders. This behavior is also reflected in the transfer functions shown in Fig.~\ref{Figure18}, where the higher-order branches become increasingly concentrated within a narrower interval of impact parameters as the effective separation between the two barriers decreases.

In the asymmetric configurations with horizons, we recover the standard structure featuring a single photon sphere for both the bounded and unbounded asymmetric black-bounce geometries [see Fig.~\ref{Figure19}], which remain very close to the classical Reissner--Nordström case, as confirmed by their transfer functions in Fig.~\ref{Figure20}. In the horizonless case, however, the regularized inner region becomes accessible to null geodesics. Nevertheless, unlike the symmetric horizonless configurations, the asymmetric geometries do not develop the same multiple-barrier structure. Consequently, they do not generate a hierarchy of additional photon rings: depending on the parameters, the geometry may support a single photon ring or no photon ring at all. This behavior parallels the gravitational-perturbation sector, where the horizonless asymmetric configurations remain free of the echo structure characteristic of the symmetric case.

\subsection{CORRESPONDENCE}\label{seccorres}
Considering a static and spherically symmetric geometry, the authors in Ref. \cite{Cardoso:2008bp} established an analytical relation between the quasinormal modes and quantities associated with photon propagation. In what follows, we recall this construction taking as a starting point the line element (\ref{eq1}). The Lyapunov exponent characterizes the rate of divergence or convergence of nearby trajectories in phase space and, for the case of unstable circular geodesics, is given by \cite{Cardoso:2008bp}
\begin{equation}
\lambda = \sqrt{\frac{V_r^{''}(r_{ps})}{2\dot{t}^2}}; \quad V_r(r) = \left(\frac{dr}{d \bar{\lambda}} \right)^2.    
\end{equation}
 where $t$ is the coordinate time and $V_r$ is the potential governing radial motion. Considering circular null geodesics, $V_r(r_{ps})=V_r'(r_{ps})=0$, where $r_{ps}$ is the critical radius, and the conserved quantities $E = A(r) \dot{t}$ and $L = \Sigma(r)^2 \dot{\phi}$ we find that
\begin{equation}
\lambda  = \frac{1}{\sqrt{2}}\sqrt{\frac{A(r_{ps})\Sigma^2(r_{ps}) V''_r(r_{ps})}{L^2}}.    
\end{equation}
A straightforward derivation shows that the Lyapunov exponent can be written as
\begin{equation}\label{eq}
\lambda = \frac{1}{\sqrt{2}}\sqrt{\left.- \frac{\Sigma(r_{ps})^2}{A(r_{ps})}\frac{d^2}{dr^{*2}} \left(\frac{A(r)}{\Sigma(r)^2} \right) \right|_{r_{ps}}}.     
\end{equation}
On the other hand, the Wentzel–Kramers–Brillouin (WKB) approximation allows us to write the solution of Eq.~(\ref{eq3}) in the eikonal limit as
\begin{equation}
\frac{Q(r_0)}{\sqrt{2Q''(r_0)}} = i\left(n+\frac{1}{2}\right),
\end{equation}
where
\begin{equation}
Q(r) \approx w^2 - l^2\frac{A(r)}{\Sigma(r)^2}; \qquad Q''(r_0) \equiv \left.\frac{d^2Q}{dr^{*2}} \right|_{r=r_0},
\end{equation}
and $r_0$ denotes the position of the maximum of the effective potential. Since $r_0$ locates an extremum of the eikonal potential, it satisfies 
\begin{equation} 
A'(r_0)\Sigma(r_0) - 2A(r_0)\Sigma'(r_0) =0. 
\end{equation} 
This is precisely the condition defining a circular null geodesic. Therefore, the maximum of the eikonal potential coincides with the photon-sphere radius, $r_0=r_{ps}$. Expanding around $r_{ps}$, one can write 
\begin{equation}
 w_{\text{QNM}} = l \sqrt{\frac{A(r_{ps})}{\Sigma(r_{ps})^2}} - \frac{i}{\sqrt{2}}\left(n+\frac{1}{2} \right)\sqrt{\left.- \frac{\Sigma(r_{ps})^2}{A(r_{ps})}\frac{d^2}{dr^{*2}} \left(\frac{A(r)}{\Sigma(r)^2} \right) \right|_{r_{{ps}}}}.    
\end{equation}
Comparing with Eq.~(\ref{eq}), we finally obtain
\begin{equation}
w_{\text{QNM}} = \Omega_{{ps}} \ell- i \left( n + \tfrac{1}{2} \right) \lvert \lambda \rvert,    
\end{equation}
where $\Omega_{ps}$ is the angular velocity of a circular null geodesic. At this stage, it becomes clear why this analysis is not necessarily applicable to horizonless configurations, where the effective potential may develop multiple barriers, and the standard WKB approximation does not necessarily remain valid. Nevertheless, it is important to emphasize the strong similarity between the effective potentials and the dynamical responses found in both horizon and horizonless configurations for QNMs and shadows. This resemblance indicates that, although the same analytic argument cannot be straightforwardly extended to the horizonless case, a meaningful correspondence between the two phenomenologies may still exist.

\section{CONCLUSION}\label{secVII}
In this work we have analyzed axial gravitational perturbations and shadows of symmetric and asymmetric black-bounce geometries. For configurations with horizons, whether symmetric or asymmetric, the effective potential exhibits a single barrier, yielding quasinormal spectra nearly identical to those of classical black holes. This confirms that, in the presence of an event horizon, the regularization of the interior does not significantly affect the ringdown signal.

For horizonless configurations, however, the phenomenology changes considerably. In the symmetric case, the effective potential can develop multiple barriers, leading to the appearance of gravitational-wave echoes. The amplitude and separation of these echoes are controlled by the bounce parameter $a$ and the density parameter $\rho_0$: smaller values of $a$ generate additional peaks in the potential and longer-lived intermediate echoes, whereas increasing $\rho_0$ compresses the barriers and shortens the echo delay. On the other hand, asymmetric horizonless geometries remain free of echoes. Their potentials are regularized compared to the Reissner–Nordström case, and while the ringdown phase is sensitive to the internal geometry, the dominant modes are still observationally indistinguishable from those of their singular counterparts.

From the optical perspective, both symmetric and asymmetric configurations with horizons exhibit shadow potentials characterized by a single photon sphere, resulting in profiles nearly coincident with those of standard black holes. In the absence of horizons, symmetric configurations may support multiple photon rings whose number decreases with increasing $a$, while variations in $\rho_0$ primarily affect their relative luminosity and contrast. Asymmetric horizonless geometries, in contrast, may support either a single photon ring or none.

In symmetric configurations, geodesics with small impact parameters encounter an effective minimum radius that prevents them from probing certain regions of spacetime, while neighboring rays undergo severe deformations (Fig. \ref{Figure11}). This abrupt transition is a clear manifestation of a multi-peaked effective potential barrier, characterized by a sharp central peak flanked by two secondary maxima, which acts as a gravitational mirror. While green trajectories cross the critical threshold and wind deeply around the center, pink rays strike the steep central wall, reflecting into a tightly collimated beam. This phase-space separatrix disrupts the standard logarithmic deflection, creating a pronounced scattering asymmetry that translates into an abrupt brightness discontinuity and secondary photon rings with non-standard asymmetric magnification on the distant observer's image plane (Figs. \ref{Figure8}–\ref{Figure10}, \ref{Figure16}). Consequently, for small values of a (Figs. \ref{Figure10}a–\ref{Figure10}c), this collimated reflection generates a dense structure of nested concentric sub-rings around a distinct central shadow. As $a$ increases, the central maximum flattens and vanishes, allowing photons to penetrate or escape smoothly through the core. As a result, the resonant sub-ring patterns dissolve and the dark shadow expands into a broader, uniform region reminiscent of single-barrier geometries. Whether massive particles can access the  photon-excluded domains remains an open question for future work.

In the last part of the paper, by establishing the eikonal correspondence between the quasinormal frequencies and the Lyapunov exponent controlling the instability of circular photon orbits, we verified that both the ringdown and shadow sectors are governed by the same photon-sphere dynamics for configurations with horizons. For horizonless configurations, however, the emergence of multiple potential barriers prevents a straightforward extension of the standard eikonal correspondence, since the usual single-barrier WKB approximation is no longer directly applicable. Nevertheless, the close resemblance between the effective potentials and their corresponding dynamical responses in the gravitational and optical sectors suggests that a meaningful phenomenological connection between quasinormal modes and photon propagation may persist beyond the standard black-hole scenario.

\begin{acknowledgments}
A.C.L. Santos thanks the Coordenação de Aperfeiçoamento de Pessoal de Nível Superior (CAPES), Grants no 88887.822058/2023-00 and 88881.983410/2024-01, for financial support and the Department of Theoretical Physics $\&$ IFIC of the University of Valencia - CSIC for the kind hospitality during the elaboration of this work. This work is supported by the Spanish National Grant PID2023-149560NB-C21 and the Severo Ochoa Excellence Grant CEX2023-001292-S, funded by MICIU/AEI/10.13039/501100011033 (“ERDF A way
of making Europe”, “PGC Generacion de Conocimiento”) and FEDER, UE. L.A. Lessa is supported by CNPq/PDJ 151146/2025-0. R.V. Maluf would like to acknowledge Conselho Nacional de Desenvolvimento Científico e Tecnológico (CNPq), Grants PQ - 311393/2025-0.
\end{acknowledgments}

\end{document}